\documentclass[prd,floatfix,amsmath,amssymb,twocolumn,superscriptaddress,nofootinbib]{revtex4-2}

\usepackage{graphicx} 
\usepackage[pdftex,breaklinks,colorlinks,
linkcolor=Blue,
citecolor=teal,
anchorcolor=red,
urlcolor=cyan,
pdfencoding=auto]{hyperref}
\usepackage[dvipsnames]{xcolor}
\usepackage{accents}
\usepackage{physics}
\usepackage{mathtools}
\usepackage{bm}
\usepackage{mathrsfs}
\usepackage[scr=boondoxo]{mathalpha}
\usepackage{comment}

\newcommand{\beq}{\begin{equation}}
\newcommand{\eeq}{\end{equation}}

\newcommand{\Jring}{\mathring{J}}
\newcommand{\Ering}{\mathring{E}}
\newcommand{\piring}{\mathring{\pi}}
\newcommand{\phiring}{\mathring{\varphi}}

\newcommand{\tilt}{\tilde{t}}
\newcommand{\Jstar}{\accentset{\star}{J}}
\newcommand{\Omegastar}{\accentset{\star}{\Omega}}

\newcommand{\Pstar}{\accentset{\star}{P}}

\newcommand{\phistar}{\accentset{\star}{\varphi}}

\newcommand{\Estar}{\accentset{\star}{E}}
\newcommand{\Lstar}{\accentset{\star}{L}_z}
\newcommand{\Kstar}{\accentset{\star}{K}}

\newcommand{\utstar}{\accentset{\star}{u}^t}

\newcommand{\e}{\varepsilon}
\newcommand{\psio}{\psi_{(0)}}

\newcommand{\psivec}{\boldsymbol{\psi}}

\newcommand{\phiringvec}{\mathring{\boldsymbol{\varphi}}}
\newcommand{\phistarvec}{\accentset{\star}{\boldsymbol{\varphi}}}

\newcommand{\Hsix}{\mathcal{H}}

\renewcommand{\O}{\mathcal{O}}

\begin{document}

\title{Post-adiabatic dynamics and waveform generation in self-force theory:\\an invariant pseudo-Hamiltonian framework}

\author{Jack Lewis}
\affiliation{School of Mathematical Sciences and STAG Research Centre, University of Southampton, Southampton, United Kingdom, SO17 1BJ}
\author{Takafumi Kakehi}
\affiliation{Department of Physics, Kyoto University, Kyoto 606-8502, Japan}
\affiliation{Center for Gravitational Physics and Quantum Information, Yukawa Institute for Theoretical Physics, Kyoto University, Kyoto 606-8502, Japan}
\author{Adam Pound}
\affiliation{School of Mathematical Sciences and STAG Research Centre, University of Southampton, Southampton, United Kingdom, SO17 1BJ}
\author{Takahiro Tanaka}
\affiliation{Department of Physics, Kyoto University, Kyoto 606-8502, Japan}
\affiliation{Center for Gravitational Physics and Quantum Information, Yukawa Institute for Theoretical Physics, Kyoto University, Kyoto 606-8502, Japan}

\begin{abstract}
    Gravitational waveform modeling in self-force theory has reached a mature stage in recent years, with fast and accurate models emerging at both adiabatic (0PA) and first post-adiabatic (1PA) orders in a multiscale expansion. Here, we provide a gauge-invariant 1PA waveform-generation framework that involves no direct calculation of the (gauge-dependent) self-force. To achieve this, we recast the multiscale framework in a pseudo-Hamiltonian form, working on the six-dimensional phase space intrinsic to the multiscale expansion. We characterize the gauge freedom on phase space and show how a localization procedure avoids nonlocal-in-time effects in the 1PA dynamics. We find a conservative Hamiltonian structure can be naturally embedded into the complete, dissipative~1PA pseudo-Hamiltonian dynamics, giving rise to natural definitions of the conserved energy, angular momentum, and radial and polar actions. As a byproduct, we clarify that the on-shell value of the conservative Hamiltonian is equal to the mechanical energy historically predicted by the first law of binary black hole mechanics. 
\end{abstract}

\date{\today}

\maketitle

\tableofcontents

\section{Introduction}

Over the past two decades, a successful paradigm of gravitational-wave source modelling has emerged~\cite{LISAConsortiumWaveformWorkingGroup:2023arg}: post-Newtonian theory~\cite{Blanchet:2013haa} is combined with data from numerical relativity~\cite{Baumgarte:2010ndz} to build fast surrogate~\cite{Varma:2019csw}, phenomenological~\cite{Thompson:2023ase}, and effective-one-body (EOB)~\cite{Nagar:2021gss,Ramos-Buades:2023ehm} models that are able to rapidly generate accurate waveform templates for most currently observable compact binaries. However, systematic modeling errors are already a limiting factor for analysis of signals in regions of parameter space where fast models are not well calibrated~\cite{Dhani:2024jja}, and these systematics will become more severe for the next generation of detectors~\cite{Purrer:2019jcp,Hu:2022rjq,Hu:2022bji,Jan:2023raq,Owen:2023mid,Kapil:2024zdn}. This is especially true for asymmetric binaries in which one object is much smaller than the other~\cite{LISAConsortiumWaveformWorkingGroup:2023arg}; for such systems, post-Newtonian theory has limited accuracy~\cite{Isoyama:2021jjd} and numerical relativity simulations become intractable~\cite{Dhesi:2021yje}. 

The prime examples of asymmetric binaries are extreme-mass-ratio inspirals (EMRIs), which are key targets for space-based detectors such as LISA~\cite{Colpi:2024xhw}. In parallel with the development of models based on post-Newtonian theory and numerical relativity, fast and accurate EMRI models based on self-force theory~\cite{Barack:2018yvs,Pound:2021qin} have been developed~\cite{Chua:2020stf,Katz:2021yft,Wardell:2021fyy,Speri:2023jte,Burke:2023lno,Nasipak:2023kuf,Khalvati:2024tzz}. These models, which are formulated in terms of small-mass-ratio expansions, are unlike the prevailing modeling paradigm in that their mechanism for rapid waveform generation follows from a first-principles method of solving the Einstein equations, rather than requiring a secondary layer of effective or surrogate modeling.\footnote{However, surrogates based on self-force waveforms have nevertheless been devised for small-mass-ratio systems~\cite{Rifat:2019ltp,Islam:2022laz}. EOB models, which capture information about the test-particle limit by construction~\cite{Buonanno:1998gg,Damour:2007xr,Taracchini:2014zpa}, have also incorporated information from self-force theory~\cite{Damour:2009sm,Akcay:2012ea,Antonelli:2019fmq,vandeMeent:2023ols}, particularly when targeting EMRIs~\cite{Nagar:2022fep}.} 

This native waveform-generation framework is facilitated by the quasi-periodicity and separation of time scales in an asymmetric binary: the system exhibits tri-periodicity on the fast, orbital time scale, with three distinct periods of motion that slowly evolve over the much longer radiation-reaction time scale. This quasi-periodicity makes it possible to formulate the small-mass-ratio expansion of the Einstein equations in a multiscale form~\cite{Hinderer:2008dm,Miller:2020bft,Pound:2021qin,Miller:2023ers,Mathews:2025nyb,Wei:2025lva}, which divides the problem into triperiodic field equations on the fast, orbital time scale, coupled to ordinary differential equations (ODEs) governing the system's time evolution~\cite{Miller:2020bft,Pound:2021qin,Mathews:2025nyb}. The fast-time field equations can be solved on a grid of binary parameter values as an offline step. The online waveform generation then comprises a rapid evolution through the binary phase space, which can be performed with sufficient speed for gravitational wave data analysis~\cite{Chua:2020stf,Katz:2021yft}. 

Our overarching aim in this paper is to cast this multiscale construction in a (pseudo)-Hamiltonian form.

\subsection{Multiscale waveform generation}

In the multiscale framework, the waveform inherits the quasi-periodicity of the binary dynamics. Each spherical-harmonic mode of the waveform is then a sum of oscillatory factors with slowly varying amplitudes~\cite{Pound:2021qin}:\footnote{Our notation here differs from Ref.~\cite{Pound:2021qin} in two ways: $\phiring^i$ is denoted $\varphi_i$ in Ref.~\cite{Pound:2021qin}, and $\piring_i$ is denoted $p^i_\varphi$ there.}
\begin{multline}\label{multiscale waveform}
    h_{lm} = \sum_{\bm{k}\in\mathbb{Z}^2}\Bigl[\e \mathring h^{(1)}_{lm\bm{k}}(\piring_i) \\
    +\e^2\mathring h^{(2)}_{lm\bm{k}}(\piring_i,\delta M_A) + \O(\e^3)\Bigr]e^{-i k_i\phiring^i},
\end{multline}
where $\e:=1$ counts powers of the small mass ratio. Here $\phiring^i=(\phiring^r,\phiring^\theta,\phiring^\phi)$ represent the radial, polar, and azimuthal phases of the smaller, secondary object's orbit around the primary black hole, and the mode numbers $k_i=(k_r,k_\theta,k_\phi)$ are divided into the azimuthal mode number $k_\phi=m$ and $\bm{k}=(k_r,k_\theta)$. $\piring_i=(\mathring{p},\mathring{e},\mathring{\iota})$ are a set of three independent, slowly evolving orbital parameters (semi-latus rectum $\mathring{p}$, eccentricity $\mathring{e}$, and maximum inclination $\mathring{\iota}$), and 
$\delta M_A=(\delta M, \delta S)$ are small, evolving corrections to the primary black hole's mass and spin; we suppress dependence on the black hole's constant, background mass and spin $M$ and $S$. The waveform's time dependence is then governed by a set of simple ODEs describing the orbit,
\begin{align}
    \frac{d\phiring^i}{dt} &= \Omega^i_{(0)}(\piring_k) + \e\Omega^i_{(1)}(\piring_k,\delta M_A) + \O(\e^2),\label{dphidt}\\
    \frac{d\piring_i}{dt} &= \e\left[F_i^{(0)}(\piring_k) + \e F_i^{(1)}(\piring_k,\delta M_A) + \O(\e^2)\right],\label{dpidt}
\end{align}
alongside evolution equations for the perturbations of the primary black hole parameters, 
\beq\label{dMdt}
\frac{d\delta M_A}{dt} = \e F^{(1)}_A(\piring_k) + \O(\e^2).
\eeq
Here and throughout this paper, we assume a nonspinning secondary; see Refs.~\cite{Skoupy:2023lih,Drummond:2023wqc,Piovano:2024yks,Mathews:2025nyb}, for example, for the addition of secondary spin.

The accuracy of the multiscale waveform is typically assessed by changing to ``slow time'' $\tilde t:=\e t$ as the independent variable in Eqs.~\eqref{dphidt}--\eqref{dMdt}. Doing so, we immediately see the orbital phases (and hence the waveform phases) admit an asymptotic expansion of the form
\beq\label{phase expansion}
\phiring^i = \frac{1}{\e}\left[\phiring^i_{(0)}(\tilde t) + \e \phiring^i_{(1)}(\tilde t) + \O(\e^2)\right].
\eeq
A leading-order approximation to the evolution, comprising $\Omega^i_{(0)}$ and $F_i^{(0)}$, is referred to as adiabatic (0PA); this suffices to capture the leading-order phase $\phiring^i_{(0)}$. An approximation that includes subleading terms through to $\Omega^i_{(n)}$, $F_i^{(n)}$, and $F^{(n)}_A$ is referred to as an $n$th post-adiabatic order ($n$PA) approximation; this captures the phase through to $\phiring^i_{(n)}$. 

Note that in Eqs.~\eqref{dphidt}--\eqref{dMdt}, numerical labels correspond to the post-adiabatic order at which terms enter. In Eq.~\eqref{multiscale waveform}, on the other hand, numeric labels correspond to the order at which the Einstein equations are solved. The latter corresponds to the common `$n$SF' order counting: from the solution to the first-order field equation (linear in $\e$), one can calculate the first-order self-force; from the solution to the second-order field equation (quadratic in $\e$), one can calculate the second-order self-force; and so on.

Work toward a 0PA model for fully generic, precessing and eccentric orbits around a Kerr black hole is well underway~\cite{Fujita:2020zxe,Isoyama:2021jjd,Hughes:2021exa}, and such a model should soon be available as part of the open-source FastEMRIWaveforms software package~\cite{Katz:2021yft}. However, we can see from Eq.~\eqref{phase expansion} that a 0PA model incurs significant, $\O(\e^0)$ errors in the phases $\phiring^i$, implying 1PA precision is required for EMRI science~\cite{Burke:2023lno}. 

As of this writing, 1PA models are much more limited than 0PA ones: the most generic 1PA model available is restricted to quasicircular orbits of a spinning secondary around a slowly spinning primary~\cite{Wardell:2021fyy,Mathews:2024}. One of our goals in this paper is to help expedite the construction of 1PA models for more generic binary configurations.

\subsection{Pseudo-Hamiltonian methods, balance laws, and conserved quantities}

At 0PA order in the multiscale expansion, a `pseudo-Hamiltonian' description~\cite{Fujita:2016igj,Isoyama:2018sib} of the 1SF dynamics proved to be a powerful tool in streamlining the formalism and deriving practical, efficient formulas for the orbital evolution.\footnote{Here we follow Ref.~\cite{Blanco:2022mgd}'s terminology in distinguishing between a Hamiltonian and a pseudo-Hamiltonian.} Unlike an ordinary Hamiltonian, which is a simple function on phase space and is restricted to conservative dynamics, the pseudo-Hamiltonian is a function of multiple points in phase space, or of a point and a curve in phase space, and it encodes the complete dynamics, including all dissipative effects. Our overarching aim in this paper is to extend the pseudo-Hamiltonian description to 1PA order. 

The pseudo-Hamiltonian formalism was particularly advantageous in simplifying derivations of 0PA `balance laws'. Here by `balance law' we mean a practical formula for $F^{(0)}_i$ in terms of field amplitudes at the black hole horizon and future null infinity, regardless of whether the combination of field amplitudes has a physical or geometrical interpretation as a flux (though see Ref.~\cite{Grant:2024ivt} for work on establishing such an interpretation). These balance laws have been a key tool in building 0PA waveform models. Their derivation has been based on an important simplification: the forcing function $F_i^{(0)}$ can be calculated solely from the dissipative first-order-in-$\e$ self-force~\cite{Hinderer:2008dm}. This in turn implies $F_i^{(0)}$ can be computed from the radiative, time-antisymmetric modes of the first-order-in-$\e$ metric perturbation~\cite{Mino:2003yg}. As a consequence, $F_i^{(0)}$ can be computed directly from the solution to the first-order Teukolsky equation~\cite{Sago:2005fn} (see also~\cite{Galtsov:1982hwm,Akcay:2019bvk}), without requiring the reconstruction of the complete first-order metric perturbation~\cite{Merlin:2016boc,vandeMeent:2017bcc,Toomani:2021jlo,Wardell:2024yoi} or the extraction of the regular field that exerts the complete (conservative plus dissipative) first-order self-force~\cite{Barack:2009ux,Barack:2018yvs}. 

The original derivation of $F^{(0)}_i$ in terms of Teukolsky mode amplitudes, in Ref.~\cite{Sago:2005fn}, did not utilize a pseudo-Hamiltonian. However, Ref.~\cite{Isoyama:2018sib} showed the pseudo-Hamiltonian formalism enables a radically simplified derivation. Specifically, deriving a `balance law' for the Carter constant $K$ is far simpler when using a pseudo-Hamiltonian method (while the evolution of $E$ and $L_z$ can be derived equally straightforwardly from other methods~\cite{Akcay:2019bvk}). The forcing function $F^{(0)}_i(\piring_j)$ for $\piring_i$ is then obtained from the relationship between $\piring_i$ and $P_i=(E,L_z,K)$; see, e.g., Ref.~\cite{Pound:2021qin} for that relationship.

We expect similar simplifications to arise at 1PA order. Calculating the 1PA functions $\Omega^i_{(1)}$ and $F_i^{(1)}$ requires the complete first-order self-force but only the dissipative part of the second-order self-force~\cite{Hinderer:2008dm,Pound:2021qin}. It might therefore be possible to compute the necessary second-order contributions directly from a solution to the second-order Teukolsky equation, in analogy with the results at 0PA~\cite{Spiers:2023cip}. The pseudo-Hamiltonian framework offers a powerful tool for investigating this possibility. Our work in this paper should lay the necessary groundwork for that investigation.
 
A pseudo-Hamiltonian formalism is also a natural setting to explore the gauge freedom in the multiscale expansion of the Einstein equations. The simple waveform structure~\eqref{multiscale waveform} emerges from a clean separation,  at the level of the orbital equations of motion~\eqref{dphidt}--\eqref{dpidt}, between the system's oscillatory behavior and its secular evolution. At the level of the orbit, the separation is achieved by transforming to action-angle-type variables using the method of (near-identity) averaging transformations~\cite{VanDeMeent:2018cgn,Pound:2021qin,Lynch:2021ogr,Lynch:2023gpu,Mathews:2025nyb}. These transformations are best understood in terms of the orbital phase space, which is a symplectic manifold that can be studied with (pseudo-)Hamiltonian methods. A primary goal of this paper is to clarify how the multiscale expansion meshes with a pseudo-Hamiltonian description on phase space and to characterize the multiscale framework's gauge freedom on that space (which, we emphasise, is distinct from the usual freedom to perform small diffeomorphisms on spacetime). Our discussion of this gauge freedom complements and extends the recent analysis in Ref.~\cite{Mathews:2025nyb}.

In our study of phase-space gauge freedom, we highlight the utility of action-angle-type variables as a means of localizing the pseudo-Hamiltonian. The pseudo-Hamiltonian is constructed from the metric perturbation sourced by the particle, which generically depends on the particle's entire past history~\cite{Poisson:2011nh,Barack:2018yvs}. This leads to a pseudo-Hamiltonian that is a nonlocal-in-time function on phase space, akin to the nonlocal Hamiltonian encountered in post-Newtonian theory~\cite{Damour:2014jta,Galley:2015kus,Damour:2016abl,Bernard:2016wrg,Bernard:2017ktp,Blanchet:2017rcn}. However, using action-angle variables, together with the stationary phase approximation, allows us to eliminate this nonlocal character, in a manner reminiscent of Ref.~\cite{Damour:2016abl}.

This localization also allows us to easily construct an ordinary Hamiltonian for the conservative dynamics, in turn allowing us to identify the mechanical energy and angular momentum, which had not (to our knowledge) been highlighted previously in self-force theory. Our result for the 1SF conservative Hamiltonian is equivalent to earlier ones~\cite{Fujita:2016igj,Blanco:2022mgd}, but we believe our method of deriving it is illuminating. In particular, we derive it consistently within the complete, dissipative dynamics,  through 1PA order, showing how it is embedded within the full 1PA evolution equations. Along the same lines, we also explain how to define and work with gauge-invariant action variables in the presence of dissipation. 

Our analysis additionally makes contact with longstanding questions related to the first law of binary black hole mechanics~\cite{LeTiec:2011ab,Gralla:2012dm,Blanchet:2012at,LeTiec:2013iux,LeTiec:2015kgg,Fujita:2016igj,Blanchet:2017rcn,Ramond:2022ctc,Gonzo:2024xjk}. Specifically, we emphasise that the on-shell value of the 1SF Hamiltonian is precisely the binding energy predicted by the first law~\cite{LeTiec:2011dp,Isoyama:2014mja}, which was used to complete the fourth-post-Newtonian-order conservative binary dynamics~\cite{Bini:2013zaa} and inform EOB~\cite{Damour:2012ky}, among other applications; see the review in~\cite{LISAConsortiumWaveformWorkingGroup:2023arg}. The first-law binding energy is also currently used as a core input for the only extant 1PA waveform model~\cite{Wardell:2021fyy,Albertini:2022rfe}. However, numerical calculations of the Bondi mass suggest that the binding energy defined from the Bondi mass differs slightly from the first-law value~\cite{Pound:2019lzj}, and there has been lingering uncertainty as to whether the first-law binding energy is valid in the presence of dissipation and whether it satisfies a balance law equivalent to the Bondi mass-loss formula~\cite{Madler:2016xju}. Recent~\cite{Trestini:2025nzr} and forthcoming~\cite{Grant:FirstLaw} work establish that, in fact, the binding energy defined from the Bondi mass \emph{must} disagree with the first-law binding energy, and that the energy-balance arguments used in 1PA waveform generation must consequently be modified. Our analysis further illuminates this breakdown of energy-balance arguments, which we return to in the Conclusion.

By enabling computations of the 1SF mechanical energy, angular momentum, and radial and polar actions, we also bring self-force theory into a form more similar to other standard approaches to the two-body problem, particularly post-Newtonian, post-Minkowskian, and EOB theory.

Finally, our derivation of the complete 1PA dynamics in terms of invariant action angles allows us to reformulate the waveform generation scheme~\eqref{multiscale waveform}--\eqref{dMdt} directly in terms of invariant quantities that are geometrically defined on the binary phase space.

\subsection{Outline}

We begin in Sec.~\ref{sec:review} with descriptions of self-force theory at second order and the multiscale expansion of the Einstein equations. Our summary differs from prior versions in that it is adapted to a phase-space description of the two-body problem. It also provides a more satisfactory treatment of the primary black hole's evolution. In Sec.~\ref{sec:pseudo-Hamiltonian} we formulate the 1PA dynamics in pseudo-Hamiltonian form. Our formulation differs from prior ones~\cite{Fujita:2016igj,Isoyama:2018sib,Blanco:2022mgd} in order to better mesh with the multiscale expansion of the field equations. Section~\ref{sec:localized Hamiltonian for bound orbits} explores the gauge freedom on phase space within the class of gauges compatible with a multiscale expansion, and in that class of gauges it derives a local pseudo-Hamiltonian description of the dynamics, eliminating nonlocal phase-space dependence. Section~\ref{sec:conservative sector} shows how a conservative 1SF Hamiltonian structure is embedded in the full 1PA dynamics, with a natural conserved energy, angular momentum, and action variables. Section~\ref{sec:gauge choices} highlights two useful gauge choices and summarizes the gauge-invariant 1PA waveform generation. Some technical details are relegated to appendices.

Throughout the paper, we use geometric units with $G=c=1$ and a mostly positive metric signature $(-+++)$.


\section{Second-order self-force theory and 1PA waveform generation}\label{sec:review}

Before presenting the pseudo-Hamiltonian formalism, we summarize self-force theory at second perturbative order. We begin with a modified version of the theory's self-consistent formulation~\cite{Pound:2009sm,Pound:2015tma,Miller:2020bft}. Although this contains a large amount of review material, it provides an essential update to the self-consistent formulation by recasting it in a phase-space description. It additionally provides a robust justification for the multiscale expansion of the Einstein equations. We conclude the section by outlining that multiscale expansion and the resulting 1PA waveform-generation framework.

\subsection{Self-consistent self-force in a phase-space formulation}\label{sec:self-consistent formalism}

\subsubsection{A particle coupled to gravity}

As usual in self-force calculations, we consider the secondary, orbiting mass $\mu$ as the source of perturbations on the background Kerr metric $g_{\alpha\beta}$ of the primary black hole, of mass $M$, leading to a metric of the form
\begin{equation}\label{g expansion}
    {\sf g}_{\alpha\beta} = g_{\alpha\beta} + \e h^{(1)}_{\alpha\beta} + \e^2 h^{(2)}_{\alpha\beta} + {\cal O}(\e^3).
\end{equation}
At least through second perturbative order, the secondary can be treated as a point particle~\cite{Upton:2021oxf}, and the perturbations $h^{(n)}_{\alpha\beta}$ diverge at its position. However, assuming the secondary is nonspinning, it obeys the geodesic equation in a different, \emph{effective} metric that is smooth at its position~\cite{Detweiler:2002mi,Pound:2012nt,Harte:2011ku,Pound:2017psq}:
\begin{equation}\label{g effective}
    \tilde g_{\alpha\beta} = g_{\alpha\beta} + \e h^{{\rm R}(1)}_{\alpha\beta} + \e^2 h^{{\rm R}(2)}_{\alpha\beta} +{\cal O}(\e^3),
\end{equation}
where the regular perturbations $h^{{\rm R}(n)}_{\alpha\beta}$ are smooth, and where $\tilde g_{\alpha\beta}$ satisfies the vacuum Einstein equation~\cite{Pound:2012dk}.  
Explicitly, the particle's trajectory, which we write in coordinates as $x^\alpha_p$, satisfies the geodesic equation 
\begin{equation}\label{EOMv2}
   \tilde u^\beta\tilde\nabla_\beta\tilde u^\alpha ={\cal O}(\e^3).
\end{equation}
Here the four-velocity $\tilde u^\alpha:=dx^\alpha_p/d\tilde\tau$, proper time $\tilde\tau$, and covariant derivative $\tilde\nabla_\alpha$ are all defined with respect to the effective metric $\tilde g_{\alpha\beta}$. 

Equation~\eqref{EOMv2} can be equivalently written as the equation of an accelerated curve in the background spacetime~\cite{Pound:2015tma},
\begin{align}
    u^\beta\nabla_\beta u^\alpha &= -\frac{1}{2}P^{\alpha\mu}\left(g_{\mu}{}^\beta-h^{{\rm R}\; \beta}_{\mu}\right)\!\left(2\nabla_\gamma h^{\rm R}_{\delta\beta}-\nabla_\beta h^{\rm R}_{\gamma\delta}\right)u^\gamma u^\delta \nonumber\\
    &=:f^\alpha,\label{EOM}
\end{align}
omitting the order symbol ${\cal O}(\e^3)$ for brevity. Here $u^\alpha:=dx^\alpha_p/d\tau$, and proper time $\tau$, the covariant derivative $\nabla_\alpha$, and index raising are all defined with respect to $g_{\alpha\beta}$ and its inverse $g^{\alpha\beta}$. $P^{\alpha\beta}:=g^{\alpha\beta}+u^\alpha u^\beta$ projects orthogonally to the trajectory, and the right-hand side of the equation, which we have denoted $f^\alpha$, is referred to as the self-force (per unit mass $\mu$). For clarity, we will always use $g^{\alpha\beta}$, rather than the inverse of $\tilde g_{\alpha\beta}$, to raise indices, and $g_{\alpha\beta}$ to lower them. We have also defined $ h^{\rm R}_{\alpha\beta}:=\sum_n \e^n  h^{{\rm R}(n)}_{\alpha\beta}$ as the total regular field.

Just as the particle's trajectory is geodesic in the effective metric, its stress-energy tensor is also that of a point mass in the effective metric, referred to as the Detweiler stress-energy tensor~\cite{Detweiler:2011tt,Upton:2021oxf}:
\begin{align}\label{Detweiler T}
    T_{\alpha\beta} = \frac{1}{\mu} \int_\gamma \tilde p_\alpha \tilde p_\beta\, \frac{\delta^4(x^\mu-x^\mu_p(\tilde\tau))}{\sqrt{-\tilde g}}d\tilde\tau.
\end{align}    
Here we have introduced the particle's momentum
\beq
\tilde p_\alpha := \mu\,\tilde g_{\alpha\beta}\tilde u^\beta
\eeq
and $\tilde g$ as the determinant of $\tilde g_{\alpha\beta}$. We deliberately write $T_{\alpha\beta}$ in terms of momentum $\tilde p_\alpha$ rather than four-velocity $\tilde u^\alpha$ in order to introduce the particle's eight-dimensional (8D) mechanical phase space (the spacetime manifold's cotangent bundle), on which $(x^\alpha_p,\tilde p_\alpha)$ are cotangent coordinates~\cite{CannasdaSilva}. The particle's trajectory $\gamma$, which is integrated along in Eq.~\eqref{Detweiler T}, is then a phase-space trajectory,
\beq
\gamma:=\{(x^\alpha_p(\tilde\tau),\tilde p_\alpha(\tilde\tau))|\tilde\tau\in\mathbb{R}\}.
\eeq
Since the phase space is a cotangent bundle, it is automatically a symplectic manifold with the canonical symplectic form $dx^\alpha_p\wedge d\tilde p_\alpha$~\cite{CannasdaSilva}. 

The metric perturbations $h^{(n)}_{\alpha\beta}$ necessarily depend on the particle's trajectory that sources them. In the traditional self-consistent expansion~\cite{Pound:2009sm}, one treats each of the metric perturbations $h^{(n)}_{\alpha\beta}$ as functionals of the \emph{exact}, $\e$-dependent trajectory satisfying Eq.~\eqref{EOM}, without ever series-expanding the trajectory in powers of $\e$. Here, we specifically treat them as functions of the \emph{phase-space} trajectory: $h^{(n)}_{\alpha\beta}=h^{(n)}_{\alpha\beta}(x^\mu;\gamma)$. If we consider the exact metric as a function of $x^\mu$, $\e$, and $\gamma$, then the expansion in Eq.~\eqref{g expansion} represents an expansion in powers of $\e$ at fixed values of both the spacetime coordinates and the phase-space trajectory. This differs slightly (but meaningfully) from traditional descriptions of the self-consistent formulation, where $x^\alpha_p(\tau)$ and all its $\tau$ derivatives would be held fixed in the small-$\e$ expansion.

To obtain more concrete equations for $h^{(n)}_{\alpha\beta}$, we impose the Lorenz gauge condition,
\begin{equation}\label{Lorenz gauge}
\nabla_\beta\bar h^{\alpha\beta} = 0,
\end{equation}
where $\bar h_{\alpha\beta}:= h_{\alpha\beta}-\frac{1}{2}g_{\alpha\beta}g^{\mu\nu}h_{\mu\nu}$ is the trace-reversed metric perturbation, and we have introduced the total field $h_{\alpha\beta}:=\sum_n \e^n h^{(n)}_{\alpha\beta}$. We will recall the motivation for this gauge-fixing momentarily. First, we note that the Einstein equation 
\beq
G_{\alpha\beta}[g+h]=8\pi T_{\alpha\beta}, 
\eeq
when gauge fixed and expanded in powers of the field $h_{\alpha\beta}$, takes the form~\cite{Pound:2009sm}%
\begin{multline}\label{full EFE}%
-\frac{1}{2}E_{\alpha\beta}[\bar h] + \delta^2G_{\alpha\beta}[h,h]\\ + \delta^3G_{\alpha\beta}[h,h,h] 
    + \ldots 
    = 8\pi T_{\alpha\beta},
\end{multline}
where we have used the background, vacuum field equation $G_{\alpha\beta}[g]=0$, and we have defined 
\beq
E_{\alpha\beta}[\bar h]:=\Box \bar h_{\alpha\beta} + 2 R_\alpha{}^\mu{}_\beta{}^\nu \bar h_{\mu\nu}.
\eeq
The first term in Eq.~\eqref{full EFE} is the linearized Einstein tensor in the Lorenz gauge, $\delta^2 G_{\alpha\beta}$ is quadratic in the perturbation, and so on. The quantities $\Box:=g^{\mu\nu}\nabla_\mu\nabla_\nu$ and $R_{\alpha\mu\beta\nu}$ are the d'Alembertian and Riemann tensor of the background metric. 

Equation~\eqref{full EFE} is referred to as a \emph{relaxed} Einstein equation because it can be solved for any $T_{\alpha\beta}$, meaning in particular that it can be solved for any particle trajectory $\gamma$. This contrasts with the ordinary, non-gauge-fixed Einstein equation, which constrains matter degrees of freedom by virtue of the Bianchi identity. Since Eq.~\eqref{full EFE} is defined for all $\gamma$, we are able to define the expansion $h_{\alpha\beta}=\sum_n \e^n h^{(n)}_{\alpha\beta}(x^\mu;\gamma)$ for all $\gamma$. This in turn implies that we can equate coefficients of explicit powers of $\e$ in Eq.~\eqref{full EFE}, dividing it into a sequence of equations for subsequent $h^{(n)}_{\alpha\beta}$:
\begin{align}
    E_{\alpha\beta}[\bar h^{(1)}] &= -16\pi T^{(1)}_{\alpha\beta},\label{EFE1}\\
    E_{\alpha\beta}[\bar h^{(2)}] &= -16\pi T^{(2)}_{\alpha\beta} + 
    2\delta^2 G_{\alpha\beta}[h^{(1)},h^{(1)}].\label{EFE2}
\end{align}
 Here
\begin{align}
     T^{(1)}_{\alpha\beta} &= \frac{1}{\mu} \int_\gamma \tilde p_\alpha \tilde p_\beta\, \frac{\delta^4(x^\mu-x^\mu_p(\tilde\tau))}{\sqrt{-g}}d\tilde\tau,\\
     T^{(2)}_{\alpha\beta} &= -\frac{1}{2\mu} \int_\gamma \tilde p_\alpha \tilde p_\beta\, h_{{\rm R}(1)}^{\gamma\delta}g_{\gamma\delta}\frac{\delta^4(x^\mu-x^\mu_p(\tilde\tau))}{\sqrt{-g}}d\tilde\tau
\end{align}
are the coefficients in the expansion of the Detweiler stress-energy tensor~\eqref{Detweiler T} at fixed $\gamma$, which follows from the expansion of $\sqrt{-\tilde g}$.

Due to the divergence of $h^{(1)}_{\alpha\beta}$ on the particle's worldline, the quadratic source $\delta^2G_{\alpha\beta}$ in Eq.~\eqref{EFE2} is highly singular and does not have a unique definition as a distribution. However, Ref.~\cite{Upton:2021oxf} provided a canonical distributional definition of $\delta^2G_{\alpha\beta}$ such that the retarded solution to Eq.~\eqref{EFE2} yields the correct, physical metric. We implicitly adopt that definition here.

\subsubsection{Retarded solution}

The retarded solution to the first-order field equation~\eqref{EFE1} can be written in terms of a retarded Green's function,
\begin{equation}\label{eq:hpp}
   h^{\rm pp}_{\alpha\beta}(x;\gamma) = \frac{1}{\mu}\int_\gamma G^{\rm ret}_{\alpha\beta}{}^{\alpha'\beta'}(x,x')\tilde p_{\alpha'}\tilde p_{\beta'}d\tilde \tau,
\end{equation}
with primed quantities evaluated at $x'^\mu=x^\mu_p(\tilde\tau)$. For convenience, we have adopted the convention
\begin{equation}
    \Box\bar G^{\rm ret}_{\alpha\beta}{}^{\,\alpha'\beta'}+2R_\alpha{}^\mu{}_\beta{}^\nu \bar G^{\rm ret}_{\mu\nu}{}^{\,\alpha'\beta'} = -16\pi \delta_{\alpha}^{\alpha'}\delta_{\beta}^{\beta'} \frac{\delta^4(x^\rho-x'^\rho)}{\sqrt{-g}},\label{Greens}
\end{equation}
where a bar denotes trace reversal on the unprimed indices. From this point onward, we emphasize that arguments after a semicolon, such as $\gamma$, denote a nonlocal functional dependence on a function of time.

Traditionally, a solution of the type~\eqref{eq:hpp} has been taken to be the entirety of $h^{(1)}_{\alpha\beta}$. However, during the binary's evolution, the primary black hole evolves due to absorption of radiation. As discussed in Ref.~\cite{Miller:2020bft}, this implies we should add to $h^{\rm pp}_{\alpha\beta}$ a perturbation $h^{\rm BH}_{\alpha\beta}$ representing the linear contribution from the black hole's evolution. We then write the complete solution to Eq.~\eqref{EFE1} as
\begin{equation}
    h^{(1)}_{\alpha\beta}(x;\gamma,\delta M_A) = h^{\rm pp}_{\alpha\beta}(x;\gamma) + h^{\rm BH}_{\alpha\beta}(x;\delta M_A).
\end{equation}
Here $\delta M_A=(\delta M,\delta S)$ are the evolving corrections to the black hole's mass and spin introduced in Eq.~\eqref{multiscale waveform}. They are functions of advanced time $v$ along the horizon, and they are exact, as opposed to being the linear terms in a small-$\e$ expansion, meaning 
\beq
\delta M_A := M^{\rm BH}_{A} - M_A,
\eeq
where $M^{\rm BH}_{A}$ are the black hole's physical parameters and $M_A$ are the parameters of the background metric $g_{\alpha\beta}$. The parameters evolve according to the instantaneous flux of energy and angular momentum across the horizon: 
\beq\label{Mdot flux}
\frac{dM^{\rm BH}_A}{dv}=\frac{d\delta M_A}{dv} = F_A(v,\e),
\eeq
with fluxes $F_A$ that are constructible from the metric perturbation $h_{\alpha\beta}$ on the horizon using formulas from Refs.~\cite{Ashtekar:2004cn,Ashtekar:2021kqj}, for example.

Reference~\cite{Miller:2020bft} provided one prescription for constructing $h^{\rm BH}_{\alpha\beta}$. However, that method suffers from some conceptual shortcomings. In Appendix~\ref{app:BH perturbation} we present a more satisfactory construction, leading to
\begin{multline}\label{dg Kirchhoff}
    h^{\rm BH}_{\alpha\beta}(x;\delta M_A) = \frac{1}{16\pi}\int_\Gamma \Bigl(G^{\rm ret}_{\alpha\beta}{}^{\,\alpha'\beta'}\nabla_{\gamma'}\delta g_{\alpha'\beta'} \\[-5pt]
    - \delta g_{\alpha'\beta'}\nabla_{\gamma'}G^{\rm ret}_{\alpha\beta}{}^{\,\alpha'\beta'}\Bigr) d\Sigma^{\gamma'}, 
\end{multline}
where $\Gamma$ is a timelike surface surrounding the horizon (the `stretched horizon'~\cite{Price:1986yy}), and $d\Sigma^{\gamma'}$ is its surface element induced from $g_{\alpha\beta}$. The `source' $\delta g_{\alpha\beta}$ is $\frac{\partial g_{\alpha\beta}}{\partial M_A}\delta M_A$ in the Lorenz gauge, as described in Appendix~\ref{app:BH perturbation}. Equation~\eqref{dg Kirchhoff} mirrors~\eqref{eq:hpp}, representing the metric perturbation as an integral over the entire past history of the black hole's worldtube.

At second order, the retarded solution to the field equation~\eqref{EFE2} takes an analogous form:
\begin{align}\label{eq:h2ret}
   h^{(2)}_{\alpha\beta}&(x;\gamma,\delta M_A) 
   \nonumber\\
   &= \int G^{\rm ret}_{\alpha\beta}{}^{\,\alpha'\beta'}\biggl(T^{(2)}_{\alpha'\beta'} - \frac{1}{8\pi}\delta^2 G_{\alpha'\beta'}[h^{(1)},h^{(1)}]\biggr) dV',
\end{align}
with $dV'=\sqrt{-g'}d^4x'$. This solution automatically includes appropriate terms quadratic in $\delta M_A$. Because $\delta M_A$ is the exact difference between the physical and the background black hole parameters, $h^{(2)}_{\alpha\beta}$ should not include another linear contribution (which would be proportional to some $\delta^2 M_A$).

\subsubsection{Singular and regular fields}

So far we have only described the physical, retarded metric perturbations, while the particle's trajectory is governed by the regular field $h^{{\rm R}}_{\alpha\beta}$. 

At first order, we can write the regular field $h^{{\rm R}(1)}_{\alpha\beta}$ in terms of the Detweiler-Whiting decomposition of the retarded Green's function~\cite{Detweiler:2002mi,Poisson:2011nh},
\begin{equation}\label{Gret=GS+GR}
    G^{\rm ret}_{\alpha\beta\alpha'\beta'} = G^{\rm S}_{\alpha\beta\alpha'\beta'}+G^{\rm R}_{\alpha\beta\alpha'\beta'}.
\end{equation}
The singular Green's function, $G^{\rm S}_{\alpha\beta\alpha'\beta'}$, satisfies Eq.~\eqref{Greens}, is symmetric,
\begin{equation}
    G^{\rm S}_{\alpha\beta\alpha'\beta'}(x,x') = G^{\rm S}_{\alpha'\beta'\alpha\beta}(x',x),
\end{equation}
and is singular at coincidence, $x^\mu=x'^\mu$. 
The regular two-point function, $G^{\rm R}_{\alpha\beta\alpha'\beta'}$, is a homogeneous solution to the wave equation~\eqref{Greens} and is smooth at coincidence. From $G^{\rm R}_{\alpha\beta\alpha'\beta'}$ and $G^{\rm S}_{\alpha\beta\alpha'\beta'}$, we define the Detweiler-Whiting singular and regular fields,
\begin{align}
h^{{\rm S}(1)}_{\alpha\beta}(x;\gamma) &:=\frac{1}{\mu} \int_{\gamma} G^{\rm S}_{\alpha\beta}{}^{\,\alpha'\beta'}(x,x') \tilde p_{\alpha'}\tilde p_{\beta'}d\tilde\tau',\label{h1S}\\
h^{{\rm R,pp}}_{\alpha\beta}(x;\gamma) &:=\frac{1}{\mu} \int_{\gamma} G^{\rm R}_{\alpha\beta}{}^{\,\alpha'\beta'}(x,x') \tilde p_{\alpha'}\tilde p_{\beta'}d\tilde\tau',\label{hRpp}
\end{align}
along with the total regular field
\begin{align}\label{hR1}
h^{{\rm R}(1)}_{\alpha\beta} = h^{{\rm R,pp}}_{\alpha\beta} + h^{\rm BH}_{\alpha\beta}.
\end{align}
Note that $h^{\rm BH}_{\alpha\beta}$ contributes only to the first-order regular field, not to $h^{{\rm S}(1)}_{\alpha\beta}$. This follows from first-principles derivations of the equations of motion~\cite{Poisson:2011nh}, which show that the particle's trajectory is governed by $h^{{\rm R}(1)}_{\alpha\beta}=h^{(1)}_{\alpha\beta}- h^{{\rm S}(1)}_{\alpha\beta}$, with $h^{{\rm S}(1)}_{\alpha\beta}$ given by Eq.~\eqref{h1S}. 

At second order, there is no known Green's function decomposition of the retarded solution into appropriate singular and regular fields. However,  local definitions of $h^{{\rm S}(2)}_{\alpha\beta}$ and $h^{{\rm R}(2)}_{\alpha\beta}$ are known, derived from the method of matched asymptotic expansions~\cite{Pound:2012dk,Pound:2014xva}. This allows one to calculate $h^{{\rm R}(2)}_{\alpha\beta}$ using a puncture scheme. We define a puncture field~$h^{{\cal P}(2)}_{\alpha\beta}$ that (i) agrees with $h^{{\rm S}(2)}_{\alpha\beta}$ up to order $|x^\alpha-x^\alpha_p|^2$ in a local expansion around the particle, and (ii) is made to go to zero at some finite distance from the particle, such that it has compact support around $x^\alpha_p$. Explicit covariant expressions for the puncture are given in Ref.~\cite{Pound:2014xva}. The residual field 
\beq
h^{{\cal R}(2)}_{\alpha\beta}:=h^{(2)}_{\alpha\beta}-h^{{\cal P}(2)}_{\alpha\beta} 
\eeq
is then the retarded solution to 
\begin{align}\label{EFER2}
    E_{\alpha\beta}[h^{{\cal R}(2)}] &= -16\pi S^{{\rm eff}(2)}_{\alpha\beta}
\end{align}
where
\beq
S^{{\rm eff}(2)}_{\alpha\beta}:=T^{(2)}_{\alpha\beta} 
-\frac{1}{16\pi}\left(2\delta^2G_{\alpha\beta}[h^{(1)},h^{(1)}] - E_{\alpha\beta}[h^{{\cal P}(2)}]\right)
\eeq
is referred to as the effective source. The puncture cancels the singularity in the physical source, leaving an $S^{{\rm eff}(2)}_{\alpha\beta}$ that is integrable at $x^\alpha_p$. In terms of this source, we have
\beq\label{eq:h2Res}
 h^{{\cal R}(2)}_{\alpha\beta}(x;\gamma,\delta M_A) = \int G^{\rm ret}_{\alpha\beta}{}^{\,\alpha'\beta'}S^{{\rm eff}(2)}_{\alpha'\beta'}dV'.
\eeq
Note that $h^{{\cal P}(2)}_{\alpha\beta}$ depends on $h^{{\rm R}(1)}_{\alpha\beta}$ and therefore on $\delta M_A$.

In the equation of motion~\eqref{EOM}, we can readily replace $h^{{\rm R}(2)}_{\alpha\beta}$ with $h^{{\cal R}(2)}_{\alpha\beta}$. This follows from 
\beq
h^{{\cal P}(2)}_{\alpha\beta}-h^{{\rm S}(2)}_{\alpha\beta}={\cal O}\!\left(\left|x^\alpha-x^\alpha_p\right|^2\right), 
\eeq
which implies 
\beq
h^{{\cal R}(2)}_{\alpha\beta}\bigr|_\gamma=h^{{\rm R}(2)}_{\alpha\beta}\bigr|_\gamma \quad \text{and}\quad  \partial_\mu h^{{\cal R}(2)}_{\alpha\beta}\bigr|_\gamma=\partial_\mu h^{{\rm R}(2)}_{\alpha\beta}\bigr|_\gamma\,. 
\eeq
Second-order self-force theory has almost exclusively been formulated in terms of punctures and residual fields~\cite{Detweiler:2011tt,Pound:2012dk,Gralla:2012db}, and all concrete calculations at second order have been based on this approach~\cite{Pound:2019lzj,Warburton:2021kwk,Miller:2020bft,Miller:2023ers}; no direct integration of Eq.~\eqref{EFE2} has been attempted.

\subsubsection{Symmetric and radiative fields}

To derive practical formulas for the 0PA forcing functions $F_i^{(0)}(\piring_k)$ in Eq.~\eqref{dpidt} in terms of asymptotic mode amplitudes, one typically begins by expressing $F_i^{(0)}(\piring_k)$ in terms of the radiative piece of the Green's function~\cite{Galtsov:1982hwm,Sago:2005fn,Isoyama:2018sib,Akcay:2019bvk}.

In analogy with Eq.~\eqref{Gret=GS+GR}, we can write~\cite{Poisson:2011nh}
\begin{equation}\label{Gret=Gsym+Grad}
    G^{\rm ret}_{\alpha\beta\alpha'\beta'} = G^{\rm sym}_{\alpha\beta\alpha'\beta'}+G^{\rm rad}_{\alpha\beta\alpha'\beta'},
\end{equation}
where 
\begin{equation}
G_{\alpha\beta\alpha'\beta'}^{\rm rad} := \frac{1}{2}G_{\alpha\beta\alpha'\beta'}^{\rm ret}- \frac{1}{2}G_{\alpha\beta\alpha'\beta'}^{\rm adv}
\end{equation}
is antisymmetric, satisfying
\begin{equation}\label{Grad symmetry}
    G_{\alpha\beta\alpha'\beta'}^{\rm rad}(x,x') = -G_{\alpha'\beta'\alpha\beta}^{\rm rad}(x',x),
\end{equation}
and 
\begin{equation}
G_{\alpha\beta\alpha'\beta'}^{\rm sym} := \frac{1}{2}G_{\alpha\beta\alpha'\beta'}^{\rm ret}+ \frac{1}{2}G_{\alpha\beta\alpha'\beta'}^{\rm adv}
\end{equation}
is symmetric, satisfying
\begin{equation}\label{Gsym symmetry}
    G_{\alpha\beta\alpha'\beta'}^{\rm sym}(x,x') = G_{\alpha'\beta'\alpha\beta}^{\rm sym}(x',x).
\end{equation}
$G_{\alpha\beta\alpha'\beta'}^{\rm sym}$ satisfies the Green's-function equation~\eqref{Greens} and is singular at coincidence, 
while $G^{\rm rad}_{\alpha\beta\alpha'\beta'}$ is a homogeneous solution to Eq.~\eqref{Greens} and is smooth at coincidence.

$G^{\rm R}_{\alpha\beta\alpha'\beta'}$ can likewise be decomposed into symmetric and radiative pieces,
\begin{equation}\label{GR=Grad+GRsym}
    G_{\alpha\beta\alpha'\beta'}^{\rm R} = G_{\alpha\beta\alpha'\beta'}^{\rm R, sym}+G_{\alpha\beta\alpha'\beta'}^{\rm rad},  
\end{equation}
where $G_{\alpha\beta\alpha'\beta'}^{\rm R,sym}:=G_{\alpha\beta\alpha'\beta'}^{\rm sym}-G_{\alpha\beta\alpha'\beta'}^{\rm S}$ satisfies
\begin{equation}\label{GRsym symmetry}
    G_{\alpha\beta\alpha'\beta'}^{\rm R,sym}(x,x') = G_{\alpha'\beta'\alpha\beta}^{\rm R,sym}(x',x).
\end{equation}
In flat spacetime, $G_{\alpha\beta\alpha'\beta'}^{\rm R,sym}$ vanishes, and $G_{\alpha\beta\alpha'\beta'}^{\rm R}$ reduces to $G_{\alpha\beta\alpha'\beta'}^{\rm rad}$.

These decompositions imply 
\begin{equation}\label{hRrad=hrad}
 h^{{\rm R,rad}(1)}_{\alpha\beta} = h^{{\rm rad}(1)}_{\alpha\beta} = h^{{\rm rad,pp}}_{\alpha\beta}+h^{{\rm rad,BH}}_{\alpha\beta}.
\end{equation}
Similarly, 
\beq\label{hRrad2=hrad2}
h^{{\cal R},{\rm rad}(2)}_{\alpha\beta} = h^{{\rm rad}(2)}_{\alpha\beta},
\eeq
or
\begin{align}
  \hspace{-5pt}h^{{\cal R},{\rm rad}(2)}_{\alpha\beta} &:= \int \!G^{\rm rad}_{\alpha\beta}{}^{\,\alpha'\beta'}S^{{\rm eff}(2)}_{\alpha'\beta'}dV',\nonumber\\
  &\hphantom{:}= 
\int \! G^{\rm rad}_{\alpha\beta}{}^{\,\alpha'\beta'}\nonumber\\[-3pt]
  &\qquad  \times\biggl(T^{(2)}_{\alpha'\beta'}-\frac{1}{8\pi}\delta^2G_{\alpha'\beta'}[h^{(1)},h^{(1)}]\biggr)dV'.\label{hrad2=intSeff}
\end{align}
Equation~\eqref{hRrad=hrad} follows immediately from the fact that the radiative part of $G_{\alpha\beta\alpha'\beta'}^{\rm R}$ is identical to the radiative part of $G_{\alpha\beta\alpha'\beta'}^{\rm ret}$. Equation~\eqref{hRrad2=hrad2} follows from the fact that
\begin{align}\label{intEhP}
\int \!G^{\rm rad}_{\alpha\beta}{}^{\,\alpha'\beta'}E_{\alpha'\beta'}[h^{{\cal P}(2)}]dV'=0.
\end{align}
To derive Eq.~\eqref{intEhP}, first integrate by parts and use the fact that $G^{\rm rad}_{\alpha\beta}{}^{\,\alpha'\beta'}$ is a homogeneous solution to obtain the identity
\begin{multline}\label{intEhP Greens identity}
\int \!G^{\rm rad}_{\alpha\beta}{}^{\,\alpha'\beta'}E_{\alpha'\beta'}[h^{{\cal P}(2)}]dV' \\
= \frac{1}{16\pi}\int_{\partial V} \Bigl(G^{\rm rad}_{\alpha\beta}{}^{\,\alpha'\beta'}\nabla_{\gamma'}h^{{\cal P}(2)}_{\alpha'\beta'} \\[-2pt]
    - h^{{\cal P}(2)}_{\alpha'\beta'}\nabla_{\gamma'}G^{\rm rad}_{\alpha\beta}{}^{\,\alpha'\beta'}\Bigr) d\Sigma^{\gamma'},
\end{multline}
where $\partial V$ is the boundary of the spacetime (e.g., the boundary of the Kerr exterior). Since $h^{{\cal P}(2)}_{\alpha\beta}$ has spatially compact support around the particle, the integral on the right-hand side of Eq.~\eqref{intEhP Greens identity} only receives a contribution from points in the far future and infinite past, where the support of $h^{{\cal P}(2)}_{\alpha\beta}$ intersects the initial surface in the infinite past or the final surface in the infinite future. Since $G^{\rm rad}_{\alpha\beta}{}^{\,\alpha'\beta'}$ decays with separation, the integral contributes nothing when the field point $x^\alpha$ is not at asymptotically early or late times. Hence, the source $E_{\alpha\beta}[h^{{\cal P}(2)}]$ in Eq.~\eqref{EFER2} contributes nothing to the radiative field, and we obtain Eq.~\eqref{intEhP} and finally Eq.~\eqref{hrad2=intSeff}.

In a spacetime with a time-reflection symmetry, the retarded and advanced Green's functions are also related by~\cite{Mino:2003yg} 
\beq\label{TGret = Gadv}
{\cal T}G^{\rm ret}_{\alpha\beta\alpha'\beta'} = G^{\rm adv}_{\alpha\beta\alpha'\beta'},
\eeq
where ${\cal T}$ denotes time reversal. In Schwarzschild (and all other static spacetimes), time reversal corresponds to $t\to - t$. In Kerr (and other stationary, axisymmetric spacetimes~\cite{Wald:1984rg,Harmark:2004rm}), it corresponds to $t\to-t$, $\phi\to-\phi$. In each case, this reversal is an isometry of the background spacetime. Equation~\eqref{TGret = Gadv} implies
\begin{align}
{\cal T}G^{\rm rad}_{\alpha\beta\alpha'\beta'} &= -G^{\rm rad}_{\alpha\beta\alpha'\beta'},\label{TGrad = -Grad}\\
{\cal T}G^{\rm sym}_{\alpha\beta\alpha'\beta'} &= +G^{\rm sym}_{\alpha\beta\alpha'\beta'}
\end{align}
in these spacetimes.

\subsubsection{Summary and nonlocality on phase space}\label{sec:summary}

Second-order self-force theory can be summarized as the field equations~\eqref{EFE1}--\eqref{EFE2} coupled to the particle's equation of motion~\eqref{EOM} and the black hole's evolution equation~\eqref{Mdot flux}. Since the field equations~\eqref{EFE1}--\eqref{EFE2} are constraint-preserving~\cite{Pound:2012nt}, the `matter' equations~\eqref{EOM} and~\eqref{Mdot flux} (which are analogous to conservation of stress-energy) suffice to enforce the gauge condition~\eqref{Lorenz gauge}. 

As emphasized in early literature on the self-consistent expansion~\cite{Pound:2009sm}, the gauge condition $\nabla^\beta\bar h_{\alpha\beta}=0$ imposed on the full metric perturbation does not imply $\nabla^\beta\bar h^{(n)}_{\alpha\beta}=0$ for each $n$. Instead, we should consider%
\begin{subequations}%
\begin{align}
\nabla^\beta\bar h_{\alpha\beta}&:=Z_\alpha(x,\e;\gamma,\delta M_A)\\
&\hphantom{:}=\e Z^{(1)}_\alpha(x;\gamma,\delta M_A) \nonumber\\
&\quad\ + \e^2 Z^{(2)}_\alpha(x;\gamma,\delta M_A) + \O(\e^2)  
\end{align}%
\end{subequations}%
and set each $Z^{(n)}_\alpha(x;\gamma,\delta M_A)=0$. Our phase-space formulation clarifies precisely what is being held fixed in these expansions: we hold $x^\alpha_p(\tilde\tau,\e)$, $\tilde p_\alpha(\tilde\tau,\e)$, and $\delta M_A(v,\e)$ fixed, never expanding their $\e$ dependence, but we do expand time derivatives  $d^n x^\alpha_p/d\tilde\tau^n$, $d^n\tilde p_\alpha/d\tilde\tau^n$, and $d^n\delta M_A/dv^n$ in powers of $\e$ at fixed $(x^\alpha_p,\tilde p_\alpha,\delta M_A)$. This is effectively an order reduction, ensuring that the particle's acceleration does not depend on itself and higher time derivatives, for example. However, in our scheme it is a simple consequence of the guiding principle that all quantities are to be expanded for small $\e$ while holding the mechanical degrees of freedom $(\gamma,\delta M_A)$ fixed. The ``matter'' equations~\eqref{EOM} and~\eqref{Mdot flux} can then be derived directly from the gauge conditions $Z^{(n)}_{\alpha}=0$ evaluated on the particle's trajectory and the black hole's horizon.

The coupled equations~\eqref{EFE1}--\eqref{EFE2}, \eqref{EOM}, and~\eqref{Mdot flux} are a mix of partial differential equations (PDEs) for the field degrees of freedom ($h^{(n)}_{\alpha\beta}$) and ODEs for the `matter' degrees of freedom ($x^\alpha_p$, $\tilde p_\alpha$, and $\delta M_A$). In Eqs.~\eqref{eq:hpp}, \eqref{dg Kirchhoff}, and \eqref{eq:h2ret}, we `integrate out' the field degrees of freedom, expressing them as functionals of the matter degrees of freedom. The matter equations of motion~\eqref{EOM} and~\eqref{Mdot flux} then become integro-differential, meaning the dynamics on phase space becomes nonlocal in time.\footnote{We loosely refer to the ten-dimensional space spanned by $x^\alpha_p$, $\tilde p_\alpha$, and $\delta M_A$ as a phase space. However, unlike the 8D space spanned by $x^\alpha_p$ and $\tilde p_\alpha$, it is not (to our knowledge) a symplectic manifold.} In Sec.~\ref{sec:localized Hamiltonian for bound orbits} we will show how this dynamics becomes localized.

Here we restricted our summary to the Lorenz gauge. We will continue to assume the use of this gauge for concreteness. However, the equation of motion~\eqref{EOM}, and corresponding definitions of the regular field, is known to hold in all gauges used in practice~\cite{Pound:2013faa,Shah:2015nva,Pound:2015fma,Pound:2017psq,Thompson:2018lgb}. The decomposition of the Green's function also straightforwardly extends to other gauges in which the gauge-fixed field equations share the same principal part~\cite{Pound:2012dk} (though the decomposition is \emph{not} known in the radiation gauge~\cite{Casals:2024ynr}, which is the most widely used gauge for GSF calculations in Kerr spacetime~\cite{Keidl:2010pm,Shah:2012gu,vandeMeent:2017bcc}). We will, in any case, ultimately arrive at gauge-invariant formulas.

Similarly, we have assumed a Kerr background, which is our primary interest. However, most of the material in this section is equally valid in any globally hyperbolic vacuum background, and much of our later analysis will apply in any such background spacetime that is also stationary and in which the geodesic equation is integrable. In contexts other than the black hole binary problem, $\delta M_A$ can be replaced by self-consistently evolving corrections to any other background parameters.

\subsection{Multiscale expansion through 1PA order}\label{sec:multiscale expansion}

The self-consistent framework is useful for deriving governing equations in self-force theory (equations of motion, punctures, etc.). It also provides a useful starting point for other expansions because it is valid in any vacuum background spacetime and on all spatial and temporal scales. However, it has never been numerically implemented except in a scalar-field problem~\cite{Diener:2011cc,Wittek:2024gxn,Wittek:2024pis}. Instead, most practical implementations have been based on a multiscale framework. Here we review the framework as presented in Refs.~\cite{Pound:2019lzj,Mathews:2025nyb}; see also Refs.~\cite{Miller:2020bft,Miller:2023ers} for more detailed explications of special cases. 

\subsubsection{Orbital motion}

The multiscale expansion of the Einstein equations, and the waveform-generation scheme it leads to, begins with a choice of global time $s$. One typically chooses $s$ to be a hyperboloidal-type time, such that slices of constant $s$ penetrate the future horizon and extend to future null infinity~\cite{Miller:2020bft,Miller:2023ers,Cunningham:2024dog}. For simplicity, we assume this $s$ reduces to Boyer-Lindquist time $t$ along the particle's worldline, to advanced time $v$ along the horizon, and retarded time $u$ at future null infinity. The slices can be everywhere spacelike or contain null segments~\cite{Miller:2020bft,Miller:2023ers}. Our spacetime coordinates are then $x^\alpha=(s,x^i)$, and we write the spacetime manifold as ${\cal M} = \mathbb{R}\times \Sigma$, where $x^i$ are coordinates on $\Sigma$. We ultimately work on the cotangent bundle of $\Sigma$ as a 6D phase space (which is again a symplectic manifold because it is a cotangent bundle).

One must consider the field equations and equations of motion together, as a coupled problem. But it is useful to begin with the orbital motion; the multiscale ansatz for the metric is then motivated by the existence of phase-space coordinates $(\phiring^i,\piring_i)$ for which the particle's orbital equations take the form~\eqref{dphidt}--\eqref{dpidt}. These coordinates cleanly separate the dynamics into fast, periodic variables $\phiring^i$ and slow, adiabatic variables $\piring_i$. Reference~\cite{Pound:2019lzj} shows how to find such coordinates, beginning with the method of osculating geodesics~\cite{Pound:2007th,Warburton:2013yj,Gair:2010iv} and then applying the method of near-identity averaging transformations~\cite{VanDeMeent:2018cgn}.  

We first introduce a quasi-Keplerian parametrization of the motion in terms of orbital elements $\pi_i=(p,e,z_{\rm max})$ and phases $\psi^i = (\psi^r,\psi^\theta,\phi_p)$. It will also be useful to separate the azimuthal motion from the polar and radial motion, introducing boldface symbols $\bm{x}_p=(r_p,\theta_p)$ and $\bm{\psi}=(\psi^r,\psi^\theta)$, and also introducing the orbital elevation $z_p=\cos\theta_p$. The quasi-Keplerian parametrization of the coordinate trajectory and velocity is then 
\begin{align}
    r_p(\bm{\psi},\pi_i) &= \frac{p\, M}{1+e \cos\psi^r},\label{eq:r(psi,pi)}\\
    z_p(\bm{\psi},\pi_i) &= z_{\rm max} \cos\psi^\theta,\label{eq:z(psi,pi)}
\end{align}
and 
\begin{equation}\label{eq:xdot(psi,pi)}
\dot x^i_p(\bm{\psi},\pi_j) = \frac{\partial x^i_p}{\partial \psi^j}\omega^j_{(0)}(\bm{\psi},\pi_k).
\end{equation}
Here and throughout this paper, an overdot denotes differentiation with respect to $t$, as in $\dot x^i_p:=dx^i_p/dt$. In Eq.~\eqref{eq:xdot(psi,pi)}, $\omega^j_{(0)}$ is the ``frequency'' $\dot\psi^j$ of a Kerr geodesic:
\beq
\left(\frac{d\psi^i}{dt}\right)_{\rm geo} = \omega^j_{(0)}(\bm{\psi},\pi_k) = \frac{\mathscr{f}^j(\bm{\psi},\pi_k)}{\mathscr{f}^t(\bm{\psi},\pi_k)}
\eeq
with $\mathscr{f}^t$, $\mathscr{f}^r$, $\mathscr{f}^\theta$, and $\mathscr{f}^\phi$ given by Eqs.~(205), (216), (217), and (206) of Ref.~\cite{Pound:2021qin}, respectively.\footnote{We have changed notation relative to Ref.~\cite{Pound:2021qin} by placing indices up on phases and frequencies, such that $\mathscr{f}^\alpha$ here corresponds to $\mathscr{f}_\alpha$ in Ref.~\cite{Pound:2021qin}. This change is motivated by wanting coordinates on configuration space to have indices up and momenta to have indices down.} Note that for orbital equations we write the time parameter as $t$ rather than $s$ because we specialised our choice of $s$ to reduce to $t$ along the particle's trajectory.

Equations~\eqref{eq:r(psi,pi)}--\eqref{eq:xdot(psi,pi)} are referred to as ``osculation conditions''. The right-hand side in each case is identical to the quasi-Keplerian parametrization of a Kerr geodesic. If the motion were geodesic in Kerr (i.e., if the self-force vanished), we would have $d\pi_i/dt=0$ and $d\psi^i/dt=\omega^i_{(0)}$. However, in the present context, Eqs.~\eqref{eq:r(psi,pi)}--\eqref{eq:xdot(psi,pi)} are more appropriately viewed as a coordinate transformation on the particle's orbital phase space, $(x^i_p,\dot x^i_p)\mapsto (\psi^i,\pi_i)$. This phase space (the cotangent bundle of $\Sigma$, as mentioned above) is 6D rather than 8D as a consequence of our use of $t$ as the time parameter: since $x^\alpha_p(t)=(t,x^i_p(t))$ and $\dot x^\alpha_p(t)=(1,\dot x^i_p(t))$, the evolution equations for $t_p(t)$ and $\dot t_p(t)$ are vacuous. The $t$ component of the four-velocity, $u^t=dt_p/d\tau$, which is required to evaluate the self-force~\eqref{EOM} and other quantities, is determined from $u^\alpha u_\alpha =-1$, which implies 
\beq
u^t  = \left[-\left(g_{tt}+2 g_{ti}\dot x^i_p + g_{ij}\dot x^i_p\dot x^j_p\right)\right]^{-1/2}.
\eeq

Given the quasi-Keplerian parameterization~\eqref{eq:r(psi,pi)}--\eqref{eq:xdot(psi,pi)}, the phase-space trajectory is governed by equations of the form
\begin{align}
    \frac{d\psi^i}{dt}&=\omega_{(0)}^i(\psivec,\pi_j)+\delta\omega^i(t,\psi^i,\pi_j,\e;\gamma,\delta M_A),\label{osculating psi} \\
    \frac{d\pi_i}{dt}&=g_i(t,\psi^i,\pi_j,\e;\gamma,\delta M_A). \label{osculating pi}
\end{align}
The ``frequency''  correction $\delta \omega^i$ and forcing functions $g_i$ are linear combinations of the spatial components of the self-force~\eqref{EOM}, which we can write as
\begin{align}
    \delta\omega^i &= A^i_{\ j}(\bm{\psi},\pi_i) f^j(t,\psi^k,\pi_k,\e;\gamma,\delta M_A),\label{domega=Af} \\
    g_i &= B_{ij}(\bm{\psi},\pi_i) f^j(t,\psi^k,\pi_k,\e;\gamma,\delta M_A).\label{g=Bf}
\end{align}
The coefficients $A^i_{\ j}$ and  $B_{ij}$ can be read off from Eqs.~(289)--(295) in Ref.~\cite{Pound:2021qin}, noting that those equations must be divided by $dt/d\lambda=\mathscr{f}^t$ to convert $\lambda$ derivatives to $t$ derivatives, where $\lambda$ is a time parameter referred to as Mino time~\cite{Mino:2003yg}.

If we substitute $h^{\rm R}_{\alpha\beta}=\sum_n \e^n h^{{\rm R}(n)}_{\alpha\beta}(x;\gamma,\delta M_A)$ into Eq.~\eqref{EOM}, we obtain an expansion for the self-force,
\begin{multline}\label{f expansion}
f^i = \e f^i_{(1)}(t,\psi^j,\pi_j;\gamma,\delta M_A) \\+ \e^2 f^i_{(2)}(t,\psi^j,\pi_j;\gamma,\delta M_A) + \O(\e^3).
\end{multline}
Equations~\eqref{osculating psi} and \eqref{osculating pi}, with Eqs.~\eqref{domega=Af} and \eqref{g=Bf}, then become
\begin{align}
    \frac{d\psi^i}{dt}&=\omega_{(0)}^i(\psivec,\pi_j)+\e \omega_{(1)}^i(t,\psi^j,\pi_j;\gamma,\delta M_A)+\mathcal{O}(\e^2), \label{osculating psi v2} \\
    \frac{d\pi_i}{dt}&=\e g^{(0)}_i(t,\psi^j,\pi_j;\gamma,\delta M_A)\nonumber\\
    &\qquad\qquad +\e^2 g^{(1)}_i(t,\psi^j,\pi_j;\gamma,\delta M_A)+\mathcal{O}(\e^3), \label{osculating pi v2}
\end{align}
with $\omega^i_{(1)}=A^i_{\ j}f^j_{(1)}$ and $g_i^{(n)}=B_{ij}f^j_{(n+1)}$. As in the Introduction, numeric labels in Eq.~\eqref{f expansion} correspond to the powers of $\e$ in the metric that generates the force, while numeric labels in Eqs.~\eqref{osculating psi v2} and \eqref{osculating pi v2} denote the post-adiabatic order at which each term contributes in the multiscale expansion. 

The multiscale expansion is based on an averaging transformation $(\psi^i,\pi_i)\mapsto (\phiring^i,\piring_i)$ that puts the orbital evolution equations in the form~\eqref{dphidt}--\eqref{dpidt}. We write this transformation as
\begin{align}
    \psi^i(\phiring^j,\piring_j,\e)&=\psio^i(\phiring^j,\piring_j)\nonumber\\
    &\qquad \quad +\e \psi_{(1)}^i(\phiringvec,\piring_j,\delta M_A)+\mathcal{O}(\e^2), \label{psi to phiring} \\
    \pi_i(\phiring^j,\piring_j,\e)&=\piring_i+\e \pi^{(1)}_i(\phiringvec,\piring_j,\delta M_A)\nonumber\\
    &\qquad\quad +\e^2 \pi^{(2)}_i(\phiringvec,\piring_j,\delta M_A)+\mathcal{O}(\e^3),
    \label{pi to piring}
\end{align}
where $\phiringvec=(\phiring^r,\phiring^\theta)$. The zeroth-order term in the phase transformation divides into a linear piece and an oscillatory one,
\beq\label{psi0=phi+Dpsi}
\psio^i(\phiring^j,\piring_j) = \phiring^i +\Delta \psi^i(\phiringvec,\piring_j).
\eeq
We will particularly highlight the azimuthal case,
\beq\label{phip=phiring+Dphi}
\phi^{(0)}_p = \phiring^\phi + \Delta\phi_p(\phiringvec,\piring_j).
\eeq
All of the functions $\Delta \psi^i$, $\psi^i_{(n>0)}$, and $\pi_i^{(n)}$ are $2\pi$-periodic in each of $\phiring^r$ and $\phiring^\theta$. The periodic functions are chosen to eliminate all oscillations in the equations of motion; i.e., they are chosen to ensure there is no $\phiring^i$ dependence on the right-hand sides of Eqs.~\eqref{dphidt} and~\eqref{dpidt}. This process of removing oscillations is detailed in Sec.~\ref{sec:localized Hamiltonian for bound orbits} below.

The functional forms in the transformations~\eqref{psi to phiring} and~\eqref{pi to piring} differ from the generic functional forms in Eqs.~\eqref{osculating psi v2} and \eqref{osculating pi v2} in two important ways. First, there is no dependence on $t$ or on the azimuthal phase $\phiring^\phi$, while functions in Eqs.~\eqref{osculating psi v2} and \eqref{osculating pi v2} might generically depend on $t$ and $\psi^\phi=\phi_p$ through their dependence on $h^{\rm R}_{\alpha\beta}$. Second, rather than being nonlocal, integral functionals of $\gamma$ and $\delta M_A$, they only depend on the values of $\phiring^i$, $\piring_i$, and $\delta M_A$ at time $t$; i.e., they are local transformations on phase space. These properties follow from the form of the self-force in the multiscale expansion, 
\beq\label{f multiscale}
f^i = \e \mathring{f}^i_{(1)}(\phiringvec,\piring_j,\delta M_A) + \e^2 \mathring{f}^i_{(2)}(\phiringvec,\piring_j,\delta M_A) + \O(\e^3).
\eeq
This form, in turn, follows from the multiscale expansion of the metric, which we outline below.

\subsubsection{Metric and field equations}

The essential idea in the multiscale expansion of the metric is that all time dependence in the metric can be encoded in a dependence on the phase-space variables. As a function of $(s,x^i,\e)$, we can then write 
\beq
h_{\alpha\beta}=h_{\alpha\beta}[x^i,\phiring^i(s,\e),\piring_i(s,\e),\delta M_A(s,\e),\e]\,,
\eeq
such that $h_{\alpha\beta}$ can be treated as a function on an 11D product manifold: the Cartesian product of (i) 3D ``space'' $\Sigma$ spanned by $x^i$, (ii) the particle's 6D orbital phase space spanned by $(x^i_p,\tilde p_i)$, and (iii) the 2D space spanned by $\delta M_A$. We additionally assume the metric, treated as a function on this product space, admits a regular expansion for small $\e$:
\beq\label{multiscale h}
h_{\alpha\beta}=\sum_{n\geq1}\e^n \mathring h^{(n)}_{\alpha\beta}(x^i,\phiring^i,\piring_i,\delta M_A).
\eeq
Since $\phiring^i$ are periodic variables, the coefficients $\mathring h^{(n)}_{\alpha\beta}$ can be expanded in discrete Fourier series,
\begin{align}\label{multiscale h Fourier}
    \mathring h^{(n)}_{\alpha\beta} &=\sum_{\vec{k}\in\mathbb{Z}^3} \mathring h^{(n,\vec{k})}_{\alpha\beta}(x^i,\piring_i,\delta M_A)e^{-i k_i\phiring^i},
\end{align}
with $\vec{k} = (k_r,k_\theta,k_\phi)$. Finally, because of the axial symmetry of the background, the azimuthal angle $\phi$ only enters the Einstein equations through the delta function $\delta(\phi-\phi_p)$ in $T_{\alpha\beta}$. Given Eq.~\eqref{phip=phiring+Dphi}, this implies the only dependence on $\phiring^\phi$ in the Einstein equations is in the form $(\phi-\phiring^\phi)$, such that the expansion~\eqref{multiscale h Fourier} can be written as
\begin{align}\label{multiscale h phi}
    \mathring h^{(n)}_{\alpha\beta}
    &= \sum_{\bm{k},m} \mathring h^{(n,\bm{k},m)}_{\alpha\beta}(\bm{x},\piring_i,\delta M_A)e^{-i [\bm{k}\cdot\bm{{\phiring}} - m(\phiring^\phi-\phi)]},
\end{align}
with $\bm{k}=(k_r,k_\theta)$ and $m=k_\phi$.

When substituting Eq.~\eqref{multiscale h} into the Einstein equation, we apply the chain rule, 
\beq
\frac{\partial}{\partial s} = \frac{d\phiring^i}{ds}\frac{\partial}{\partial\phiring^i}+\frac{d\piring_i}{ds}\frac{\partial}{\partial\piring_i} + \frac{d\delta M_A}{ds}\frac{\partial}{\partial\delta M_A}.
\eeq
Equations~\eqref{dphidt}--\eqref{dMdt} then imply the expansion
\beq
\nabla_\alpha = \nabla^{(0)}_\alpha + \e s_\alpha \left(\vec{\partial}_{\cal V}+\Omega^i_{(1)}\frac{\partial}{\partial\phiring^i}\right) + {\cal O}(\e^2),
\eeq
where $s_\alpha:=\partial_\alpha s$ and 
the zeroth-order covariant derivative is 
\beq
\nabla^{(0)}_\alpha = e^i_\alpha\frac{\partial}{\partial x^i} + s_\alpha \Omega^i_{(0)}\frac{\partial}{\partial\phiring^i} + \text{Christoffel terms}.\label{nabla0}
\eeq
Here $e^i_\alpha:=\partial_\alpha x^i$ are basis covectors on the slices of constant $s$, $\vec{\cal V}=(F^{(0)}_i,F^{(0)}_A)$ is the leading-order velocity through parameter space, and 
\beq\label{param D}
\vec{\partial}_{\cal V}:= F_i^{(0)}\frac{\partial}{\partial\piring_i} + F_A^{(0)}\frac{\partial}{\partial\delta M_A}
\eeq
is a directional derivative in the parameter space. 

On the right-hand side of the Einstein equation, we require the expansion of the stress-energy tensor. We first rewrite the Detweiler stress-energy tensor~\eqref{Detweiler T} in terms of coordinate time $t$:
\begin{align}\label{Detweiler T-t}
    T_{\alpha\beta} =  \frac{\tilde p_\alpha \tilde p_\beta}{\tilde g^{t\mu}\tilde p_\mu} \frac{\delta^3(x^i-x^i_p(t))}{\sqrt{-\tilde g}},
\end{align} 
where we used $dt/d\tilde\tau=\tilde u^t = \tilde g^{t\alpha}\tilde p_\alpha/\mu$. 
Given Eqs.~\eqref{eq:r(psi,pi)}--\eqref{eq:xdot(psi,pi)} and~\eqref{psi to phiring}--\eqref{pi to piring}, the particle's coordinate trajectory takes the form
\begin{align}
    x^i_p &= x^i_{(0)}(\phiring^j,\piring_j) + \e x^i_{(1)}(\phiring^j,\piring_j) + \O(\e^2),\label{x(phi,pi)}\\        
    \dot x^i_p &= \dot x^i_{(0)}(\phiring^j,\piring_j) + \e \dot x^i_{(1)}(\phiring^j,\piring_j) + \O(\e^2).\label{xdot(phi,pi)}
\end{align}
This, together with the expansion of the metric, implies
\beq\label{p(phi,pi)}
\tilde p_\alpha = p^{(0)}_\alpha(\phiring^j,\piring_j) + \e \tilde p^{(1)}_\alpha(\phiring^j,\piring_j) + \O(\e^2).
\eeq
Putting these expansions together, with the expansion of $\tilde g^{t\mu}$, yields
\beq
T_{\alpha\beta}=\sum_{n\geq1}\e^n \mathring{T}^{(n)}_{\alpha\beta}(x^i,\phiring^i,\piring_i,\delta M^A).
\eeq

Substituting the above  expansions into the Einstein equation~\eqref{full EFE}, we obtain a sequence of field equations for the coefficients in Eq.~\eqref{multiscale h}:
\begin{align}
    E^{(0)}_{\alpha\beta}[\mathring{\bar h}^{(1)}] &= -16\pi \mathring{T}^{(1)}_{\alpha\beta},\label{multiscale EFE 1}\\
    E^{(0)}_{\alpha\beta}[\mathring{\bar h}^{(2)}] &= -16\pi \mathring{T}^{(2)}_{\alpha\beta} + 2\delta^2 G^{(0)}_{\alpha\beta}[\mathring h^{(1)},\mathring h^{(1)}] - E^{(1)}_{\alpha\beta}[\mathring{\bar h}^{(1)}],\label{multiscale EFE 2}
\end{align}
where the source
\begin{multline}
E^{(1)}_{\mu\nu}[\mathring{\bar h}] = s_\alpha s^\alpha F_i^{(0)}\partial_{\phiring^i}\mathring{\bar h}_{\mu\nu} + 2 s^\alpha\nabla^{(0)}_\alpha\vec{\partial}_{\cal V}\mathring{\bar h}_{\mu\nu} \\
 +\bigl(\nabla^{(0)}_\alpha s^\alpha\bigr)\vec{\partial}_{\cal V} \mathring{\bar h}_{\mu\nu}\,,\label{E1}
\end{multline}
arises from the slow evolution of the spacetime. If we substitute the Fourier expansion~\eqref{multiscale h Fourier}, then the field equations~\eqref{multiscale EFE 1} and \eqref{multiscale EFE 2} become 2D PDEs in $\bm{x}$ for the Fourier mode coefficients $\mathring h^{(n,\bm{k},m)}_{\alpha\beta}$.

\subsubsection{Summary and gauge freedom}\label{subsubsec:gauge freedom}

The waveform generation scheme contained in Eqs.~\eqref{multiscale waveform}--\eqref{dMdt} requires as input the mode amplitudes $\mathring h^{(n)}_{lm\bm{k}}$, frequency corrections $\Omega^i_{(1)}$, and forcing functions $F^{(0)}_i$, $F^{(1)}_i$, and $F^{(1)}_A$. Each of those inputs can be precomputed as a function of $\piring_i$ from the coefficients $\mathring h^{(n,\bm{k},m)}_{\alpha\beta}$, which are obtained by solving the field equations~\eqref{multiscale EFE 1} and \eqref{multiscale EFE 2}. We refer to Refs.~\cite{Pound:2019lzj,Miller:2020bft,Miller:2023ers} for further details.

However, we note that the functions $\mathring h^{(n)}_{lm\bm{k}}$, $\Omega^i_{(n>0)}$, and $F^{(n>0)}_i$ are not unique; they can be modified by a small transformation of the phase space coordinates $(\phiring^i,\piring_i)$. This gauge freedom arises because the transformation $(\psi^i,\pi_i)\mapsto (\phiring^i,\piring_i)$ in Eqs.~\eqref{psi to phiring} and \eqref{psi to phiring} is not unique. The transformation is only required to eliminate oscillatory functions from the equations of motion, leaving arbitrary the averaged parts of the transformation. Concretely, the functions $\langle \pi^{(n)}_i\rangle$ and $\langle \Delta\psi^i\rangle$ are freely specified. 
Here and below,
\beq\label{average def}
\langle\cdot\rangle:=\frac{1}{(2\pi)^3}\oint \cdot \,d^3\phiring\,,
\eeq
denotes an average over the 3-torus. However, in practice, it reduces to an average over the 2-torus spanned by $(\phiring^r,\phiring^\theta)$ because $\phiring^\phi$ does not enter into the dynamics. 

In Ref.~\cite{Pound:2021qin}, this gauge freedom was fixed by requiring (i) $\Omega^i_{(n>0)}=0$, such that $\piring_i$ are related to the physical frequencies $\Omega^i:=d\phiring^i/dt$ by the Kerr geodesic relationship $\Omega^i=\Omega^i_{(0)}(\piring_j)$, and (ii)  $\phiring^i$ vanishes at $\psi^i=0$, such that the transformation preserves the origin of the phases $\phiring^i=0$. Since $\psi^r$ and $\psi^\theta$ vanish at radial and polar turning points (at fixed $\pi_i$), condition (ii) ensures that $\phiring^r=0$ and $\phiring^\theta=0$ correspond to those same turning points (at fixed $\piring_i$). We explore this gauge freedom more fully in the bulk of this paper.

As first noted in Ref.~\cite{Hinderer:2008dm}, the 1PA evolution involves the complete first-order self-force and the dissipative part of the second-order self-force. Concretely, the 1PA frequency corrections $\Omega_{(1)}^i$ (if they are nonzero) involve the conservative part of the first-order self-force $f^\alpha_{(1)}$, and the 1PA forcing functions $F^{(1)}_i$ involve the conservative part of $f^\alpha_{(1)}$ and the dissipative part of $f^\alpha_{(2)}$. Here conservative and dissipative pieces are defined by their symmetry under reversal around turning points: in terms of the quasi-Keplerian phases $\bm{\psi}$,
\begin{align}
f^\alpha_{(n)\rm diss}(\bm{\psi}) &= \frac{1}{2}f^\alpha_{(n)}(\bm{\psi}) - \frac{1}{2} \epsilon^\alpha f^\alpha_{(n)}(-\bm{\psi}),\\
f^\alpha_{(n)\rm con}(\bm{\psi}) &= \frac{1}{2} f^\alpha_{(n)}(\bm{\psi}) + \frac{1}{2}\epsilon^\alpha f^\alpha_{(n)}(-\bm{\psi}),
\end{align}
where $\epsilon^\alpha=(-1,1,1,-1)$, there is no summation over $\alpha$, and the reversals are at fixed $\pi_i$. Similarly, in terms of the  variables $(\phiring^i,\piring_i)$, 
\begin{align}
f^\alpha_{(n)\rm diss}(\phiringvec) &= \frac{1}{2}f^\alpha_{(n)}(\phiringvec) - \frac{1}{2} \epsilon^\alpha f^\alpha_{(n)}(2\phiringvec_O-\phiringvec),\label{fdiss}\\
f^\alpha_{(n)\rm con}(\phiringvec) &= \frac{1}{2} f^\alpha_{(n)}(\phiringvec) + \frac{1}{2}\epsilon^\alpha f^\alpha_{(n)}(2\phiringvec_O-\phiringvec),\label{fcon}
\end{align}
where $\phiringvec_O$ corresponds to the value of $\phiringvec$ at the origin $\bm{\psi}=0$. From Eq.~\eqref{psi0=phi+Dpsi}, this depends on the choice of $\Delta\bm{\psi}$, as it must satisfy $\phiringvec_O = -\Delta \psivec(\phiringvec_O,\piring_j)$. If we decompose $\Delta\psi^i$ into a purely oscillatory part (with zero average) and an average piece, as in
\beq
\Delta\psi^i = \Delta\psi^i_{\rm osc} + \langle \Delta\psi^i\rangle,
\eeq
we can choose $\phiringvec_O=\bm{0}$ by choosing $\langle \Delta\psi^i\rangle = - \Delta\psi^i_{\rm osc}(\bm{0},\piring_j)$; this is the choice made in Ref.~\cite{Pound:2021qin}.

Finally, we observe that our exposition assumed the ansatz~\eqref{multiscale h} in order to explain the existence of the preferred class of coordinates $(\phiring^i,\piring_i)$. In this sense, the multiscale expansion is an internally consistent ansatz rather than a derived result. However, one can also derive the multiscale expansion of the metric directly from the self-consistent expansion. This is made possible by allowing the transformations~\eqref{psi to phiring} and \eqref{pi to piring} to be nonlocal on phase space (see, e.g., Refs.~\cite{Blanco:2022mgd,Blanco:2024fte} for examples of such transformations). Once the equations of motion are put in the form~\eqref{dphidt}--\eqref{dpidt}, the multiscale expansion~\eqref{multiscale h}, with~\eqref{multiscale h Fourier}, follows from applying a stationary phase approximation to the self-consistent solution~\eqref{eq:hpp}, \eqref{dg Kirchhoff}, and~\eqref{eq:h2ret}. We describe that application of the stationary phase approximation (which appears here for the first time) in Appendix~\ref{sec:multiscale from self-consistent}. Similar derivations will be core parts of our analyses later in this paper.


\section{Pseudo-Hamiltonian description of self-forced dynamics}
\label{sec:pseudo-Hamiltonian}

In this section we present the pseudo-Hamiltonian description of the self-consistent dynamics. Besides the extension to second order, our treatment differs from that of Refs.~\cite{Fujita:2016igj,Isoyama:2018sib} in that we reduce the dynamics to the 6D phase space with our preferred choice of time $t$. As noted in Sec.~\ref{sec:multiscale expansion}, this 6D phase space arises naturally in the modern multiscale formulation of the field equations and waveform generation. In Ref.~\cite{Kakehi:2024bnh}, two of us describe the pseudo-Hamiltonian formalism with a generic choice of time parameter.

\subsection{Pseudo-Hamiltonian on an 8D phase space}\label{sec:8D pseudo-Hamiltonian}

Since the particle's self-forced motion is geodesic in the effective metric $\tilde g_{\alpha\beta}$, its equation of motion~\eqref{EOMv2} can be written as Hamilton's equations 
\begin{equation}
    \frac{dx^\alpha_p}{d\tilde\tau} = \frac{\partial H}{\partial\tilde p_\alpha} \quad \text{and} \quad \frac{d\tilde p_\alpha}{d\tilde\tau} = -\frac{\partial H}{\partial x^\alpha_p}\label{8D EOMs}\,,
\end{equation}
with the standard test-mass (pseudo-)Hamiltonian
\beq
H = \frac{1}{2\mu}\tilde g^{\alpha\beta}\tilde p_\alpha \tilde p_\beta.
\eeq
Here 
\begin{equation}
\tilde g^{\alpha\beta} = g^{\alpha\beta} - \e h^{\alpha\beta}_{\rm R(1)}-\e^2 \left(h^{\alpha\beta}_{\rm R(2)}-\frac{1}{2}g_{\gamma\delta}h^{\alpha\gamma}_{\rm R(1)}h^{\delta\beta}_{\rm R(1)}\right) + {\cal O}(\e^3),
\end{equation}
is the inverse of $\tilde g_{\alpha\beta}$. We emphasise that Eqs.~\eqref{8D EOMs} are identical to the original equation of motion~\eqref{EOMv2}, and in particular, these Hamilton's equations include all dissipative effects in the dynamics.

Following Refs.~\cite{Fujita:2016igj,Isoyama:2018sib}, we split $H$ into a background term and an interaction term: 
\beq
H=H_{(0)}(x^i_p,\tilde p_\alpha)+ H_{\rm int}(x^\alpha_p,\tilde p_\alpha,\e; \gamma,\delta M_A),
\eeq
with
\begin{align}
    H_{(0)} &:= \frac{1}{2\mu}g^{\alpha\beta}(x^i_p)\tilde p_{\alpha}\tilde p_{\beta}\label{H0(x,p)}\,,
\end{align}  
and
\beq
    H_{\rm int} = \sum_{n\geq1}\e^n H_{(n)}(x^\alpha_p,\tilde p_\alpha;\gamma,\delta M_A),\label{Hint(x,p)}
\eeq
where
\begin{align}
 H_{(1)} &= -\frac{1}{2\mu}h^{\alpha\beta}_{{\rm R}(1)}(x^\mu_p; \gamma,\delta M_A)\tilde p_{\alpha}\tilde p_{\beta},\label{H1(x,p)}\\
  H_{(2)} &= -\frac{1}{2\mu}\left(h^{\alpha\beta}_{{\rm R}(2)} - \frac{1}{2}g_{\gamma\delta}h^{\alpha\gamma}_{{\rm R}(1)}h^{\delta\beta}_{{\rm R}(1)}\right)\tilde p_{\alpha}\tilde p_{\beta}.\label{H2(x,p)}
\end{align}

We refer to $H$ as a pseudo-Hamiltonian, loosely following Ref.~\cite{Blanco:2022mgd}, because it depends on the entire phase-space trajectory $\gamma$, not only on a point in phase space. We emphasise that the derivatives in Hamilton's equations~\eqref{8D EOMs} do \emph{not} act on this functional dependence on $\gamma$; acting on that dependence would yield equations of motion that differ from the original equation~\eqref{EOMv2} (an equation which is a direct consequence of the Einstein field equations~\cite{Pound:2017psq}). This means our pseudo-Hamiltonian description differs from a nonlocal Hamiltonian description, in which the partial derivatives in Hamilton's equations become functional derivatives that do act on the nonlocal phase-space dependence~\cite{Bernard:2016wrg}. One consequence of this is that we are able to evaluate the source trajectory on shell.

Our pseudo-Hamiltonian description also differs from the one in Ref.~\cite{Blanco:2022mgd}. In that reference, which introduced the ``pseudo'' nomenclature, Hamilton's equations are made local through order reduction. $H_1$ is treated as a function of two points in phase space, which in our case we would denote $Q=(x^\alpha_p,\tilde p_\alpha)$ and $Q'=(x'^\alpha_p,\tilde p'_\alpha)$. The order reduction is performed by replacing the self-accelerated trajectory $\gamma$ in Eq.~\eqref{hRpp} with a zeroth-order (background geodesic) trajectory emanating from $Q'$. In Hamilton's equations, after taking the partial derivatives at $Q$,  one then evaluates at coincidence, $Q'=Q$. We can view our Eq.~\eqref{8D EOMs} in the same manner, with $\gamma$ emanating from $Q'$ and evaluation at $Q'=Q$ taking place after differentiation. Concretely, we can write $H=H(Q,Q')$ and write Eq.~\eqref{8D EOMs} more explicitly as
\begin{equation}
    \frac{dx^\alpha_p}{d\tilde\tau} = \left[\frac{\partial H}{\partial\tilde p_\alpha}\right] \quad \text{and} \quad \frac{d\tilde p_\alpha}{d\tilde\tau} = -\left[\frac{\partial H}{\partial x^\alpha_p}\right],\label{8D EOMs with coincidence}
\end{equation}
where the square brackets denote evaluation at $Q'=Q$. However, we stress that in our self-consistent pseudo-Hamiltonian description, $\gamma$ is always the self-consistently generated trajectory, as described in Sec.~\ref{sec:review}, and Hamilton's equations are integro-differential. Equivalently, we can view $\gamma$ as a fixed trajectory, and the square brackets denote evaluation at a point $Q=Q'$ on the trajectory; this is the more natural perspective stemming from the original equations of motion~\eqref{EOM} or~\eqref{EOMv2}, in which one differentiates a field and then evaluates the derivative on the particle's worldline. In any case, at this stage, one can view Eq.~\eqref{8D EOMs with coincidence} less as a set of equations to be solved and more as a property of solutions to the Einstein equation for given exact trajectories $\gamma$. Only after our localization procedure in later sections will we reduce the equations of motion to ordinary differential equations.

\subsection{Pseudo-Hamiltonian on a 6D phase space}\label{sec:6D pseudo-Hamiltonian}

We could work directly from the 8D equations of motion. However, to mesh with the multiscale expansion, it will be more useful to work with a 6D phase space, with $t$ reduced to a parameter along trajectories through phase space. Concretely, we reduce the phase space to the 6D submanifold defined by the on-shell condition $H=-\mu/2$, with $\tilde p_t$ the new pseudo-Hamiltonian. Explicitly, solving
\begin{equation}
    \label{phase space reduction}
    H(t,x^i_p,\tilde p_t,\tilde p_i;\gamma,\delta M_A)=-\frac{\mu}{2}
\end{equation}
yields
\begin{equation}
\tilde p_t=-\Hsix(t,x^i_p,\tilde p_i;\gamma_{\rm 6D},\delta M_A),
\end{equation}
with
\begin{equation}\label{Hsix}
    \Hsix = \frac{1}{\tilde g^{tt}}\left[\tilde g^{ti}\tilde p_i-\sqrt{\left(\tilde g^{ti}\tilde p_i\right)^2-\tilde g^{tt}\left(\tilde g^{ij}\tilde p_i \tilde p_j+\mu^2\right)}\right].
\end{equation}
Here $\gamma_{\rm 6D}$ is  the 6D trajectory
\beq
\gamma_{\rm 6D}:=\{(x^i_p(t),\tilde p_i(t))\,|\, t\in\mathbb{R}\}.
\eeq

As mentioned previously, the 6D phase space is the cotangent bundle of $\Sigma$. On this phase space, $(x^i_p,\tilde p_i)$ are cotangent coordinates, and $dx^i_p\wedge d\tilde p_i$ is a canonical symplectic form.\footnote{An observant reader might note that some physical intuition is lost in this reduction to six dimensions because $\tilde p_i$ is not bijectively related to the ``spatial velocity'' $\tilde u^i$ by $\tilde p_i =\mu\tilde g_{ij}\tilde u^j$. Instead, the two are related by $\tilde p_i =\mu(\tilde g_{ij}\tilde u^j+\tilde g_{ti}\tilde u^t)$. However, the cotangent vector $\tilde p_i dx^i_p$ on $\Sigma$ still suffices to define cotangent coordinates $(x^i_p,\tilde p_i)$ and the canonical symplectic form~\cite{CannasdaSilva}.} The 6D self-forced trajectories obey (pseudo-)Hamilton's equations,
\begin{equation}
\label{eq:eom_6D coincidence}
    \frac{dx^i_p}{dt}=\left[\pdv{\Hsix}{\tilde p_i}\right],\;\;\;\;\;\frac{d\tilde p_i}{dt}=-\left[\pdv{\Hsix}{x^i_p}\right].
\end{equation}
As before, the derivatives do not act on the functional dependence on $\gamma_{\rm 6D}$. The square brackets now denote evaluation at ${\cal Q}={\cal Q}'$, with ${\cal Q}=(x^i_p,\tilde p_i)$ now a point in the 6D phase space and ${\cal Q}'=(x'^i_p,\tilde p'_i)$ a point on $\gamma_{\rm 6D}$.

Equations~\eqref{eq:eom_6D coincidence} follow immediately from Eqs.~\eqref{8D EOMs} and~\eqref{phase space reduction}. For example, differentiating $H(t,x^i_p,-\Hsix(\tilde p_i,x^i_p),\tilde p_i)=-\mu/2$ with respect to $\tilde p_i$, we obtain 
\beq
-\frac{\partial H}{\partial \tilde p_t}\frac{\partial\Hsix}{\partial\tilde p_i} +\frac{\partial H}{\partial\tilde p_i} =0, 
\eeq
which we can rewrite as 
\beq
\frac{dt}{d\tilde \tau}\frac{\partial\Hsix}{\partial \tilde p_i} =\frac{dx^i_p}{d\tilde \tau}\,,
\eeq
using Eq.~\eqref{8D EOMs}. Multiplying by $\frac{d\tilde \tau}{dt}$ and evaluating at coincidence, we obtain Eq.~\eqref{eq:eom_6D coincidence}. 

We can derive the expansion of $\Hsix$ by substituting
\beq
\Hsix = \Hsix_{(0)}(x^i_p,\tilde p_i) + \sum_{n\geq1}\e^n \Hsix_{(n)}(t,x^i_p,\tilde p_i;\gamma_{\rm 6D},\delta M_A)
\eeq
into Eq.~\eqref{phase space reduction}. Here the expansion is performed at fixed $(t,x^i_p,\tilde p_i;\gamma_{\rm 6D},\delta M_A)$. 
This differs from the expansions in Secs.~\ref{sec:self-consistent formalism} and \ref{sec:8D pseudo-Hamiltonian} in that $\tilde p_t$ is now re-expanded. Writing $H(t,x^i_p,-\Hsix,\tilde p_i)=-\mu/2$ (with $\gamma$ and $\delta M_A$ dependence suppressed), substituting the expansions of $H$ and $\Hsix$, and solving order by order in $\e$, we obtain
\begin{align}
    H_{(0)}|_{\tilde p_t=-\Hsix_0} &= -\mu/2\label{3D H0}  
\end{align}
at zeroth order, and
\begin{align}
    \Hsix_{(1)} &= \frac{H_{(1)}}{\partial_{\tilde p_t}H_{(0)}}\Bigr|_{\tilde p_t=-\Hsix_{(0)}},\label{3D H1}\\
    \Hsix_{(2)} &= \frac{H_{(2)}}{\partial_{\tilde p_t}H_{(0)}}\biggr|_{\tilde p_t=-\Hsix_{(0)}} \\
    &\quad +\frac{\Hsix_{(1)} \left(\Hsix_{(1)}\partial^2_{\tilde p_t} H_{(0)} -2 \partial_{\tilde p_t} H_{(1)}\right)}{2 \partial_{\tilde p_t}H_{(0)}}\biggr|_{\tilde p_t=-\Hsix_{(0)}}\label{3D H2}  
\end{align}
through second order. Similar expansions at first order, in a similar context, were carried out in Ref.~\cite{Blanco:2022mgd}.

Equation~\eqref{3D H0} determines that $\Hsix_{(0)}(x^i_p,\tilde p_i)$ is the background-geodesic orbital energy as a function of $(x^i_p,\tilde p_i)$, 
\beq
\Hsix_{(0)}(x^i_p,\tilde p_i) = E_{(0)}(x^i_p,\tilde p_i) := \Estar,
\eeq
where we use a star to denote variables that are related to $(x^i_p,\tilde p_i)$ by the background-geodesic relationship. Written in the form~\eqref{Hsix},
\begin{equation}\label{Hsix0}
    \Hsix_{(0)}(x^i_p,\tilde p_i) = \frac{1}{g^{tt}}\left[ g^{ti}\tilde p_i-\sqrt{\left( g^{ti}\tilde p_i\right)^2- g^{tt}\left( g^{ij}\tilde p_i \tilde p_j+\mu^2\right)}\right]\!.
\end{equation}
Equation~\eqref{3D H1} can be simplified using 
\begin{align}
\partial_{\tilde p_t}H_{(0)} &= \frac{1}{\mu}g^{t\alpha}\tilde p_\alpha,\\  \partial^2_{\tilde p_t}H_{(0)} &= \frac{1}{\mu}g^{tt},\\
\partial_{\tilde p_t}H_{(1)} &= -\frac{1}{\mu}h^{t\alpha}_{\rm R(1)}\tilde p_\alpha, 
\end{align}
which imply
\begin{align}
    \Hsix_{(1)} &= -\frac{1}{2}h^{\alpha\beta}_{\rm R(1)}\frac{\tilde p_\alpha \tilde p_\beta}{\mu\utstar},\label{H1=hRuu}\\
    \Hsix_{(2)} &= -\frac{1}{2}h^{\alpha\beta}_{\rm R(2)}\frac{\tilde p_\alpha \tilde p_\beta}{\mu\utstar} +\frac{1}{4}g_{\gamma\delta}h^{\alpha\gamma}_{\rm R(1)}h^{\beta\delta}_{\rm R(1)}\frac{\tilde p_\alpha\tilde p_\beta}{\mu\utstar}\nonumber\\
    &\quad-\frac{1}{2}h^{\alpha\beta}_{\rm R(1)} h^{t\gamma}_{\rm R(1)}\frac{\tilde p_\alpha\tilde p_\beta \tilde p_\gamma}{(\mu \utstar)^2}\nonumber\\
    &\quad +\frac{1}{8}g^{tt} h^{\alpha\beta}_{\rm R(1)} h^{\gamma\delta}_{\rm R(1)}\frac{p_\alpha\tilde p_\beta \tilde p_\gamma \tilde p_\delta}{(\mu \utstar)^3},\label{H2=hRuu}    
\end{align}
with 
\beq
\mu\utstar = g^{t\alpha}(\bm{x}_p)\tilde p^{(0)}_\alpha = -g^{tt}(\bm{x}_p)\Estar+g^{t\phi}(\bm{x}_p)\Lstar.
\eeq

\subsection{Osculating action-angle variables}
\label{sec:osculating action angles}

In Sec.~\ref{sec:multiscale expansion}, we obtained the preferred coordinates $(\phiring^i,\piring_i)$ by first constructing quasi-Keplerian phase-space coordinates $(\psi^i,\pi_i)$ that are related to $(x^i_p,\dot x^i_p)$ by the Kerr-geodesic relationships. Our analysis in later sections will instead begin from coordinates $(\phistar^i,\Jstar_i)$ that are related to $(x^i_p,\tilde p_i)$ by Kerr-geodesic relationships. We refer to these as `osculating action angles'. They are the same coordinates used in Ref.~\cite{Fujita:2016igj}, except on the 6D rather than 8D phase space, and the content of this section is equivalent to material in Ref.~\cite{Fujita:2016igj}.

We first recall the construction of action angles for geodesics of Kerr. Since the geodesic motion is tri-periodic, it is confined to a 3-torus in phase space, where the torus is defined by constant values of the orbital parameters $P_i=(E,L_z,K)$. We can therefore construct action-angle coordinates $(\varphi^i,J_i)$ on phase space. The action variables are defined to be
\begin{equation}
    \label{J geo}
    J_i^{(0)}(P_j)=\frac{1}{2\pi}\oint_{C_i(P_j)}p^{(0)}_k(x^j_p,P_j) dx^k_p,
\end{equation}
where the contours $C_i(P_j)$ are any representatives of the three distinct loops on the 3-torus, and $p^{(0)}_k(x^j_p,P_j)$ are the Kerr-geodesic momenta evaluated on the torus. The momenta are given by~\cite{Schmidt:2002qk}
\begin{align}
p^{(0)}_r &= \frac{1}{\Delta}\Bigl\{\left[(r^2_p+a^2)E - a L_z\right]^2 \nonumber\\
&\qquad\quad - \Delta\left[\mu^2 r^2_p+(L_z - a E)^2+Q\right]\Bigr\}^{1/2},\\
p^{(0)}_\theta &= \left\{Q - \left[(\mu^2-E^2)a^2+ \frac{L^2_z}{\sin^2 \theta_p}\right] \cos^2\theta_p\right\}^{1/2},\\
p^{(0)}_\phi &= L_z,
\end{align}
with $\Delta:=r^2_p - 2M r_p+a^2$ and $Q:=K-(L_z-a E)^2$. The geodesic actions $J^{(0)}_i$ are given in closed form in terms of elliptic integrals in Ref.~\cite{Witzany:2024ttz}. We immediately see that $J^{(0)}_\phi=L_z$. 

The canonical transformation to action-angle variables is generated by Hamilton's characteristic function 
\beq
W_{(0)}(x^i_p,J_i)=S_r(r_p,J_i)+S_\theta(\theta_p,J_i)+ L_z \phi_p, 
\eeq
which satisfies the time-independent Hamilton-Jacobi equation
\begin{equation}
    \Hsix_{(0)}\!\left(x^i_p,\pdv{W_{(0)}}{x^i_p}\right)=E.
\end{equation}
Here $\Hsix_{(0)}\!\left(x^i_p,p_i\right)$ is given by Eq.~\eqref{Hsix0}, and the characteristic function reads more explicitly
\begin{multline}\label{W}
    W_{(0)}(x^i_p,J_k)=L_z \phi_p + \int^{r_p} p^{(0)}_r(r',P_i(J_k))\,dr'\\ + \int^{\theta_p} p^{(0)}_\theta(\theta',P_i(J_k))\, d\theta'.
\end{multline} 
The angle variables canonically conjugate to $J_i$ are then derivatives of this type-2 generating function, 
\begin{equation}\label{phi = dW/dJ}
    \varphi^i=\pdv{W_{(0)}}{J_i},
\end{equation}
and $J_i$ are implicitly defined by $p^{(0)}_i(x^j,P_j(J_k))=\partial W/\partial x^i_p$. 
The geodesic equation in Kerr becomes
\begin{align}
    \frac{d\varphi^i}{dt}&=\pdv{\Hsix_{(0)}}{J^{(0)}_i}=\Omega^i_{(0)}(J^{(0)}_j), \label{phigeo}\\
    \frac{dJ^{(0)}_i}{dt}&=-\pdv{\Hsix_{(0)}}{\varphi^i}=0.\label{Jgeo}
\end{align}

We now define osculating action-angle variables $(\phistar^i,\Jstar_i)$ by using the Kerr-geodesic map $(x^i_p,\tilde p_i)\overset{\rm geo}{\mapsto} (\phistar^i,\Jstar_i)$. The osculating action variables are then 
\begin{equation}
    \label{J geo}
    \Jstar_i(x^j_p,\tilde p_j)=\frac{1}{2\pi}\oint_{C_i(\Pstar_j)}p^{(0)}_k\bigl(x'^j_p,\Pstar_j(x^k_p,\tilde p_k)\bigr) dx'^k_p,
\end{equation}
and the angle variables are
\begin{equation}\label{phistar = dW/dJstar}
    \phistar^i=\pdv{W_{(0)}}{\Jstar_i},
\end{equation}
with the Kerr-geodesic generating function $W_{(0)}=W_{(0)}(x^i_p,\Jstar_i)$. Here $\Estar=E_{(0)}(x^i_p,\tilde p_i)$ is given by Eq.~\eqref{Hsix0}, and
\beq
\Lstar=\tilde p_\phi, \quad \Kstar = K^{\alpha\beta}\tilde p_\alpha \tilde p_\beta,
\eeq
with $K^{\alpha\beta}$ the Killing tensor of Kerr. 

The definition~\eqref{phistar = dW/dJstar} leaves a residual freedom to shift $\phistar^i$ by an arbitrary function of $\Jstar_i$, $\phistar^i\to\phistar^i+\Delta\phistar^i(\Jstar_j)$, corresponding to the freedom to choose lower limits of integration in Eq.~\eqref{W}. This is equivalent to shifting the origin on each torus of constant $\Jstar_i$. For simplicity, we make the choice that $\phistar^i$ shares an origin with $\psi^i$. In particular, this implies that $\phistar^i=0$ corresponds to a turning point in the radial and polar motion (at fixed $\Jstar_i$); $\phistarvec_O=0$ in notation analogous to that of Eqs.~\eqref{fdiss} and~\eqref{fcon}. This does not represent a loss of generality because the same freedom will arise in the coordinates $\phiring^i$.

Hamilton's equations for the variables $(\phistar^i,\Jstar_i)$ read 
\begin{align}
    \frac{d\phistar^i}{dt} = \left[\frac{\partial\Hsix}{\partial\Jstar_i}\right] \quad \text{and} \quad
    \frac{d\Jstar_i}{dt} = -\left[\frac{\partial\Hsix}{\partial\phistar^i}\right].\label{eq:Jistardot}
\end{align}
One might suspect that applying the Kerr-geodesic transformation to Eq.~\eqref{eq:eom_6D coincidence} would not yield Eq.~\eqref{eq:Jistardot} because $\gamma_{\rm 6D}$ is not a Kerr geodesic trajectory. However, Eq.~\eqref{eq:Jistardot} automatically follows from Eq.~\eqref{eq:eom_6D coincidence} because we use a generating function [\eqref{W} with $p_i\to \tilde p_i$, $P_i\to \Pstar_i$, and $J_i\to \Jstar_i$], for the transformation, which ensures the new coordinates are canonical in the sense of satisfying the pseudo-Hamilton equations~\eqref{eq:Jistardot}; this is true for any ${\cal H}$, so long as the generating function only depends on the phase-space coordinates and not on the source trajectory $\gamma_{\rm 6D}$ (such that the Jacobian of the transformation commutes with the coincidence limit $[\cdot]$ in Hamilton's equations). 

Note that in these coordinates, ${\cal H}_{(0)}={\cal H}_{(0)}(\Jstar_i)$, meaning $\frac{\partial{\cal H}_{(0)}}{\partial\phistar^i}=0$ and $\frac{\partial {\cal H}_{(0)}}{\partial\Jstar_i}=\Omega^i_{(0)}(\Jstar_j)$, where $\Omega^i_{(0)}(\Jstar_j)$ are the Kerr geodesic frequencies as functions of $\Jstar_i$. Our equations of motion then become 
\begin{align}
    \frac{d\phistar^i}{dt} &= \Omega^i_{(0)}(\Jstar_j) + \e\Omegastar^i_{(1)}(\phistar^j,\Jstar_j)+{\cal O}(\e^2),\label{eq:phistardot expanded}\\
    \frac{d\Jstar_i}{dt} &= \e \left[G^{(0)}_i(\phistar^j,\Jstar_j) + \e G^{(1)}_i(\phistar^j,\Jstar_j) + {\cal O}(\e^2)\right],\label{eq:Jstardot expanded}
\end{align}
where 
\begin{align}
\Omegastar^i_{(n)}(\phistar^j,\Jstar_j) &= \left[\frac{\partial\Hsix_{(n)}}{\partial\Jstar_i}\right],\label{OmegaStar_n}\\
G^{(n)}_i(\phistar^j,\Jstar_j) &= -\left[\frac{\partial\Hsix_{(n+1)}}{\partial\phistar^i}\right].\label{G_n}
\end{align}
Here the numeric labels in parentheses denote the post-adiabatic order at which each term enters a multiscale expansion of the Einstein equation~\cite{Hinderer:2008dm,Pound:2021qin}.

Equations~\eqref{eq:phistardot expanded} and~\eqref{eq:Jstardot expanded} will be the starting point of Sec.~\ref{sec:localized Hamiltonian for bound orbits}.

\section{Localized pseudo-Hamiltonian description of 1PA dynamics}
\label{sec:localized Hamiltonian for bound orbits}

The equations~\eqref{eq:phistardot expanded} and~\eqref{eq:Jstardot expanded} have the same general form as equations~\eqref{osculating psi v2} and \eqref{osculating pi v2}. To set the stage for what follows, we underline two features of these equations. First, they are not in the form~\eqref{dphidt}--\eqref{dpidt} suitable for a multiscale expansion of the field equations because the frequency corrections $\Omegastar^i_{(n)}$ and forcing functions $G^{(n)}_i$ depend on the angle variables $\phistar^i$. In other words, these equations have not yet separated the system's periodic motions from its slow evolution. Second, they are gauge dependent: $(\phistar^i,\Jstar_i)$ are related in a fixed way to $(x^i_p,\tilde p_i)$. A spacetime gauge transformation $x^\alpha\to x^\alpha - \e\xi^\alpha+\ldots$ alters $x^i_p$ and $\tilde p_i$, inducing a change in the variables $(\phistar^i,\Jstar_i)$.

In this section, we find the most general averaging transformation to coordinates $(\phiring^i,\Jring_i)$ satisfying equations of the general form~\eqref{dphidt}--\eqref{dpidt}, analogous to the transformation~\eqref{psi to phiring}--\eqref{pi to piring}. We use these coordinates, together with a stationary phase approximation, to eliminate the nonlocal dependence on $\gamma_{\rm 6D}$ in the 1PA dynamics. The outcome is a pseudo-Hamiltonian that is an explicit local function of two points in phase space, ${\cal Q}=(\phiring^i,\Jring_i)$ and ${\cal Q}'=(\phiring'^i,\Jring'_i)$. We then use this local form to analyse the conservative and dissipative sectors of the dynamics. In our analysis, we largely eliminate the gauge dependence of the variables, an issue we return to in Secs.~\ref{sec:invariance of Jtilde} and~\ref{sec:gauge choices}.

Averaging transformations also underlie the description of 1SF conservative Hamiltonian dynamics in Ref.~\cite{Fujita:2016igj}. However, our treatment differs from the treatment there in several ways:
\begin{enumerate}
    \item We include dissipation, we carry the analysis to 1PA order, and we work on the 6D phase space appropriate for the multiscale expansion of the field equations.
    \item We do not require our transformation to be canonical (and in fact, it \emph{cannot} be canonical). Instead we only require compatibility with a multiscale expansion. We return to the question of canonical coordinates in Secs.~\ref{sec:conservative sector} and~\ref{sec:gauge choices}. Ultimately, in those sections we specialize to a non-canonical transformation that brings the system from coordinates that are canonical in the \emph{pseudo}-Hamiltonian system to action-angle coordinates that are canonical in a conservative, \emph{Hamiltonian} system. 
    \item We do not require our transformation to be induced by a gauge transformation $x^\alpha\to x^\alpha-\e \xi^\alpha + \ldots$ on spacetime. Reference~\cite{Fujita:2016igj} required their (8D) coordinates $(\phistar^\alpha,\Jstar_\alpha)$ to be related to $(x^\alpha_p,\tilde p_\alpha)$ by the map to Kerr-geodesic action-angles, $(x^\alpha_p,\tilde p_\alpha)\overset{\rm geo}{\mapsto} (\phistar^\alpha,\Jstar_\alpha)$. Their phase-space coordinates could then only be modified through a transformation of the spacetime coordinates. We avoid that restriction, instead keeping the spacetime coordinates fixed and changing their relationship to the phase-space coordinates; this is the more typical approach, ``re-parameterizing the orbit'',  in studies of self-forced dynamics (e.g., in~\cite{Barack:2011ed,VanDeMeent:2018cgn,Pound:2021qin,Barack:2022pde,Drummond:2022efc,Mathews:2025nyb}). It has the considerable advantage of not affecting the choice of spacetime gauge used in solving the field equations. 
\end{enumerate}

\subsection{Perturbed action angles}
\label{sec:perturbed_action_angles}
Our first goal is to transform to new variables $(\phiring^i,\Jring_i)$ satisfying equations of the form
\begin{align}
    \frac{d\phiring^i}{dt} &= \Omega^i_{(0)}(\Jring_j) + \e \Omega^i_{(1)}(\Jring_j)+{\cal O}(\e^2),\label{eq:phiringdot}\\
    \frac{d\Jring_i}{dt} &= \e \left\{\mathring{G}^{(0)}_i(\Jring_j) + \e\mathring{G}^{(1)}_i(\Jring_j) + {\cal O}(\e^2)\right\}.\label{eq:Jringdot}
\end{align}
These coordinates cleanly separate rapid oscillations from secular evolution in the full system that includes dissipation. We refer to them as perturbed action-angle coordinates, though, as mentioned above, we do not require them to be canonically conjugate.

To obtain the new variables, we adopt an ansatz
\begin{align}
\phiring^i &= \phistar^i+\Delta\phiring^i(\Jring_j)+\e \phiring^i_{(1)}(\phistar^j,\Jstar_j)+{\cal O}(\e^2),\label{phistar to phiring}\\ 
\Jring_i &= \Jstar_i + \e \Jring^{(1)}_i(\phistar^j,\Jstar_j) +\e^2 \Jring_i^{(2)}(\phistar^j,\Jstar_j)+{\cal O}(\e^3),\label{Jstar to Jring}
\end{align}
where functions of $\phistar^i$ are $2\pi$-periodic; this ensures that the dynamics and the spacetime metric are $2\pi$-periodic in each angle $\phiring^i$. Note we could allow the order-$\e^0$ term in Eq.~\eqref{Jstar to Jring} to be an arbitrary smooth function of $\Jstar$, but it will be convenient to restrict to variables $\Jring_i$ that are a small deformation of the geodesic actions. We could also allow the zeroth-order term in Eq.~\eqref{phistar to phiring} to be any smooth function that increases by $2\pi$ when each $\phistar^i$ increases by $2\pi$. However, to mesh with the multiscale expansion of the field equations, $d\phiring^i/dt$ must be approximately the geodesic orbital frequency, as in Eq.~\eqref{eq:phiringdot}, which restricts the transformation to the form~\eqref{phistar to phiring}. If we were to ignore dissipation, then the zeroth-order term $\Delta\phiring^i(\Jring_j)$ would correspond to a choice of origin on each torus defined by $\Jring_i=\text{constant}$. In the presence of dissipation, this term evolves dynamically, effectively moving the origin of the angular coordinates as the system continuously evolves from one torus to the next~\cite{Pound:2021qin,Isoyama:2021jjd,Mathews:2025nyb}.  

Equations~\eqref{phistar to phiring} and \eqref{Jstar to Jring} imply
\begin{align}
    \frac{d\phiring^i}{dt} &= \frac{d\phistar^i}{dt} + \e\frac{\partial\Delta\phiring^i}{\partial\Jring_j}\mathring{G}^{(0)}_j(\Jring_j) + \e\frac{\partial\phiring^i_{(1)}}{\partial\phistar^j}\Omega^j_{(0)} + {\cal O}(\e^2),\label{phiring dot v0}\\
    \frac{d\Jring_i}{dt} &= \frac{d\Jstar_i}{dt} + \e\frac{\partial\Jring_i^{(1)}}{\partial\phistar^j}\Omega^j_{(0)}  +\e^2 \Biggl(\pdv{\Jring_i^{(1)}}{\phistar^j}\Omegastar^j_{(1)}+\pdv{\Jring_i^{(1)}}{\Jstar_j}G_j^{(0)}\nonumber\\
   &\qquad\qquad\qquad\qquad\quad +\pdv{\Jring_i^{(2)}}{\phistar^j}\Omega^j_{(0)}\Biggr) +{\cal O}(\e^3).\label{Jring dot v0}
\end{align}
Here we have used Eq.~\eqref{eq:Jringdot} to evaluate $d\Delta\phiring^i/dt$, and all quantities are functions of $(\phistar^i,\Jstar_i)$ unless indicated otherwise. Substituting Eqs.~\eqref{eq:phistardot expanded} and~\eqref{eq:Jstardot expanded} for $d\phistar^i/dt$ and $d\Jstar_i/dt$, with Eqs.~\eqref{phistar to phiring} and \eqref{Jstar to Jring}, we get%
\begingroup\allowdisplaybreaks%
\begin{align}
    \frac{d\phiring^i}{dt} &= \Omega^i_{(0)} + \e\Biggl(\Omegastar^i_{(1)} +\frac{\partial\Delta\phiring^i}{\partial\Jring_j}\mathring{G}^{(0)}_j\nonumber\\*
    &\qquad\qquad\quad - \Jring^{(1)}_j\frac{\partial\Omega^i_{(0)}}{\partial\Jring_j} + \frac{\partial\phiring^i_{(1)}}{\partial\phiring^j}\Omega^j_{(0)}\Biggr) + {\cal O}(\e^2),\label{phiring dot}\\
    \frac{d\Jring_i}{dt} &= \e \left(G^{(0)}_i + \frac{\partial\Jring_i^{(1)}}{\partial\phiring^j}\Omega^j_{(0)}\right)\nonumber \\*
    &\quad+ \e^2\Biggr\{G_i^{(1)}+\pdv{\Jring_i^{(1)}}{\phiring^j}\Omegastar^j_{(1)}+\pdv{\Jring_i^{(1)}}{\Jring_j}G_j^{(0)}+\pdv{\Jring_i^{(2)}}{\phiring^j}\Omega^j_{(0)}\nonumber \\*
    &\qquad-\left(\phiring^j_{(1)}\frac{\partial}{\partial \phiring^j}+\Jring_j^{(1)}\frac{\partial}{\partial \Jring_j}\right)\!\left(G_i^{(0)}+\frac{\partial\Jring_i^{(1)}}{\partial\phiring^j}\Omega^j_{(0)}\right)\!\Biggr\} \nonumber\\*
    &\quad+{\cal O}(\e^3). \label{Jring dot}
\end{align}%
\endgroup%
On the right, and throughout the remainder of this section, all functions of $(\phistar^i,\Jstar_i)$ are now evaluated at $(\phistar^j_{(0)},\Jring_j)$, where 
\begin{equation}
\phistar^i_{(0)}:=\phiring^i-\Delta\phiring^i.
\end{equation}

Eliminating oscillations from the right-hand side of Eqs.~\eqref{phiring dot} and \eqref{Jring dot} requires
\begin{align}
    \frac{\partial\phiring^i_{(1)}}{\partial\phiring^j}\Omega^j_{(0)} &= -\Biggl(   \Omegastar^i_{(1)} - \Jring^{(1)}_j\frac{\partial\Omega^i_{(0)}}{\partial\Jring_j}+\frac{\partial\Delta\phiring^i}{\partial\Jring_j}\mathring{G}^{(0)}_j\nonumber\\
    &\qquad-\left\langle \Omegastar^i_{(1)} - \Jring^{(1)}_j\frac{\partial\Omega^i_{(0)}}{\partial\Jring_j}+\frac{\partial\Delta\phiring^i}{\partial\Jring_j}\mathring{G}^{(0)}_j\right\rangle\Biggr),\label{phi1 oscillatory}\\
    \frac{\partial\Jring_i^{(1)}}{\partial\phiring^j}\Omega^j_{(0)} &= - \left(G^{(0)}_i-\left\langle G^{(0)}_i\right\rangle\right)\label{J1 oscillatory}.
\end{align}
We similarly choose the oscillatory part of $\Jring^{(2)}_i$ to cancel all oscillations in the order-$\e^2$ term in Eq.~\eqref{Jring dot}.

By eliminating the oscillations from Eqs.~\eqref{phiring dot} and~\eqref{Jring dot}, we reduce them to 
\begin{align}
    \frac{d\phiring^i}{dt} &= \Omega^i_{(0)}(\Jring_j) + \e\Omega^i_{(1)}(\Jring_j) + {\cal O}(\e^2),\label{phidot=Omega expansion}\\
    \frac{d\Jring_i}{dt} &= \e \left\langle G^{(0)}_i\right\rangle + \e^2\left(\left\langle G_i^{(1)}\right\rangle +K_i\right) + {\cal O}(\e^3),
\end{align}
where
\begin{equation}\label{eq:Omega1def}
    \Omega^i_{(1)}:=\left\langle\Omegastar^i_{(1)}\right\rangle  +\frac{\partial\Delta\phiring^i}{\partial\Jring_j} \left\langle G^{(0)}_j\right\rangle - \left\langle\Jring^{(1)}_j\right\rangle\frac{\partial\Omega^i_{(0)}}{\partial\Jring_j}
\end{equation}
and
\begin{multline}\label{Kdef}
    K_i := \Biggl\langle\pdv{\Jring_i^{(1)}}{\phiring^j}\Omegastar^j_{(1)}
    +\pdv{\Jring_i^{(1)}}{\Jring_j}G_j^{(0)}-\pdv{\langle G_i^{(0)}\rangle}{\Jring_j}\Jring_j^{(1)}\Biggr\rangle.
\end{multline}
In obtaining the latter, we have used the fact that $\partial\langle G^{(0)}_i\rangle/\partial\phiring^j=0$. 

Expressed in terms of the pseudo-Hamiltonian, these equations read
\begin{align}
    \frac{d\phiring^i}{dt} 
    &= \Omega^i_{(0)}(\Jring_j) + \e\Biggl\{\left\langle\left[\frac{\partial\Hsix_{(1)}}{\partial\Jring_i}\right]\right\rangle -\frac{\partial\Delta\phiring^i}{\partial\Jring_j} \left\langle\left[\frac{\partial\Hsix_{(1)}}{\partial\phiring^j}\right]\right\rangle \nonumber\\
    &\qquad\qquad\qquad\quad -\left\langle\Jring^{(1)}_j\right\rangle\frac{\partial\Omega^i_{(0)}}{\partial\Jring_j}\Biggr\} + {\cal O}(\e^2),\label{dphidt generic}\\
    \frac{d\Jring_i}{dt}
    &= -\e\left\langle\left[\frac{\partial\Hsix_{(1)}}{\partial\phiring^i}\right]\right\rangle - \e^2\left(\left\langle\left[\pdv{\Hsix_{(2)}}{\phiring^i}\right]\right\rangle-K_i\right)\nonumber\\
    &\qquad\qquad\qquad\qquad\qquad\qquad\qquad\quad +{\cal O}(\e^3),\label{dJdt generic}
\end{align}
We stress that at this stage, $\Delta\phiring^i$ and $\left\langle\Jring^{(1)}_j\right\rangle$ are arbitrary functions of $\Jring_i$. We will discuss choices of these functions in Sec.~\ref{sec:conservative and dissipative}.

Finally, since the stress-energy tensor and punctures entering the field equations are explicit functions of $(x^i_p,\tilde p_i)$, we also require $x^i_p$ and $\tilde p_i$ as functions of $(\phiring^i,\Jring_i)$ [i.e., the analogue of Eqs.~\eqref{x(phi,pi)} and \eqref{p(phi,pi)}]. Given that $x^i_p = x^i_{(0)}(\phistar^j,\Jstar_j)$, $\tilde p_i = p_i^{(0)}(\phistar^j,\Jstar_j)$, we can immediately substitute Eqs.~\eqref{phistar to phiring} and \eqref{Jstar to Jring} to obtain
\begin{align}
    x^i_p &= x^i_{(0)}(\phistar^j_{(0)},\Jring_j) - \e\left( \phiring^j_{(1)}\frac{\partial x^i_{(0)}}{\partial \phistar^j_{(0)}}+\Jring_j^{(1)}\frac{\partial x^i_{(0)}}{\partial\Jring_j}\right)\! + \O(\e^2),\label{x(phi,J)}\\
    \tilde p_i &= p^{(0)}_i(\phistar^j_{(0)},\Jring_j) - \e\left(\phiring^j_{(1)}\frac{\partial p_i^{(0)}}{\partial \phistar^j_{(0)}}+\Jring_j^{(1)}\frac{\partial p_i^{(0)}}{\partial\Jring_j}\right)\! + \O(\e^2).\label{p(phi,J)}
\end{align}
The geodesic functions $x^i_{(0)}(\phistar^j_{(0)},\Jring_j)$ and $\tilde p_i^{(0)}(\phistar^j_{(0)},\Jring_j)$ are not known in closed form, but they can be obtained as a Fourier series in $\phistar^j_{(0)}$~\cite{Drasco:2003ky,Pound:2021qin}. Alternatively, since these functions are typically only ever needed within integrals over the phases $\phiring^i$, the integration variables can be changed to variables for which the geodesic functions are known analytically---for example, the Keplerian variables $(\psi^i,\pi_i)$ discussed in Sec.~\ref{sec:multiscale expansion} 
or Mino-time action-angles~\cite{Fujita:2009bp,Pound:2021qin,Lynch:2024hco}.

\subsection{Phase-space localization}
\label{sec:localization}

The equations of motion~\eqref{dphidt generic} and \eqref{dJdt generic} have eliminated dependence on the angle variables, meaning the right-hand sides are non-oscillatory. However, the right-hand sides still contain a nonlocal dependence on the phase-space trajectory. In this section, we show how the nonlocality collapses to a local function by virtue of the stationary phase approximation.

We can write out this localization starting with $\Hsix_1$ in Eq.~\eqref{H1=hRuu}:
\begin{align}
\Hsix_{(1)} &= -\frac{1}{2\mu} h^{\alpha\beta}_{\rm R(1)}(t,x^i_p)\frac{\tilde p_\alpha\tilde p_\beta}{\utstar}\\
&= -\frac{1}{2\mu^2} \int^{\infty}_{-\infty}dt' \frac{\tilde p_\alpha\tilde p_\beta}{\utstar} G^{\alpha\beta\alpha'\beta'}_{\rm R}(t-t',x^i_p,x^{i'}_p)\frac{\tilde p_{\alpha'}\tilde p_{\beta'}}{(\utstar)'}.
\end{align}
We write the Green's function in terms of its Fourier transform,
\begin{multline}\label{GR Fourier}
G^{\alpha\beta\alpha'\beta'}_{\rm R}(t-t',x^i_p,x^{i'}_p) \\= \int_{-\infty}^\infty d\omega\, e^{-i\omega (t-t')}\hat G_{\rm R}^{\alpha\beta\alpha'\beta'}(\omega,x^i_p,x^{i'}_p),
\end{multline}
to obtain
\begin{multline}\label{calH1}
\Hsix_{(1)} = -\frac{1}{2\mu^2} \int^{\infty}_{-\infty}\int^{\infty}_{-\infty}dt'd\omega\, e^{-i\omega (t-t')}\\
\times\frac{\tilde p_\alpha\tilde p_\beta}{\utstar} \hat G^{\alpha\beta\alpha'\beta'}_{\rm R}(\omega,x^i_p,x^{i'}_p)\frac{\tilde p_{\alpha'}\tilde p_{\beta'}}{(\utstar)'}.
\end{multline}
We next use the fact that $(x^{i}_p,\tilde p_i)$ are $2\pi$-periodic ($\e$-dependent) functions of each $\phiring^j$, such that we can expand in a discrete Fourier series:
\begin{multline}\label{Gkk}
\frac{\tilde p_\alpha\tilde p_\beta}{\mu^2\utstar} \hat G^{\alpha\beta\alpha'\beta'}_{\rm R}(\omega,x^i_p,x^{i'}_p)\frac{\tilde p_{\alpha'}\tilde p_{\beta'}}{\mu^2(\utstar)'} \\
= \sum_{\vec{k},\vec{k}'}\hat G_{\vec{k},\vec{k}'}(\omega,\Jring_i,\Jring'_i,\e)e^{i(k_i\phiring^i-k'_i\phiring'^i)},
\end{multline}
with 
\begin{multline}\label{Ghat expansion}
\hat G_{\vec{k},\vec{k}'}(\omega,\Jring_i,\Jring'_i,\e) \\
\ \ \ = \hat G^{(0)}_{\vec{k},\vec{k}'}(\omega,\Jring_i,\Jring'_i) + \e \hat G^{(1)}_{\vec{k},\vec{k}'}(\omega,\Jring_i,\Jring'_i) + {\cal O}(\e^2).
\end{multline}
Here $\hat G^{(n)}_{k_i, k'_i}$ are the mode coefficients obtained by substituting the expansions~\eqref{x(phi,J)} and \eqref{p(phi,J)} into the left-hand side of Eq.~\eqref{Gkk} and then Fourier expanding the coefficient of $\e^n$. Equation~\eqref{Gkk} in Eq.~\eqref{calH1} gives us
\begin{multline}\label{eq:H1 double int}
\Hsix_{(1)} = -\frac{\mu^2}{2} \sum_{\vec{k},\vec{k}'}\int^{\infty}_{-\infty}\int^{\infty}_{-\infty}dt'd\omega\, e^{i[k_i\phiring^i-k'_i\phiring'^i(t')-\omega (t-t')]}\\[-5pt]
\times \hat G_{\vec{k},\vec{k}'}[\omega,\Jring_i,\Jring'_i(t'),\e]. 
\end{multline}
Functions of $\Jring_i$ and $\Jring'_i$ here are technically functions of $\Jring_i/\mu$ and $\Jring'_i/\mu$.

The integral~\eqref{eq:H1 double int} is dominated by the contribution from values of $t'$ and $\omega$ where the exponential becomes approximately constant. Since we require the approximation to be valid on the radiation-reaction timescale $\sim 1/\e$, we introduce $\tilde t := \e t$ and $\xi^i(\e t) := \e\phiring^i(t)$,\footnote{Given that we have defined $\e$ as a formal counting parameter rather than a genuinely small quantity, these rescalings are purely formal. One could instead work with $\tilde t=\mu\, t/M$ and $\xi^i(\mu\, t/M)=(\mu/M) \phiring^i(t)$.} such that 
\begin{multline}
\hspace{-10pt}\Hsix_{(1)} = -\frac{\mu^2}{2\e} e^{ik_i\phiring^i}\sum_{\vec{k},\vec{k}'}\int^{\infty}_{-\infty}\int^{\infty}_{-\infty}d\tilde t'd\omega\, e^{-i[\xi_{k'}(\tilde t')+\omega (\tilde t-\tilde t')]/\e}\\
\times\hat G_{\vec{k},\vec{k}'}[\omega,\Jring_i,\Jring'_i(\tilde t'),\e], \label{Ham modes}
\end{multline}
where $\xi_{k'}:=k'_i\xi'^i$. This is an integral of the form
\begin{equation}
I = \iint g(x)e^{if(x)/\e} d^2x    
\end{equation}
with one saddle point of $f(x)$, say $x_0$, at which $(\partial_{x_1} f,\partial_{x_2}f)=0$. Such an integral admits a 2D stationary phase approximation, the general form of which can be found in \cite{Wong:doi:10.1137/1.9780898719260}. 
At leading order, this approximation reads
\begin{equation}\label{eq:2D stationary phase}
    I \approx \frac{2\pi\e g(x_0)}{|{\rm det}A(x_0)|^{1/2}} e^{if(x_0)/\e+i\pi \sigma/4},
\end{equation}
where $A=\frac{\partial^2 f}{\partial x_i\partial x_j}$ is the Hessian, and $\sigma$ is the signature of $A$ (the number of positive eigenvalues minus number of negative eigenvalues). In our case, $f(\omega,\tilde t')=-[\xi_{k'}(\tilde t')+\omega(\tilde t-\tilde t')]$, and the stationary point is $(\omega,\tilde t')=(\tilde\omega_{k'}(\tilde t),\tilde t)$, with $\tilde\omega_{k'}:= d\xi_{k'}/d\tilde t'$. The Hessian is
\begin{equation}
    A=\begin{pmatrix}\, 0 & 1\\
    \,1 & \ -\frac{d\tilde\omega_{k'}}{d\tilde t'}\end{pmatrix},
\end{equation}
which has $|{\rm det}A|=1$ and $\sigma=0$ (i.e. $A$ one positive eigenvalue and one negative). The extension of Eq.~\eqref{eq:2D stationary phase} to subleading order is explained in Appendix~\ref{app:stationary phase H1}. Applying it to Eq.~\eqref{Ham modes}, we obtain
\begin{align}
\hspace{-5pt}\Hsix_{(1)} &= - \mu^2\pi\sum_{\vec{k},\vec{k}'} e^{i[k_i\phiring^i-k'_i\phiring'^i(t)]}\Biggl\{\hat G^{(0)}_{\vec{k},\vec{k}'}[\omega'_{k'}(t),\Jring_i,\Jring'_i(t)]\nonumber\\
&\quad\qquad+\e\Biggl[\hat G^{(1)}_{\vec{k},\vec{k}'}[\omega'_{k'}(t),\Jring_i,\Jring'_i(t)]\nonumber\\
&\qquad\qquad\quad+k'_i\Omega'^i_{(1)}\partial_\omega \hat G^{(0)}_{\vec{k},\vec{k}'}[\omega'_{k'}(t),\Jring_i,\Jring'_i(t)]\nonumber\\
&\qquad\qquad\quad-i\frac{d{\omega}'_{k'}(\tilt)}{d\tilt}\pdv[2]{\hat G^{(0)}_{\vec{k},\vec{k}'}[\omega'_{k'}(t),\Jring_i,\Jring'_i(t)]}{\omega} \nonumber\\
&\qquad\qquad\quad -2i\frac{d\Jring'_i(\tilt)}{d\tilt}\pdv{\Jring'_i}\pdv{\hat G^{(0)}_{\vec{k},\vec{k}'}[\omega'_{k'}(t),\Jring_i,\Jring'_i(t)]}{\omega}\Biggr]\nonumber\\
&\qquad\qquad\qquad\qquad\qquad\qquad\qquad + \mathcal{O}(\e^2)\Biggr\}.\label{eq:H1 localized}
\end{align}
Here we have introduced
\begin{equation}
    \omega'_{k'} := k'_i\,\Omega^i_{(0)}(\Jring'_j)
\end{equation}
and expanded $\hat G_{\vec{k},\vec{k}'}(\tilde\omega_{k'},\Jring_i,\Jring'_i,\e)$ using Eqs.~\eqref{Ghat expansion} and~\eqref{phidot=Omega expansion}. 

Equation~\eqref{eq:H1 localized} expresses $\Hsix_{(1)}$ as a \emph{local} function on phase space. Concretely, it is a function of two points ${\cal Q}=(\phiring^i,\Jring_i)$ and ${\cal Q}'=(\phiring'^i,\Jring'_i)$. To the best of our knowledge, this expression and its derivation appear here for the first time. Arriving at this result is made possible by the clean identification of periodic variables and the separation of time scales. This localization is a general feature of the multiscale expansion of the Einstein equations, a point we also explore in Appendix~\ref{sec:multiscale from self-consistent}. 

We can apply the same localization procedure for $\Hsix_{(2)}$. The application of the stationary phase approximation in Appendix~\ref{sec:multiscale from self-consistent} shows that $h^{(1)}_{\alpha\beta}$ can be expanded in a Fourier series $\sum_{\vec{k}}\mathring{h}^{(1,\vec{k})}_{\alpha\beta}(x^i,\Jring_i,\delta M_A)e^{-ik_i\phiring^i}+{\cal O}(\e)$, as in Eq.~\eqref{multiscale h}, which suffices to show the source in Eq.~\eqref{EFER2} can be similarly expanded:
\begin{equation}
S^{{\rm eff}(2)}_{\alpha\beta} = \sum_{\vec{k}}\mathring{S}^{{\rm eff}(2,\vec{k})}_{\alpha\beta}(x^i,\Jring_i,\delta M_A)e^{-ik_i\phiring^i}+{\cal O}(\e).   
\end{equation}
Substituting this into Eq.~\eqref{eq:h2Res} and repeating the steps we followed for $\Hsix_1$, we obtain an expansion of the form
\begin{equation}\label{eq:H2 localized}
    \Hsix_{(2)} = \sum_{\vec{k},\vec{k}'} \Hsix^{\vec{k},\vec{k}'}_{(2)}\!(\Jring_i,\Jring'_i)e^{i[k_i\phiring^i-k'_i\phiring'^i]} + {\cal O}(\e).
\end{equation}
Since every term in $\Hsix_2$ is made up of (or sourced by) products of first-order fields, it is also possible to express $\Hsix_{(2)}$ as a function of three phase-space points, ${\cal Q}$, ${\cal Q}'$, and ${\cal Q}''$. However, for simplicity we do not distinguish between the two ``source points'': ${\cal Q}''={\cal Q}'$.

We can now write down our localized expression for the pseudo-Hamiltonian $\Hsix(\phistar^i({\cal Q},\e),\Jstar_i({\cal Q},\e);\delta M_A,\gamma_{\rm 6D})$. Substituting the transformations~\eqref{phistar to phiring} and \eqref{Jstar to Jring} and appealing to the above results, we arrive at 
\begin{multline}\label{Hsix(phiring,Jring) v1}
    \Hsix' = \Hsix_{(0)}({\cal Q}) + \e\mathring{\Hsix}'_{(1)}({\cal Q},{\cal Q}') \\
    + \e^2\mathring{\Hsix}'_{(2)}({\cal Q},{\cal Q}') + \mathcal{O}(\e^3),
\end{multline}
where the prime on $\Hsix'$ indicates that the pseudo-Hamiltonian is now a function of the ringed coordinates (not to be confused with our use of primes to denote quantities at the source point ${\cal Q}'$). 
Noting $\Hsix_{(0)}=E_{(0)}(\Jstar_i)$ and $\partial E_{(0)}(\Jring_j)/\partial\Jring_i = \Omega^i_{(0)}(\Jring_j)$, we have
\begin{align}
    \mathring\Hsix'_{(0)}({\cal Q}) &= \Ering,\\
    \mathring\Hsix'_{(1)}({\cal Q},{\cal Q}') &= \mathring\Hsix_{(1)}({\cal Q},{\cal Q}') - \Jring^{(1)}_i\Omega^i_{(0)},\\
    \mathring\Hsix'_{(2)}({\cal Q},{\cal Q}') &= \mathring\Hsix_{(2)}({\cal Q},{\cal Q}')  - \Jring^{(2)}_i\Omega^i_{(0)}\nonumber\\
    &\quad +\left(\phiring_{(1)}^j\frac{\partial\Jring^{(1)}_i}{\partial\phiring^j}+\Jring^{(1)}_j\frac{\partial\Jring^{(1)}_i}{\partial\Jring_j}\right)\Omega^i_{(0)}\nonumber\\
   &\quad - \phiring_{(1)}^i\frac{\partial \mathring\Hsix_{(1)}}{\partial\phiring^i} 
    -\Jring^{(1)}_i\frac{\partial \mathring\Hsix_{(1)}}{\partial\Jring_i},\label{H2'}
\end{align}
where $\Ering := E_{(0)}(\Jring_i)$, and functions of $(\phistar^i,\Jstar_i)$ are now evaluated at $(\phistar^i_{(0)},\Jring_i)$. Here we have defined $\mathring\Hsix_{(n)}$ such that the subleading terms in $\Hsix_{(1)}$ that arise from the expansion of the source orbit $\gamma_{\rm 6D}$ and from the multiscale expansion  are absorbed into $\mathring\Hsix_{(n>1)}$, while the expansion of functions at the unprimed point $(\phistar^i,\Jstar_i)$ are kept explicit. This ensures that we can continue to use the equations of motion~\eqref{dphidt generic} and \eqref{dJdt generic} as written, with the simple replacements $\Hsix_{(n)}\to \mathring{\Hsix}_{(n)}$. 

More explicitly, each $\mathring\Hsix_{(n)}$ is expressible as a Fourier series,
\begin{equation}\label{Hn mode sum}
     \mathring\Hsix_{(n)} = \sum_{\vec{k},\vec{k}'} \mathring\Hsix^{\vec{k},\vec{k}'}_{(n)}\!(\Jring_i,\Jring'_i)e^{i[k_i\phiring^i-k'_i\phiring'^i]}.
\end{equation}
In particular, $\mathring\Hsix_{(1)}$ is the leading term in Eq.~\eqref{eq:H1 localized}, with
\begin{equation}\label{H1kk}
    \mathring{\Hsix}^{\vec{k},\vec{k}'}_{(1)} = - \mu^2\pi\hat G^{(0)}_{\vec{k},\vec{k}'}[\omega'_{k'},\Jring_i,\Jring'_i].
\end{equation}
$\mathring\Hsix_{(2)}$ is given by the sum of the leading term in Eq.~\eqref{eq:H2 localized} and the order-$\e$ terms in Eq.~\eqref{eq:H1 localized}, \emph{excluding} the part of $\hat G^{(1)}_{\vec{k},\vec{k}'}$ arising from the expansion of the unprimed point in Eq.~\eqref{Gkk}, which is accounted for with the explicit $\phiring^i_{(1)}$ and $\Jring^{(1)}_i$ terms in Eq.~\eqref{H2'}. Although this division of terms might seem artificial, it is the natural division from the perspective of the field equations: $\mathring{\Hsix}_{(1)}$ and $\mathring{\Hsix}_{(2)}$ are given simply by Eqs.~\eqref{H1=hRuu} and \eqref{H2=hRuu} with the replacements $h^{\alpha\beta}_{\rm R(n)}\mapsto \mathring h^{\alpha\beta}_{\rm R(n)}$, where $\mathring h^{\alpha\beta}_{\rm R(n)}$ is obtained from the solution of the $n$th-order Einstein equation in the multiscale expansion.

%
%

Following these localizations, we can continue to view $\Hsix$ as a pseudo-Hamiltonian, but while the dependence on ${\cal Q}'$ used to be an integral over a phase-space trajectory emanating from ${\cal Q}'$, the dependence is now an explicit local function, with no nonlocal integral. 
In Hamilton's equations
\begin{equation}
    \frac{dx^i_p}{dt} = \left[\frac{\partial \Hsix}{\partial\tilde p_i}\right] \quad \text{and} \quad \frac{d\tilde p_i}{dt} = -\left[\frac{\partial \Hsix}{\partial x^i_p}\right],
\end{equation}
or
\begin{equation}
    \frac{d\phistar^i}{dt} = \left[\frac{\partial \Hsix}{\partial\Jstar_i}\right] \quad \text{and} \quad \frac{d\Jstar_i}{dt} = -\left[\frac{\partial \Hsix}{\partial\phistar^i}\right],\label{local star EOMs}
\end{equation}
the evaluation at coincidence now means evaluation at $(\phiring'^i,\Jring'_i)=(\phiring^i(\phistar^j,\Jstar_j,\e),\Jring_i(\phistar^j,\Jstar_j,\e))$. These Hamilton equations are now ordinary differential equations rather than integro-differential. 
It can also be useful to think of $\Hsix$ as a time-dependent pseudo-Hamiltonian object $\Hsix=\Hsix({\cal Q},t)$. We can consider the source trajectory, $\gamma_{\rm 6D}=\{(\phiring'^i(t),\Jring'_i(t))\,|\,t\in\mathbb{R}\}$, to be given, such that $\phiring'^i(t)$ and $\Jring'_i(t)$ in Eq.~\eqref{eq:H1 localized} are given functions of time (they could be, for example, obtained by solving the Einstein field equations together with the particle's equations of motion in multiscale form). From this perspective, in Hamilton's equations we differentiate with respect to ${\cal Q}$ and then evaluate at ${\cal Q}={\cal Q}'(t)$. 

Finally, we reiterate that the (now localized) equations of motion~\eqref{dphidt generic} and \eqref{dJdt generic} remain valid, with the simple replacements $\Hsix_{(n)}\to \mathring{\Hsix}_{(n)}$. We can then solve Eqs.~\eqref{phi1 oscillatory} and \eqref{J1 oscillatory} by expanding $\Jring^{(1)}_i$ and $\phiring_{(1)}^i$ in Fourier series; e.g., $\Jring^{(1)}_i = \sum_{\vec{k}}\Jring^{(1,\vec{k})}_i e^{ik_i\phiring^i}$. Given the Fourier expansion of $\mathring\Hsix_{(1)}$, we find the solution to Eq.~\eqref{J1 oscillatory} is
\begin{equation}\label{J1osc modes}
 \Jring^{(1,\vec{k})}_i = -\mu^2\pi\sum_{\vec{k}'}\frac{(k_i+k'_i)}{\omega_k} \hat G_{\vec{k}+\vec{k}',\vec{k}'}(\omega_{k'},\Jring_j,\Jring_j)
\end{equation}
for $\vec{k}\neq0$ modes. This can be equivalently written as
\begin{align}\label{J1osc no modes}
    \Jring^{(1)}_i &= \Bigl\langle \Jring^{(1)}_i\Bigr\rangle + \left[\frac{\partial}{\partial\phiring^i}{\textstyle\int} dt\, \mathring{\Hsix}_{(1)}\right],
\end{align}
where we define the oscillatory antiderivative
\begin{equation}
    {\textstyle\int} dt\, \mathring{\Hsix}_{(1)} := \sum_{\vec{k}\neq\vec{k}'}\frac{\mathring\Hsix^{\vec{k},\vec{k}'}_{(1)}}{i(\omega_k - \omega'_{k'})}e^{i(k_i\phiring^i-k_i'\phiring'^i)}.
\end{equation}

To solve Eq.~\eqref{phi1 oscillatory} for the oscillatory part of $\phiring^i_{(1)}$, for simplicity we specialize to $\Delta\phiring^i=0$. on the right side of  Eq.~\eqref{phi1 oscillatory} we then substitute
\begin{equation}
\mathring\Hsix_{(1)} = \left(\Omega^j_{(0)}\partial_{\phiring^j}+\Omega'^j_{(0)}\partial_{\phiring'^j}\right){\textstyle\int} dt\, \mathring{\Hsix}_{(1)}
\end{equation}
in the first term, and we substitute Eq.~\eqref{J1osc no modes} and the identity
\begin{equation}\label{dOjdJi=dOidJj}
\frac{\partial\Omega^j_{(0)}}{\partial\Jring_i} = \frac{\partial^2\! E_{(0)}}{\partial\Jring_i\partial\Jring_j}=\frac{\partial\Omega^i_{(0)}}{\partial\Jring_j}
\end{equation}
in the second term. This reduces the equation to
\begin{multline}
    \Omega^j_{(0)}\partial_{\phiring^j}\phiring^i_{(1)} \\= -\left[\left(\Omega^j_{(0)}\partial_{\phiring^j}+\Omega'^j_{(0)}\partial_{\phiring'^j}\right)\partial_{\Jring_i}{\textstyle\int} dt\, \mathring{\Hsix}_{(1)}\right].
\end{multline}
After appealing to Synge's rule for coincidence limits, $\partial_i[f(x,x')]=\left[\partial_i f(x,x')+\partial_{i'}f(x,x')\right]$~\cite{Poisson:2011nh}, we can strip off the derivatives to obtain
\begin{equation}\label{phi1osc no modes}
    \phiring^i_{(1)} = \left\langle \phiring^i_{(1)}\right\rangle - \left[\frac{\partial}{\partial\Jring_i} {\textstyle\int} dt\, \mathring{\Hsix}_{(1)}\right],
\end{equation}
again noting this is for the special case $\Delta\phiring^i=0$.

Equations~\eqref{J1osc no modes} and \eqref{phi1osc no modes} can be compared to the transformation in Eq.~(19) of Ref.~\cite{Blanco:2022mgd}. That equation is meant to apply only to the conservative sector, and divergent terms arise in it when applied to periodic motion. However, with some work (and some regularization of divergent terms), one can reduce it to a form analogous to ours.

\subsection{Conservative and dissipative sectors}
\label{sec:conservative and dissipative}

At linear order we can cleanly divide the dynamics into conservative and dissipative pieces using the split of the Green's function into symmetric and radiative (antisymmetric) pieces, as given in Eq.~\eqref{GR=Grad+GRsym}.  
This split immediately defines a conservative and dissipative split of the pseudo-Hamiltonian, 
\beq
\Hsix_{(1)} = \Hsix^{\rm rad}_{(1)}+\Hsix^{\rm sym}_{(1)}, 
\eeq
and a corresponding split in the equations of motion~\eqref{dphidt generic} and \eqref{dJdt generic}. In this section, we further refine the equations of motion by isolating the conservative and dissipative effects. This will ultimately allow us to cleanly identify a time-independent Hamiltonian dynamics within the full pseudo-Hamiltonian system and to find canonical coordinates for the Hamiltonian subsystem.

Our further simplifications of the equations of motion will build on our localized pseudo-Hamiltonian along with the symmetries~\eqref{Grad symmetry} and \eqref{GRsym symmetry} of the 4D time-domain Green's functions. Those symmetries imply corresponding symmetries of the Fourier-domain Green's functions,
\begin{align}
\hat G_{\rm rad}^{\alpha\beta\alpha'\beta'}(\omega,x^i,x'^i) &= -\hat G_{\rm rad}^{\alpha'\beta'\alpha\beta}(-\omega,x'^i,x^i),\\
\hat G_{\rm R,sym}^{\alpha\beta\alpha'\beta'}(\omega,x^i,x'^i) &= +\hat G_{\rm R,sym}^{\alpha'\beta'\alpha\beta}(-\omega,x'^i,x^i),
\end{align}
which in turn imply
\begin{align}
\hat G^{\rm rad}_{\vec{k},\vec{k}'}(\omega,\Jring_j,\Jring'_j) &= -\hat G^{\rm rad}_{-\vec{k}',-\vec{k}}(-\omega,\Jring'_j,\Jring_j),\label{Grad id}\\ 
\hat G^{\rm sym}_{\vec{k},\vec{k}'}(\omega,\Jring_j,\Jring'_j) &= +\hat G^{\rm sym}_{-\vec{k}',-\vec{k}}(-\omega,\Jring'_j,\Jring_j).\label{Gsym id}
\end{align}
Here and below, $\hat G^{\rm rad}_{\vec{k},\vec{k}'}$ and $\hat G^{\rm sym}_{\vec{k},\vec{k}'}$ refer to the radiative and symmetric contributions to $\hat G^{(0)}_{\vec{k},\vec{k}'}$; we omit the ``(0)'' and ``R'' labels for brevity. However, we also observe that the same identities hold for $\hat G_{\vec{k},\vec{k}'}(\omega,\Jring_j,\Jring'_j,\e)$.

We now calculate the conservative and dissipative contributions to the equations of motion. Given Eq.~\eqref{eq:H1 localized}, we have
\begin{align}
    \left\langle\left[\frac{\partial\mathring\Hsix_{(1)}}{\partial\Jring_i}\right]\right\rangle &= -\mu^2\pi\sum_{\vec{k}}\partial_{\Jring_i}\hat G^{(0)}_{\vec{k},\vec{k}}[\omega'_k,\Jring_j,\Jring'_j]\bigr|_{\Jring'=\Jring}, \label{<[dH1/dJ]> generic}  \\
    \left\langle\left[\frac{\partial\mathring\Hsix_{(1)}}{\partial\phiring^i}\right]\right\rangle &= -i\mu^2\pi\sum_{\vec{k}}k_i\hat G^{(0)}_{\vec{k},\vec{k}}[\omega_k,\Jring_j,\Jring_j],   \label{<[dH1/dphi]> generic}
\end{align}
where $\omega'_k := \Omega^i_{(0)}(\Jring'_j)k_i$, $\omega_k:=\Omega^i_{(0)}(\Jring_j)k_i$, and the combination of coincidence limit and averaging have eliminated the sum over $\vec{k}'$. After splitting the Green's function into its symmetric and antisymmetric pieces and using the identities \eqref{Grad id} and \eqref{Gsym id}, we get, for example,
\begin{subequations}
\begin{align}
    \left\langle\left[\frac{\partial\mathring\Hsix^{\rm rad}_{(1)}}{\partial\Jring_i}\right]\right\rangle &= -\mu^2\pi\sum_{\vec{k}}\partial_{\Jring_i}\hat G^{\rm rad}_{-\vec{k},-\vec{k}}[-\omega'_k,\Jring_j,\Jring'_j]\bigr|_{\Jring'=\Jring}\\
    &= \mu^2\pi\sum_{\vec{k}}\partial_{\Jring_i}\hat G^{\rm rad}_{\vec{k},\vec{k}}[\omega'_k,\Jring'_j,\Jring_j]\bigr|_{\Jring'=\Jring}\\
    &= -\left\langle\left[\frac{\partial\mathring\Hsix^{\rm rad}_{(1)}}{\partial\Jring'_i}\biggr\vert_{\omega'_k}\right]\right\rangle,\label{dHrad/dJ identity}
\end{align}
\end{subequations}
where we have relabelled $\vec{k}\to -\vec{k}$ in the first line and used Eq.~\eqref{Grad id} in the second line. Note on the right side of this identity, the derivative with respect to $\Jring'_i$ acts only on the third argument of the Green's function (i.e., it is taken with fixed $\omega'_k$). The same manipulations, using Eq.~\eqref{Gsym id}, show
\begin{equation}
    \left\langle\left[\frac{\partial\mathring\Hsix^{\rm sym}_1}{\partial\Jring_i}\right]\right\rangle = \left\langle\left[\frac{\partial\mathring\Hsix^{\rm sym}_1}{\partial\Jring'_i}\biggr\vert_{\omega'_k}\right]\right\rangle.\label{dHsym/dJ identity}
\end{equation}
We employ similar manipulations for $\left\langle\left[\partial\mathring\Hsix^{\rm rad}_{(1)}/\partial\phiring^i\right]\right\rangle$ and $\left\langle\left[\partial\mathring\Hsix^{\rm sym}_{(1)}/\partial\phiring^i\right]\right\rangle$.

Rearranging Eqs.~\eqref{dHrad/dJ identity} and~\eqref{dHsym/dJ identity} in suggestively antisymmetric and symmetric forms, and tallying the results for $\left\langle\left[\partial\mathring\Hsix^{\rm rad}_{(1)}/\partial\phiring^i\right]\right\rangle$ and $\left\langle\left[\partial\mathring\Hsix^{\rm sym}_{(1)}/\partial\phiring^i\right]\right\rangle$, we obtain
\begin{align}
    \left\langle\left[\frac{\partial\mathring\Hsix^{\rm rad}_{(1)}}{\partial\Jring_i}\right]\right\rangle &= \frac{1}{2}\left\langle\left[\frac{\partial\mathring\Hsix_{(1)}}{\partial\Jring_i}-\frac{\partial\mathring\Hsix_{(1)}}{\partial\Jring'_i}\biggr\vert_{\omega'_{k}}\right]\right\rangle,  \label{<dH1rad/dJ> antisymmetric}\\  \left\langle\left[\frac{\partial\mathring\Hsix^{\rm rad}_{(1)}}{\partial\phiring^i}\right]\right\rangle &= \left\langle\left[\frac{\partial\mathring\Hsix_{(1)}}{\partial\phiring^i}\right]\right\rangle \nonumber\\
    &= -i\mu^2\pi\sum_{\vec{k}}k_i\hat G^{\rm rad}_{\vec{k},\vec{k}}[\omega_k,\Jring_j,\Jring_j]\,,   
\end{align}
for the dissipative sector, and
\begin{align}
\left\langle\left[\frac{\partial\mathring\Hsix^{\rm sym}_{(1)}}{\partial\Jring_i}\right]\right\rangle &= \frac{1}{2}\left\langle\left[\frac{\partial\mathring\Hsix_{(1)}}{\partial\Jring_i} +\frac{\partial\mathring\Hsix_{(1)}}{\partial\Jring'_i}\biggr\vert_{\omega'_{k}}\right]\right\rangle\,,\\
\left\langle\left[\frac{\partial\mathring\Hsix^{\rm sym}_{(1)}}{\partial\phiring^i}\right]\right\rangle &=0\,,\label{<[dHsym/dphi]>}
\end{align}
for the conservative sector.

Equation~\eqref{<[dHsym/dphi]>} simplifies the equations of motion~\eqref{dphidt generic} and \eqref{dJdt generic} to
\begin{align}
    \frac{d\phiring^i}{dt} &= \Omega^i_{(0)}(\Jring_j) + \e\Biggl\{\left\langle\left[\frac{\partial\mathring\Hsix^{\rm sym}_{(1)}}{\partial\Jring_i}\right]\right\rangle + \left\langle\left[\frac{\partial\mathring\Hsix^{\rm rad}_{(1)}}{\partial\Jring_i}\right]\right\rangle \nonumber\\
    &\quad\ - \frac{\partial\Delta\phiring^i}{\partial\Jring_j} \left\langle\left[\frac{\partial\mathring\Hsix^{\rm rad}_{(1)}}{\partial\phiring^j}\right]\right\rangle -\left\langle\Jring^{(1)}_j\right\rangle\frac{\partial\Omega^i_{(0)}}{\partial\Jring_j}\Biggr\} + {\cal O}(\e^2),\label{dphidt Hsym Hrad}\\
    \frac{d\Jring_i}{dt} &=-\e\left\langle\left[\frac{\partial\mathring\Hsix^{\rm rad}_{(1)}}{\partial\phiring^i}\right]\right\rangle -\e^2\left\langle\left[\frac{\partial\mathring\Hsix_{(2)}}{\partial\phiring^i}\right]\right\rangle+\e^2\mathring K_i \nonumber\\
    &\quad + {\cal O}(\e^3),\label{}
\end{align}
where $\mathring K_i$ is given by Eq.~\eqref{Kdef} with the replacement $\Hsix_{(1)}\to \mathring\Hsix_{(1)}$. This result has two shortcomings. First, all derivatives of the pseudo-Hamiltonian are still evaluated before taking the coincidence limit; this means that none of the terms have the form of an ordinary Hamiltonian system. Second, we would expect a purely conservative right-hand side for $d\phiring^i/dt$, corresponding to the perturbed fundamental frequencies of motion, while we have a dissipative contribution that is not immediately eliminated by Eq.~\eqref{<dH1rad/dJ> antisymmetric}. Both of these features can be rectified through a choice of phase-space coordinates, using our residual freedom in $\Delta\phiring^i$ and $\left\langle\Jring^{(1)}_j\right\rangle$. 

We first note we can obtain total derivatives, outside the coincidence limit, by again appealing to Synge's rule, $\partial_i[f(x,x')]=[\partial_i f(x,x')+\partial_{i'}f(x,x')]$.
Motivated by this and the form~\eqref{<dH1rad/dJ> antisymmetric}, we choose  
\begin{multline}\label{<J1>}
\frac{\partial\Delta\phiring^i}{\partial\Jring_j} \left\langle\left[\frac{\partial\mathring\Hsix_{(1)}}{\partial\phiring^j}\right]\right\rangle  +\left\langle\Jring^{(1)}_j\right\rangle\frac{\partial\Omega^i_{(0)}}{\partial\Jring_j} 
\\= \frac{1}{2}\left\langle\left[\frac{\partial\mathring\Hsix_{(1)}}{\partial\Jring_i} -\frac{\partial\mathring\Hsix_{(1)}}{\partial\Jring'_i}\right]\right\rangle,
\end{multline}
such that
\begin{subequations}
\begin{align}
       \frac{d\phiring^i}{dt} &= \Omega^i_{(0)}(\Jring_j) + \frac{\e}{2}\left\langle\left[\frac{\partial\mathring\Hsix_{(1)}}{\partial\Jring_i}+\frac{\partial\mathring\Hsix_{(1)}}{\partial\Jring'_i}\right]\right\rangle + {\cal O}(\e^2)\label{dphi/dt symmetric}\\
       &= \Omega^i_{(0)}(\Jring_j) + \frac{\e}{2}\frac{\partial\left\langle[\mathring\Hsix_{(1)}]\right\rangle}{\partial\Jring_i} + {\cal O}(\e^2),
\end{align}
\end{subequations}
where we have used Synge's rule in the second line. Note that in Eq.~\eqref{dphi/dt symmetric}, the derivative with respect to $\Jring'_i$ is \emph{not} at fixed $\omega'_{k'}$. To confirm that we have eliminated the radiative contribution, also note that
\begin{subequations}
\begin{align}
\left\langle\left[\mathring\Hsix_{(1)}\right]\right\rangle &= - \mu^2\pi\sum_{\vec{k},\vec{k}'}\left\langle e^{i(k_i-k'_i)\phiring^i}\right\rangle\hat G^{(0)}_{\vec{k},\vec{k}'}[\omega_{k'},\Jring_i,\Jring_i]\\
&= - \mu^2\pi\sum_{\vec{k}} \hat G^{(0)}_{\vec{k},\vec{k}}[\omega_{k},\Jring_i,\Jring_i].
\label{eq:bound_averaging}
\end{align}
\end{subequations}
The symmetry~\eqref{Grad id} then implies 
\begin{equation}
    \left\langle\left[\mathring\Hsix_{(1)}\right]\right\rangle= \left\langle\left[\mathring\Hsix^{\rm sym}_{(1)}\right]\right\rangle,\label{<[H1]>=<[H1sym]>}
\end{equation}
meaning the dissipative contribution now vanishes from $d\phiring^i/dt$.

With our carefully chosen coordinates on phase space, we are now left with
\begin{align}
    \frac{d\phiring^i}{dt} &= \Omega^i_{(0)}(\Jring_j) + \frac{\e}{2}\frac{\partial\left\langle[\mathring\Hsix^{\rm sym}_{(1)}]\right\rangle}{\partial\Jring_i} + {\cal O}(\e^2),\label{dphidt final}\\
    \frac{d\Jring_i}{dt} &=-\e\left\langle\left[\frac{\partial\mathring\Hsix^{\rm rad}_{(1)}}{\partial\phiring^i}\right]\right\rangle \nonumber\\
    &\quad -\e^2\left(\!\left\langle\left[\frac{\partial\mathring\Hsix_{(2)}}{\partial\phiring^i}\right]\right\rangle- \mathring K_i\right) + {\cal O}(\e^3)\label{dJdt final}.
\end{align}
Here we have the expected split between conservative and dissipative sectors: the conservative dynamics, associated with the symmetric part of the Green's function, determines the correction to the geodesic frequencies; the dissipative dynamics, associated with the radiative part of the Green's function, then governs the slow evolution of those frequencies. 

Like $\Jstar_i$, our variables $\Jring_i$ are still gauge dependent due to the residual freedom left by Eq.~\eqref{<J1>}. However, the gauge freedom is greatly reduced. Specifically, bringing the equations of motion to the form~\eqref{dphidt final} and \eqref{dJdt final} completely fixes the phase-space gauge up to the following freedom:
\begin{align}
    \phiring^i &\to \phiring^i + \Delta \phiring^i(\Jring_j),\label{phi residual freedom}\\
    \Jring_i &\to \Jring_i + \e \Delta \Jring_i(\Jring_j),\label{J residual freedom}
\end{align}
where $\Delta\Jring_i$ is chosen to cancel the change in $d\phiring^i/dt$. To make this explicit, we examine the changes in Eq.~\eqref{dphidt final}, noting the change in \eqref{dJdt final} is order $\e^2$. We have
\begin{align}
    \frac{d\phiring^i}{dt} \to \frac{d\phiring^i}{dt} &= \Omega^i_{(0)}(\Jring_j) + \frac{\e}{2}\frac{\partial\left\langle[\mathring\Hsix^{\rm sym}_1]\right\rangle}{\partial\Jring_i} \nonumber\\
    &\quad + \e\left\{\frac{\partial\Delta\phiring^i}{\partial\Jring_j}\left(\frac{d\Jring_j}{dt}\right)^{\!\!(1)} - \frac{\partial\Omega^i_{(0)}}{\partial\Jring_j}\Delta\Jring_j\right\} \nonumber\\
    &\quad + {\cal O}(\e^2),\label{dphidt residual}
\end{align}
where $\left(\frac{d\Jring_j}{dt}\right)^{\!(1)}$ is the coefficient of $\e$ in Eq.~\eqref{dJdt final}. Choosing 
\begin{equation}
\Delta\Jring_i = \left(\frac{\partial\Omega^j_{(0)}}{\partial\Jring_i}\right)^{\!-1}\frac{\partial\Delta\phiring^j}{\partial\Jring_k}\left(\frac{d\Jring_k}{dt}\right)^{\!\!(1)}\,,\label{Delta J and Delta phi}
\end{equation}
eliminates the terms in curly brackets, preserving the form of Eq.~\eqref{dphidt final}. Thus the only residual freedom is in $\Delta\phiring^i(\Jring_j)$, which corresponds to the choice of origin on each torus of constant $\Jring_i$. Equivalently, we can invert Eq.~\eqref{Delta J and Delta phi} to fix $\Delta\phiring^i$ in terms of $\Delta \Jring_i$. Regardless of how this residual freedom is used, it leaves the frequencies (as functions of the action variables) invariant; this is a trivial consequence of it leaving Eq.~\eqref{dphidt final} unchanged.

The residual gauge freedom has a simple physical interpretation: it corresponds to a simultaneous shift in time and angles. If we change our choice of time by an amount $\Delta t$, then $\Jring_i$ changes by $\e \Delta t\, d\Jring_i/dt$. The condition that the terms in curly brackets cancel in Eq.~\eqref{dphidt residual}  then becomes 
\begin{equation}
    \frac{\partial(\Omega^i_{(0)}\Delta t)}{\partial\Jring_i} = \frac{\partial\Delta \phiring^i}{\partial\Jring_i},
\end{equation}
or simply $\Delta\phiring^i = \Omega^i_{(0)}\Delta t$, where we have assumed $\Delta t$ is independent of the phase-space coordinates. In other words, under a redefinition of $t$, we can keep the orbital frequencies unchanged  (as functions of $\Jring_i$) through a redefinition of the angles. This freedom only exists in the presence of dissipation, as it involves the evolution of $\Jring_i$. It is then equivalent to a fundamental gauge freedom in the multiscale expansion: the freedom to choose the foliation of spacetime into time slices.

\subsection{Stationary, axisymmetric spacetimes}\label{sec:SAS spacetimes}

Our derivations in this section have not relied specifically on the Kerr spacetime. Our results are valid in any globally hyperbolic, stationary, vacuum background spacetime in which the geodesic equation is integrable; although we presented explicit Kerr formulas (e.g., for the geodesic momenta) in preceding sections, we have not relied on them. We have only relied on background-geodesic integrability (ensuring the existence of osculating action-angle variables $\phistar^i$ and $\Jstar_i$) and background stationarity and hyperbolicity (ensuring that retarded and advanced Green's functions and Detweiler-Whiting two-point functions exist and only depend on time in the combination $t-t'$). However, it is worth examining the simplifications that occur when we specialize to stationary, axisymmetric spacetimes with the time-reversal symmetry~\eqref{TGret = Gadv}. This allows us to recover traditional results from Refs.~\cite{Mino:2003yg,Hinderer:2008dm}, in which the conservative and dissipative effects associated with the symmetric and radiative two-point functions become identified with appropriate symmetry or anti-symmetry under time reversal. 

In a stationary, axisymmetric spacetime, each of our Green's functions and two-point functions only depend on $\phi$ and $\phi'$ in the combination $(\phi-\phi')$. The symmetry~\eqref{TGret = Gadv} under time reflection ($t\to-t,\phi\to-\phi$) then implies
\begin{multline}\label{Grad reversal}
    \hat G^{{\rm rad}}_{\alpha\beta\alpha'\beta'}(\omega,\phi-\phi',\bm{x},\bm{x}')\\=-\epsilon_\alpha\epsilon_\beta\epsilon_{\alpha'}\epsilon_{\beta'}\hat G^{{\rm rad}}_{\alpha\beta\alpha'\beta'}(-\omega,\phi'-\phi,\bm{x},\bm{x}'),
\end{multline}
and
\begin{multline}\label{Gsym reversal}
    \hat G^{{\rm sym}}_{\alpha\beta\alpha'\beta'}(\omega,\phi-\phi',\bm{x},\bm{x}')\\=+\epsilon_\alpha\epsilon_\beta\epsilon_{\alpha'}\epsilon_{\beta'}\hat G^{{\rm sym}}_{\alpha\beta\alpha'\beta'}(-\omega,\phi'-\phi,\bm{x},\bm{x}'),
\end{multline}
where  $\epsilon_\alpha=+1$ if $\alpha=r,\theta$, $\epsilon_\alpha=-1$ if $\alpha=t,\phi$, and there is no summation over repeated indices. We omit the ``R'' label on the symmetric two-point function and recall our notation $\bm{x}=(r,\theta)$.

\subsubsection{Consequences of time-reversal symmetry}

To simplify the following analysis, we define
\begin{multline}\label{Ghat(phase space)}
\hat G(\omega,\phiring^i,\Jring_i,\phiring'^i,\Jring'_i,\e) \\
:= 
\frac{\tilde p_\alpha\tilde p_\beta}{\mu^2\utstar} \hat G^{\alpha\beta\alpha'\beta'}_{\rm R}(\omega,x^i_p,x^{i'}_p)\frac{\tilde p_{\alpha'}\tilde p_{\beta'}}{\mu^2(\utstar)'},
\end{multline}
where $\tilde p_\alpha=\tilde p_\alpha(\phiringvec,\Jring_i,\e)$, $\utstar=\utstar(\phiringvec,\Jring_i,\e)$, and $x^i_p=x^i_p(\phiring^j,\Jring_j,\e)$. 
To further simplify the analysis, we also choose $\Delta\phiring^i$ such that $\phiringvec=0$ lies at a mutual turning point of the $(r,\theta)$ motion, as is the case for the quasi-Keplerian phases $\psi^i$; in other words, in the notation of Eqs.~\eqref{fdiss} and~\eqref{fcon}, we choose $\phiringvec_O=0$.\footnote{We emphasize that this is a choice of coordinate origin in phase space. We do not require any particular on-shell trajectory to pass through a common turning point.}   Recalling that we chose $\phistar^i$ to vanish at turning points, we  can write this specification as
\beq
\Delta\phiring^i(\Jring_j)=0.
\eeq
At zeroth order, the positions and momenta then satisfy natural time-reversal identities:
\begin{align}
    x^i_{(0)}(\phiring^j,\Jring_j) &= \epsilon_i x^i_{(0)}(-\phiring^j,\Jring_j),\\ 
    \tilde p^{(0)}_\alpha(\phiringvec,\Jring_j) &=-\epsilon_\alpha \tilde p^{(0)}_\alpha(-\phiringvec,\Jring_j),\\
    u^t_{(0)}(\phiringvec,\Jring_i) &= u^t_{(0)}(-\phiringvec,\Jring_i).
\end{align} 
By combining these identities with the identities~\eqref{Grad reversal} and~\eqref{Gsym reversal}, we immediately find
\begin{align}
    \hat G^{\rm rad}_{(0)}(\omega,\phiring^i,\Jring_i,\phiring'^i,\Jring'_i) &= -\hat G^{\rm rad}_{(0)}(-\omega,-\phiring^i,\Jring_i,-\phiring'^i,\Jring'_i),\\
    \hat G^{\rm sym}_{(0)}(\omega,\phiring^i,\Jring_i,\phiring'^i,\Jring'_i) &= +\hat G^{\rm sym}_{(0)}(-\omega,-\phiring^i,\Jring_i,-\phiring'^i,\Jring'_i).
\end{align}
At the level of modes, these identities imply
\begin{align}
    \hat G^{\rm rad}_{\vec{k},\vec{k}'}(\omega,\Jring_j,\Jring'_j) &= -\hat G^{\rm rad}_{-\vec{k},-\vec{k}'}(-\omega,\Jring_j,\Jring'_j),\label{T Grad_kk}\\
    \hat G^{\rm sym}_{\vec{k},\vec{k}'}(\omega,\Jring_j,\Jring'_j) &= +\hat G^{\rm sym}_{-\vec{k},-\vec{k}'}(-\omega,\Jring_j,\Jring'_j),\label{T Gsym_kk}
\end{align}
which, when combined with Eqs.~\eqref{Grad id} and~\eqref{Gsym id}, in turn imply
\begin{align}
    \hat G^{(0)}_{\vec{k},\vec{k}'}(\omega,\Jring_i,\Jring'_i) &= \hat G^{(0)}_{\vec{k}',\vec{k}}(\omega,\Jring_i',\Jring_i)
\end{align}
for both the symmetric and radiative pieces.

Next, we write the first-order pseudo-Hamiltonian as
\begin{multline}\label{H1 modes}
    \mathring\Hsix_{(1)}(\phiring^i,\Jring_i,\phiring'^i,\Jring'_i) \\
    = -\mu^2\pi \sum_{\vec{k},\vec{k}'}\hat G^{(0)}_{\vec{k},\vec{k}'}(\omega'_{k'},\Jring_j,\Jring'_j)e^{i(k_i\phiring^i-k'_i\phiring'^i)}.
\end{multline}
Equations~\eqref{T Grad_kk} and~\eqref{T Gsym_kk} then immediately imply
\begin{align}
    \mathring\Hsix^{\rm rad}_{(1)} &= \frac{1}{2}\mathring\Hsix_{(1)}(\phiring^j,\Jring_j,\phiring'^j,\Jring'_j) \nonumber\\
    &\quad - \frac{1}{2}\mathring\Hsix_{(1)}(-\phiring^j,\Jring_j,-\phiring'^j,\Jring'_j),\label{Hrad T symmetry}\\
    \mathring\Hsix^{\rm sym}_{(1)} &= \frac{1}{2}\mathring\Hsix_{(1)}(\phiring^j,\Jring_j,\phiring'^j,\Jring'_j) \nonumber\\
    &\quad + \frac{1}{2}\mathring\Hsix_{(1)}(-\phiring^j,\Jring_j,-\phiring'^j,\Jring'_j).\label{Hsym T symmetry}
\end{align}
These equalities are equivalent to classic results of Refs.~\cite{Mino:2003yg,Hinderer:2008dm}. They state that in stationary, axisymmetric spacetime with time-reversal symmetry, the conservative and dissipative effects defined from the symmetric and radiative two-point functions are completely equivalent to the conservative and dissipative effects defined from reversing phases around turning points: the dissipative sector as defined from the radiative two-point function coincides with the dissipative force as defined by Eq.~\eqref{fdiss}, and the conservative sector as defined from the symetric two-point function coincides with the conservative force as defined by Eq.~\eqref{fcon}. However, we emphasize that we have only shown this identification at first perturbative order.

We can make use of these symmetries to illuminate the conservative-dissipative split of the equations of motion in the previous section and clarify the gauge condition~\eqref{<J1>}. Since the average of an antisymmetric function trivially vanishes, we can infer from Eq.~\eqref{Hrad T symmetry} that  $\left\langle\left[\partial_{\Jring_i}\mathring\Hsix^{\rm rad}_{(1)}\right]\right\rangle=0$. Given our specification $\Delta\phiring^i=0$, 
Eq.~\eqref{dphidt Hsym Hrad} reduces to
\begin{multline}
    \frac{d\phiring^i}{dt} = \Omega^i_{(0)}(\Jring_j) + \e\Biggl\{\left\langle\left[\frac{\partial\mathring\Hsix^{\rm sym}_{(1)}}{\partial\Jring_i}\right]\right\rangle -\left\langle\Jring^{(1)}_j\right\rangle\frac{\partial\Omega^i_{(0)}}{\partial\Jring_j}\Biggr\} \\+ {\cal O}(\e^2).
\end{multline}
Unlike in the generic case in the previous section, here we see that for a spacetime with time-reversal symmetry, the dissipative sector drops out of $d\phiring^i/dt$ without requiring a specific gauge choice. The gauge condition~\eqref{<J1>} now serves solely to put  $d\phiring^i/dt$ in the form of a total derivative with respect to $\Jring_i$. The condition now reduces to
\begin{equation}\label{<J1> condition with T symmetry}
\left\langle\Jring^{(1)}_j\right\rangle\frac{\partial\Omega^i_{(0)}}{\partial\Jring_j} = \frac{1}{2}\left\langle\left[\frac{\partial\mathring\Hsix^{\rm sym}_{(1)}}{\partial\Jring_i} -\frac{\partial\mathring\Hsix^{\rm sym}_{(1)}}{\partial\Jring'_i}\right]\right\rangle.
\end{equation}
Using Eq.~\eqref{dHsym/dJ identity}, we can write the right-hand side more explicitly as
\begin{multline}
    \left\langle\left[\frac{\partial\mathring\Hsix^{\rm sym}_1}{\partial\Jring_i} -\frac{\partial\mathring\Hsix^{\rm sym}_1}{\partial\Jring'_i}\right]\right\rangle \\
    = \mu^2\pi\sum_{\vec{k}}\frac{\partial\omega_k}{\partial \Jring_i}\partial_\omega\hat G^{\rm sym}_{\vec{k},\vec{k}}(\omega_k,\Jring_j,\Jring_j).
\end{multline}
We can also rewrite the left-hand side using Eq.~\eqref{dOjdJi=dOidJj}. These steps put Eq.~\eqref{<J1> condition with T symmetry} in the form
\begin{equation}
\frac{\partial\Omega^j_{(0)}}{\partial\Jring_i}\left\langle\Jring^{(1)}_j\right\rangle = \frac{\mu^2\pi}{2}\sum_{\vec{k}}\frac{\partial\omega_k}{\partial \Jring_i}\partial_\omega\hat G^{\rm sym}_{\vec{k},\vec{k}}(\omega_k,\Jring_j,\Jring_j).
\end{equation}
In this form, the equation can be solved by inspection:
\begin{equation}\label{<J1> solution}
    \left\langle \Jring^{(1)}_i\right\rangle = \frac{\mu^2\pi}{2}\sum_{\vec{k}}k_i\partial_\omega\hat G^{\rm sym}_{\vec{k},\vec{k}}(\omega_k,\Jring_j,\Jring_j).
\end{equation}

\subsubsection{Consequences of axisymmetry}

In addition to the above consequences of time-reversal symmetry, there are also more direct consequences of axisymmetry. 

Since $\hat G^{\rm R}_{\alpha\beta\alpha'\beta'}$ has the form $\hat G^{\rm R}_{\alpha\beta\alpha'\beta'}(\omega,\phi-\phi',\bm{x},\bm{x}')$, we immediately deduce the standard result that the dynamics is independent of $\phi_p$: the coincidence limit in $[\partial\Hsix/\partial x^i_p]$ and $[\partial\Hsix/\partial\tilde p_i]$ trivially eliminates all $\phi_p$ dependence because $[\phi_p-\phi'_p]=0$.\footnote{Note this does \emph{not} imply $[\partial\Hsix/\partial \phi_p]=0$, since the derivative is taken before the coincidence limit.} As a consequence, since $\phi_p=\phistar^\phi + \Delta\phi_p(\phistarvec,\Jstar_i)$, all functions in the equations of motion~\eqref{eq:phistardot expanded} and \eqref{eq:Jstardot expanded} and in the near-identity transformations~\eqref{phistar to phiring} and \eqref{Jstar to Jring} are independent of $\phistar^\phi$.

We can also show that the background's axisymmetry implies $d\tilde p_\phi/dt=0$ in the conservative sector; i.e., $\tilde p_\phi = \Jstar_\phi$ is a constant if dissipation is neglected. This contrasts with the non-axisymmetric case, where all three $\Jstar_i$ variables are oscillatory even in the conservative sector. To establish this result, we start from the fact that $\Hsix_{(0)}(x^i_p,\tilde p_i)$ is independent of $\phi_p$, which implies
\begin{equation}\label{dp_phi/dt}
    \frac{d\tilde p_\phi}{dt} = -\left[\frac{\partial\Hsix}{\partial\phi_p}\right] = -\e \left[\frac{\partial\Hsix_{(1)}}{\partial\phi_p}\right] + {\cal O}(\e^2).
\end{equation}
Next we treat $\Hsix_{(1)}$ as a function of the ringed variables and use the fact that the only ringed variable which depends on $\phi_p$ is $\phiring^\phi = \phi_p - \Delta\phi_p(\bm{x}_p,\tilde p_\phi) +{\cal O}(\e)$. Hence, 
\begin{equation}
    \frac{d\tilde p_\phi}{dt} = -\e\left[\frac{\partial\mathring\Hsix_{(1)}}{\partial\phiring^\phi}\right] + {\cal O}(\e^2).
\end{equation}
To evaluate the right-hand side, we note that the integrand~\eqref{Ghat(phase space)} in $\Hsix_{(1)}$ has the form 
\beq
\hat G(\omega,\phiring^\phi-\phiring'^\phi,\phiringvec,\Jring_i,\phiringvec',\Jring'_i,\e).
\eeq
This implies that in the pseudo-Hamiltonian \eqref{H1 modes}, all modes vanish except those with $k_\phi=k'_\phi$. Hence, we can rewrite the pseudo-Hamiltonian as
\begin{align}
    &\mathring\Hsix_{(1)}(\phiring^\phi-\phiring'^\phi,\phiringvec,\Jring_i,\phiringvec',\Jring'_i) \nonumber\\
    &= -\mu^2\pi \!\!\sum_{\bm{k},\bm{k}',m}\!\!\hat G^{(0)}_{\bm{k},\bm{k}',m}(\omega'_{k'},\Jring_j,\Jring'_j)e^{i(\bm{k}\cdot\phiringvec-\bm{k}'\cdot\phiringvec')+im(\phiring^\phi-\phiring'^\phi)},
\end{align}
where $\bm{k}\cdot\phiringvec=k_r\phiring^r+k_\theta\phiring^\theta$ and $\omega'_{k'}=\bm{k}'\cdot\bm{\Omega}_{(0)}(\Jring'_i)  + m\Omega^\phi_{(0)}(\Jring'_i)$. The $\phiring^\phi$ derivative, at coincidence, evaluates to
\begin{equation}
    \left[\partial_{\phiring^\phi}\mathring\Hsix_{(1)}\right] = -i\mu^2\pi \!\!\sum_{\bm{k},\bm{k}',m}\!\! m\,\hat G^{(0)}_{\bm{k},\bm{k}',m}(\omega_{k'},\Jring_j,\Jring_j)e^{i(\bm{k} -\bm{k}')\cdot\phiringvec}.
\end{equation}
Relabeling $\bm{k}\to -\bm{k}'$, $\bm{k}'\to -\bm{k}$, and $m\to-m$, and then appealing to Eq.~\eqref{Gsym id} shows that $\left[\partial_{\phiring^\phi}\mathring\Hsix^{\rm sym}_{(1)}\right]=-\left[\partial_{\phiring^\phi}\mathring\Hsix^{\rm sym}_{(1)}\right]$. Therefore,
\begin{equation}
      \left[\partial_{\phiring^\phi}\mathring\Hsix^{\rm sym}_{(1)}\right] = 0,
\end{equation}
and Eq.~\eqref{dp_phi/dt} reduces to
\begin{equation}\label{dp_phi/dt=0}
\frac{d\tilde p_\phi}{dt} = {\cal O}(\e^2)    
\end{equation}
in the conservative sector. As promised, in the conservative sector in an axisymmetric background, $\tilde p_\phi = \Jstar_\phi$ is a constant.

\section{Hamiltonian description of the conservative sector}\label{sec:conservative sector}

In the previous section, we split the first-order dynamics into conservative and dissipative pieces. We now focus on the conservative sector, meaning we set $\Hsix_{(1)}=\Hsix^{\rm sym}_{(1)}$, and we ignore second-order terms in the 1PA dynamics. In this setting, we present a simple Hamiltonian description of this conservative dynamics. We are able to cleanly identify the conserved energy, angular momentum and action variables, showing that they are equal to the ``renormalized'' variables defined in Ref.~\cite{Fujita:2016igj}. We highlight the distinction between canonically conjugate variables in the Hamiltonian dynamics and canonically conjugate variables in the pseudo-Hamiltonian dynamics, emphasizing that $\tilde p_i$ is not conjugate to $x^i_p$ in the Hamiltonian dynamics. Using the existence of canonical momenta $p_i$, which \emph{are} conjugate to $x^i_p$, we provide a simple proof of the action variables' gauge invariance.

We show how these results fit into the full 1PA dynamics, with dissipation, in Sec.~\ref{sec:gauge choices}.

\subsection{Time-independent Hamiltonian dynamics}

If we specialize to the conservative sector and to first order, then the calculations of the preceding section largely go through without change. In particular, Secs.~\ref{sec:perturbed_action_angles} and~\ref{sec:conservative and dissipative} require no modification, and we observe that $d\Jring_i/dt$ vanishes, making both $\Jring_i$ and $d\phiring^i/dt$ constant. Concretely, the equations of motion~\eqref{dphidt final} and \eqref{dJdt final} reduce to
\begin{align}
    \frac{d\phiring^i}{dt} &:=\Omega^i(\Jring_j) = \Omega^i_{(0)}(\Jring_j) + \frac{\e}{2}\frac{\partial\left\langle[\Hsix_1]\right\rangle}{\partial\Jring_i} + {\cal O}(\e^2),\label{phidot conservative}\\
    \frac{d\Jring_i}{dt} &\hphantom{:}= {\cal O}(\e^2).\label{Jdot conservative}
\end{align}
We can immediately rewrite these as Hamilton's equations,
\begin{align}
    \frac{d\phiring^i}{dt} &= \frac{\partial H_{\rm 6D}}{\partial\Jring_i},\label{dphidt Hamilton}\\
    \frac{d\Jring_i}{dt} &= -\frac{\partial H_{\rm 6D}}{\partial\phiring^i},  
\end{align}
with the Hamiltonian
\begin{equation}
H_{\rm 6D} := \Ering + \frac{1}{2}\e\left\langle[\Hsix_{(1)}]\right\rangle +{\cal O}(\e^2),\label{Hamiltonian - final} 
\end{equation}
where we have recalled Eq.~\eqref{<[H1]>=<[H1sym]>}.

From this we can conclude the following: 
\begin{enumerate}
    \item The conservative dynamics is governed by a time-independent Hamiltonian $H_{\rm 6D}$. The time independence follows from Eqs.~\eqref{eq:bound_averaging} and~\eqref{Jdot conservative}. 
    \item  The coordinates $(\phiring^i,\Jring_i)$ are canonically conjugate variables in the conservative sector. $\phiring^i$ are cyclic variables, and the dynamics is $2\pi$-periodic in them (as $x^i_p$ and $\tilde p_i$ are manifestly $2\pi$-periodic functions of them). Hence, $(\phiring^i,\Jring_i)$ are action-angle variables by definition~\cite{CannasdaSilva}. Moreover, the residual gauge freedom described by Eq.~\eqref{Delta J and Delta phi} is eliminated because $d\Jring_i/dt=0$;  $\Jring_i$ is now \emph{fully fixed} by the gauge condition~\eqref{<J1>}. On the other hand, $\Delta\phiring^i(\Jring_j)$ remains freely specified, and it does not contribute a term in the equation of motion~\eqref{dphidt Hsym Hrad}; this is the standard freedom to shift the origin of the angles, as reviewed in Sec.~\ref{sec:osculating action angles}.
\end{enumerate}

Note that the correction to the Kerr geodesic Hamiltonian is proportional to the Detweiler(-Barack-Sago) averaged redshift~\cite{Detweiler:2008ft,Barack:2011ed}:
\begin{equation}
\left\langle[\Hsix_{(1)}]\right\rangle=\mu\langle z_{(1)}\rangle,\label{H1=z1}    
\end{equation}
where 
\beq
z_{(1)}:=-\frac{1}{2}z_0 h^{\rm R}_{\alpha\beta}u^\alpha u^\beta 
\eeq
is the order-$\e$ term in the redshift 
\begin{subequations}\label{z<->H}%
\begin{align}
z&:=\frac{d\tilde\tau}{dt} \\
&= z_0\sqrt{-\tilde g_{\alpha\beta}u^\alpha u^\beta} \\
&= z_0 +\frac{\e}{\mu}[\Hsix_{(1)}] + {\cal O}(\e^2),
\end{align}
\end{subequations}
with 
\beq\label{z0}
z_0:=\frac{d\tau}{dt} = \sqrt{-\left(g_{tt}+2g_{ti}\dot x^i_p+g_{ij}\dot x^i_p \dot x^j_p\right)}
\eeq
and $\Hsix_{(1)}$ given by Eq.~\eqref{H1=hRuu}. Here, recall, a dot denotes $d/dt$. In the last equality of Eq.~\eqref{z<->H}, we have made the replacement $\e h^{\rm R}_{\alpha\beta}\frac{u^\alpha u^\beta}{dt/d\tau} = \e h_{\rm R}^{\alpha\beta}\frac{\tilde p_\alpha \tilde p_\beta}{\mu^2\utstar}+{\cal O}(\e^2)$. Given Eq.~\eqref{H1=z1}, we can write the canonical angle variables' 1PA frequency corrections in terms of the redshift: 
\begin{equation}
\Omega^i_{(1)} = \frac{\mu}{2}\frac{\partial\langle z_{(1)}\rangle}{\partial\Jring_i}.
\end{equation}

Note that Eq.~\eqref{z<->H} is a formal expansion in powers of $h^{\rm R}_{\alpha\beta}$, not an expansion at fixed phase-space coordinates, and $z_0$ is not a geodesic function of $(x^i_p,\tilde p_i)$. Through its dependence on $dx^i_p/dt$, $z_0$ is instead a function of $(x^i,\tilde p_i,\e)$, not only of $(x^i,\tilde p_i)$. For that reason we deliberately write it as $z_0$ rather than $z_{(0)}$. However, it is known that the \emph{averaged} redshift can be written as~\cite{Akcay:2015pza}
\beq\label{<z>}
\langle z \rangle = \langle z_{(0)} \rangle(\Omega^i) +\e \langle z_{(1)}\rangle(\Omega^i) + {\cal O}(\e^2),
\eeq
where $\langle z_{(0)} \rangle(\Omega^i)$ \emph{is} the geodesic function of the invariant frequencies $\Omega^i$. In other words, $\langle z_{0} \rangle(\Omega^i,\e)=\langle z_{(0)} \rangle(\Omega^i) +{\cal O}(\e^2)$; on average, the redshift on an accelerated orbit is identical to the redshift on a geodesic with the same set of fundamental frequencies. We provide a proof of this in Appendix~\ref{sec:averaged redshift} 

We conclude this section with a comment on the localization procedure of  Sec.~\ref{sec:localization}. In the conservative sector, the leading-order term in the 2D stationary phase approximation becomes exact; there are no subleading terms. To see this, refer back to Eq.~\eqref{eq:H1 double int}. Since $\phiring'^i(t')=\Omega'^i t' + \phiring'^i(0)$ in the conservative sector, we can employ the identity
\begin{equation}
\int^{\infty}_{-\infty}dt' e^{-i (k'_i\Omega'^i - \omega) t'} = 2\pi\delta(\omega-k'_i\Omega'^i).
\end{equation}
Equation~\eqref{eq:H1 double int} then reduces to the leading term in the stationary phase approximation.

\subsection{Canonical momenta}

In the previous section, we expressed our Hamiltonian in terms of the action-angle variables $(\phiring^i,\Jring_i)$. We can also write it in terms of our original coordinates $(x^i_p,\tilde p_i)$. Given $\Jring_i \approx \Jstar_i + \e \Jring^{(1)}_i$ and $\partial\Estar/\partial\Jstar_i=\Omega^i_{(0)}(\Jstar_j)$, the Hamiltonian~\eqref{Hamiltonian - final} can be written as%
\begin{multline}\label{H6D(phistar,Jstar)}%
H_{\rm 6D} = \Estar(\Jstar_i) + \frac{1}{2}\e\left\langle[\Hsix_{(1)}(\phistar^i,\Jstar_i,\phistar'_i,\Jstar'_i)]\right\rangle\\ -\e \Omega^i_{(0)}(\Jstar_j)\Jring^{(1)}_i(\phistar^j,\Jstar_j)+{\cal O}(\e^2). 
\end{multline}
Since $(\phistar^i,\Jstar_i)$ are related to $(x^i_p,\tilde p_i)$ by the $\e$-independent, Kerr-geodesic map, we can equivalently write this as
\begin{multline}\label{H6D(x,ptilde)}
H_{\rm 6D} = \Estar(x^i_p,\tilde p_i) + \frac{1}{2}\e\left\langle[\Hsix_{(1)}(x^i_p,\tilde p_i,x'^i_p,\tilde p'_i)]\right\rangle\\ 
-\e\, \Omega^i_{(0)}(x^j_p,\tilde p_j)\Jring^{(1)}_i(x^j_p,\tilde p_j)+{\cal O}(\e^2). 
\end{multline}

However, with these variables, Hamilton's equations in canonical form are \emph{not} satisfied; $dx^i_p/dt \neq \partial H_{\rm 6D}/\partial \tilde p_i$ and  $d\tilde p_i/dt \neq -\partial H_{\rm 6D}/\partial x^i_p$. They are violated because $(x^i_p,\tilde p_i)$ are \emph{not} canonical coordinates in the Hamiltonian dynamics. This is an example of a general result in Ref.~\cite{Blanco:2022mgd}: if a pseudo-Hamiltonian system is also a Hamiltonian system, then coordinates which are canonical in the pseudo-Hamiltonian system, such as $(x^i_p,\tilde p_i)$, are generically \emph{not} canonical in the Hamiltonian system. This is a simple consequence of the fact that 
\begin{equation}
\left[\frac{\partial{\cal H}}{\partial q^i}\right]\neq \frac{\partial H_{\rm 6D}}{\partial q^i}
\end{equation}
for any choice of coordinates $q^i$, even when restricted to the conservative sector. In this section, we discuss canonical momenta $p_i$ that \emph{are} conjugate to $x^i_p$ in the Hamiltonian $H_{\rm 6D}$.

We will not be able to find $p_i$ explicitly; we leave that to future work. However, we are able to describe general properties of the momenta and how one might, in principle, find them. First, we note they will necessarily be related to $\tilde p_i$ by a small shift,
\begin{equation}\label{p=ptilde+dp}
    p_i = \tilde p_i +\e \delta p_i + {\cal O}(\e^2).
\end{equation}
In terms of these canonical momenta, the Hamiltonian~\eqref{H6D(x,ptilde)} will then read
\begin{multline}\label{H6D(x,p)}
H_{\rm 6D} = E_{(0)}(x^i_p,p_i) + \frac{1}{2}\e\left\langle[\Hsix_{(1)}]\right\rangle\\ 
-\e\, \Omega^i_{(0)}\Jring^{(1)}_i - \e\frac{\partial E_{(0)}}{\partial p_i} \delta p_i + {\cal O}(\e^2), 
\end{multline}
where it is understood that all functions of $(x^i_p,\tilde p_i)$ on the right are now evaluated at $(x^i_p,p_i)$. 
Since the coordinates are canonically conjugate, Hamilton's equations will be satisfied in canonical form:
\begin{equation}\label{Hamilton equations}
    \frac{dx^i_p}{dt}= \frac{\partial H_{\rm 6D}}{\partial p_i} \quad\text{and}\quad \frac{dp_i}{dt}=-\frac{\partial H_{\rm 6D}}{\partial x^i_p}.
\end{equation}
Moreover, in terms of the canonical momenta, the action variables will take the familiar form
\begin{equation}\label{J=int pdx}
    \Jring_i = \frac{1}{2\pi}\oint_{C_i}p_j dx^j_p,
\end{equation}
where the three $C_i$ are the three distinct loops on the torus of constant $\Jring_i$. We derive Eq.~\eqref{J=int pdx} below, after describing the failure of $\tilde p_i$ to be canonical and a procedure for finding $p_i$.

\subsubsection{Failure of (Hamiltonian) canonical conjugacy}

To show that $(x^i_p,\tilde p_i)$ are not canonically conjugate in the Hamiltonian system, it suffices to show that failure of canonical conjugacy for $(\phistar^i,\Jstar_i)$. Since $(\phistar^i,\Jstar_i)$ and $(x^i_p,\tilde p_i)$ are related by a generating function, one pair failing to be conjugate implies the other pair fails as well.

We first observe that if $(\phistar^i,\Jstar_i)$ \emph{were} conjugate, there would exist an infinitesimal canonical transformation relating them to $(\phiring^i,\Jring_i)$. Since every near-identity canonical transformation can be generated by a generating function, we would be able to write
\begin{equation}\label{calW}
    {\cal W}(\phistar^i,\Jring_i) = \phistar^i\Jring_i + \e {\cal W}_{(1)}(\phistar^i,\Jring_i),
\end{equation}
with 
\begin{align}
    \phiring^i &= \frac{\partial \cal W}{\partial \Jstar_i} = \phistar^i + \e\frac{\partial {\cal W}_{(1)}}{\partial \Jstar_i},\label{phiring=dW/dJring}\\
    \Jstar_i &= \frac{\partial \cal W}{\partial \phistar^i} = \Jring_i + \e\frac{\partial {\cal W}_{(1)}}{\partial \phistar^i}.\label{Jstar=dW/dphistar}   
\end{align}
This would imply that ${\cal W}_{(1)}$ is related to $\Jring^{(1)}_i$ from Eq.~\eqref{Jstar to Jring} by
\begin{equation}\label{dW1/dphi=J1}
    \frac{\partial {\cal W}_{(1)}}{\partial \phiring^i} = - \Jring^{(1)}_i
\end{equation}
(replacing starred arguments by ringed ones). If we write 
\beq
{\cal W}_{(1)}(\phiring^i,\Jring_i) =-\langle \Jring^{(1)}_i\rangle \phiring^i + \bar{\cal W}(\phiring^i,\Jring_i),
\eeq
then $\bar{\cal W}$ must satisfy
\begin{equation}
    \frac{\partial \bar{\cal W}}{\partial \phiring_i} = - \sum_{\vec k\neq0}\Jring^{(1,\vec k)}_i e^{ik_j\phiring^j}. 
\end{equation}

The presence of the $\langle \Jring^{(1)}_i\rangle\phiring^i$ term in ${\cal W}_{(1)}$ immediately suggests that the transformation cannot be canonical. Substituting the $\langle \Jring^{(1)}_i\rangle\phiring^i$ term into Eq.~\eqref{phiring=dW/dJring}, we see that it introduces a non-periodic term in the transformation of $\phiring^i$, while we know from the results of previous sections that the $\phiring^i_{(1)}$ appearing in Eq.~\eqref{phistar to phiring} is $2\pi$-periodic in each of the angles. 

We can also show that even the periodic terms cannot be generated by a generating function. By differentiating Eq.~\eqref{dW1/dphi=J1} with respect to $\phiring^j$, we obtain 
\beq
\frac{\partial^2 \bar{\cal W}_{(1)}}{\partial\phiring^j\partial \phiring^i} = - \frac{\partial\Jring^{(1)}_i}{\partial \phiring^j},
\eeq
which implies the integrability condition
\begin{equation}\label{integrability}
    \frac{\partial\Jring^{(1)}_i}{\partial \phiring^j} =     \frac{\partial\Jring^{(1)}_j}{\partial \phiring^i}.
\end{equation}
The explicit condition~\eqref{J1 oscillatory}, or its solution~\eqref{J1osc modes}, then reduces this integrability condition to
\beq
\sum_{\vec k'}k_i k'_j \hat G_{\vec{k}',\vec{k}'-\vec k}(\omega_{k'-k}) = \sum_{\vec k'}k'_i k_j \hat G_{\vec{k}',\vec{k}'-\vec k}(\omega_{k'-k}), 
\eeq
where we have relabeled $\vec k' \to \vec k' - \vec k$ and suppressed the dependence on $\Jring_i$. This equality must hold for all $\vec k\neq0$ and all $i,j$, which is plainly impossible; consider, for example, $\vec k=(1,0,0)$ and $i\neq1$, for which the left-hand side identically vanishes. 

\subsubsection{Finding the canonical momenta}

In principle, we can obtain $(x^i_p,p_i)$ from $(\phiring^i,\Jring_i)$ using a generating function
\begin{equation}\label{W=W0+eps W1}
    W(x^i_p,\Jring_i) = W_{(0)}(x^i_p,\Jring_i) + \e W_{(1)}(x^i_p,\Jring_i),
\end{equation}
where $W_{(0)}$ is the generating function for a geodesic, given (in the case of a Kerr background) by Eq.~\eqref{W}. We then have
\begin{align}
    p_i &= \frac{\partial W}{\partial x^i_p},\label{p=dW/dx}\\
    \phiring^i &= \frac{\partial W}{\partial \Jring_i}.\label{phi=dW/dJ}
\end{align}
Just as in the geodesic case, we can write 
\beq
W = \int_{x^i_0}^{x^i_p}p_j(x^k,\Jring_k)dx^j, 
\eeq
where $x^i_0$ is an arbitrary point on the torus of constant $\Jring_i$, and the integral is along any path linking $x_0^i$ to $x^i_p$ on the torus. However, unlike in the geodesic case where $W_{(0)}$ was a type-2 generating function that defined a transformation from $(x^i_p,\tilde p_i)$ to the action-angle variables, here $W$ is to be interpreted as a type-3 generating function that transforms from the action angles to the Boyer-Lindquist coordinates and their conjugate momenta. Typically, Eq.~\eqref{phi=dW/dJ} would determine $x^i_p$ as a function of $(\phiring^i,\Jring_i)$ for any choice of $W$. However, we already know the relationship $x^i_p(\phiring^i,\Jring_i,\e)$ from earlier sections; hence, Eq.~\eqref{phi=dW/dJ} instead serves to determine the $W_{(1)}$ that reproduces the known $x^i_p(\phiring^i,\Jring_i,\e)$. Equation~\eqref{p=dW/dx} then determines the canonical momenta.

Substituting the expansion~\eqref{x(phi,J)} of $x^i_p(\phiring^i,\Jring_i,\e)$ into Eq.~\eqref{W=W0+eps W1}, we obtain from Eq.~\eqref{phi=dW/dJ} that
\begin{equation}
    \phiring^i = \frac{\partial W_{(0)}}{\partial \Jring_i} + \e \left(\frac{\partial W_{(1)}}{\partial \Jring_i} +x^j_{(1)}\frac{\partial^2 W_{(0)}}{\partial x^j_{(0)}\partial\Jring_i}\right) + {\cal O}(\e^2),
\end{equation}
where all terms on the right are now evaluated at $(x^i_{(0)},\Jring_i)$. The first term is simply $\phiring^i$. Hence, $W_{(1)}$ must satisfy
\begin{equation}
    \frac{\partial W_{(1)}}{\partial \Jring_i} = -x^j_{(1)}\frac{\partial^2 W_{(0)}}{\partial x^j_{(0)}\partial\Jring_i} = - x^j_{(1)}\frac{\partial p^{(0)}_j(x^k_{(0)},\Jring_k)}{\partial\Jring_i}.\label{dW1/dJ}
\end{equation}

If we now expand the momenta as functions of the action-angles, we can write
\beq\label{p(phi,J)}
p_i = p_i^{(0)}(x^j_{(0)},\Jring_j)+\e \left(\frac{\partial W_{(1)}}{\partial x^i_{(0)}}+x^j_{(1)}\frac{\partial^2 W_{(0)}}{\partial x^j_{(0)}\partial x^i_{(0)}}\right) + {\cal O}(\e^2).
\eeq
This involves the unknown functions $W^i_{(1)}:=\partial W_{(1)}/\partial x^i_{(0)}$. Differentiating Eq.~\eqref{dW1/dJ} with respect to $x^i_{(0)}$ at fixed $\Jring_i$, we find that they must be solutions to 
\begin{equation}\label{W1 eqn}
    \frac{\partial W_{(1)}^i}{\partial \Jring_k} = -\frac{\partial}{\partial x^i_{(0)}}\Biggl(x^j_{(1)}\frac{\partial p^{(0)}_j(x^k_{(0)},\Jring_k)}{\partial\Jring_k}\Biggr).
\end{equation}
The derivative $\partial x^j_{(1)}/\partial x^i_{(0)}$ can be evaluated by noting that $x^i_{(0)}=x^i_{(0)}(\phiring^j,\Jring_j)$ defines $\phiring^i$ as a function of $x^i_{(0)}$ and $\Jring_j$, such that $\partial x^j_{(1)}/\partial x^i_{(0)}=(\partial x^j_{(1)}/\partial \phiring^k) (\partial \phiring^k/\partial x^i_{(0)})$.

The canonical momenta are now given by Eq.~\eqref{p(phi,J)}, as functions of $(\phiring^i,\Jring_i)$, in terms of the solution to Eq.~\eqref{W1 eqn}. Actually solving that equation is beyond the scope of this paper. However, we can now formally relate the canonical momenta back to $\tilde p_i$, as in Eq.~\eqref{p=ptilde+dp}, by expanding $\tilde p_i$ at fixed $(\phiring^i,\Jring_i)$. Recalling that $\tilde p_i$ is related to $(\phistar^i,\Jstar_i)$ by the Kerr-geodesic generating function, we can write
\begin{subequations}
\begin{align}
\tilde p_i &= \frac{\partial W_{(0)}(x^j_p,\Jstar_j)}{\partial x^i_p} \\
&= p_i^{(0)}(x^j_{(0)},\Jring_j) + \e\left( x^j_{(1)}\frac{\partial^2 W_{(0)}}{\partial x^j_{(0)}\partial x^i_{(0)}} - \Jring_j^{(1)}\frac{\partial^2 W_{(0)}}{\partial \Jring_j\partial x^i_{(0)}}\right)\nonumber\\
&\quad + {\cal O}(\e^2).
\end{align}
\end{subequations}
Comparing to Eq.~\eqref{p(phi,J)}, we obtain
\beq
\delta p_i(\phiring^j,\Jring_j) =  W^i_{(1)} + \Jring_j^{(1)}\frac{\partial p^{(0)}_i}{\partial \Jring_j},
\eeq
where $W^i_{(1)}$ is again the solution to Eq.~\eqref{W1 eqn}.

\subsubsection{Canonical actions in terms of canonical momenta}

With the momenta $p_i$ now (in principle) in hand, we expect that $\Jring_i$ is related to $p_i$ by the torus integral~\eqref{J=int pdx}. This is easily confirmed using the generating function and the trivial fact that the action variables can be written as
\begin{equation}
    \Jring_i = \frac{1}{2\pi}\oint_{C_i}\Jring_j d\phiring^j.
\end{equation}

We appeal to Eqs.~\eqref{p=dW/dx} and \eqref{phi=dW/dJ}, which imply 
\beq
dW = p_i dx^i_p + \phiring^i d\Jring_i. 
\eeq
Rearranging, we have
\begin{equation}
    p_i dx^i_p = \Jring_i d\phiring^i + d(W - \phiring^i \Jring_i).
\end{equation}
Next, we note that while $W$ and $\phiring^i$ each grow with the winding number of any curve on the torus, the combination $W - \phiring^i \Jring_i$ is periodic on the torus. This follows from Eq.~\eqref{phi=dW/dJ} together with the fact that all functions on the right-hand side of Eq.~\eqref{dW1/dJ} are periodic on the torus. Hence, $\oint d(W - \phiring^j \Jring_j)=0$, and we obtain $\oint_{C_i}p_j dx^j_p = \oint_{C_i} \Jring_j d\phiring^j = 2\pi \Jring_i$, as expected. 

\subsection{Gauge invariance of the action variables}
\label{sec:invariance of Jtilde}

We explained previously that in the dissipative sector, the variables $\Jring_i$ are not fully determined due to the residual freedom~\eqref{J residual freedom}. It follows that if the trajectory $\gamma_{\rm 6D}=\{(x^i_p(t),\tilde p_i(t))|t\in\mathbb{R}\}$ is calculated in two different spacetime gauges, by solving the equations of motion~\eqref{eq:eom_6D coincidence} together with the field equations in these two gauges, then the resulting $\Jring_i(t)$ will differ.

Here, we show that the $\Jring_i$ defined from the conservative dynamics, which are fully determined, are explicitly invariant under a spacetime diffeomorphism. Our derivation makes use of the canonical momenta $p_i$ but does not require us to find these $p_i$ explicitly. 

A spacetime diffeomorphism is a point transformation on the configuration space, $x^i_p\to f^i(t,x^j_p)$. Starting from the canonical coordinates $(x^i_p,p_i)$, this induces a canonical transformation $(x^i_p,p_i)\to (X^i,P_i)$, which can be described by means of a type-2 generating function~\cite{Nakahara}
\begin{align}
 \Xi(x^i_p,P_i)=P_i f^i(x^j_p)\,.
\end{align}
We suppress $t$ dependence, which will play no role, and we emphasize that the canonical momenta $P_i$ here are not the constants of motion introduced above Eq.~\eqref{J geo}; we use the same symbol only for convenience. The new coordinates are related to the old ones by
\begin{align}
X^i &= \frac{\partial\Xi}{\partial P_i} = f^i ,\\ 
p_i &= \frac{\partial\Xi}{\partial x^i_p} = P_j \frac{\partial f^j}{\partial x^i_p}. \label{p=dXi/dx}
\end{align}

In the case of a perturbative spacetime gauge transformation, we have $f^i(t,x^j) = x^i -\e \xi^i(t,x^j)$. The new momenta $P_i$ can then be found from Eq.~\eqref{p=dXi/dx} by substituting the ansatz $P_i = p_i + \e\hat\delta_\xi p_i$. We immediately conclude that the spacetime gauge transformation induces the following infinitesimal canonical transformation:
\begin{align}
 \hat \delta_\xi x^i_p
                 &=-\xi^i(x^j_p)\,,\label{dxi x}\\
 \hat \delta_\xi p_i
                 &= p_j \frac{\partial\xi^j}{\partial x^i_p}\,, \label{dxi p}
\end{align}
where $\hat\delta_\xi$ denotes the shifts of quantities evaluated along a time-parameterized trajectory, adopting the notation of \cite{Fujita:2016igj}. 

Showing the gauge invariance of the action variables is now straightforward using the formula $2\pi\Jring_i = \oint_{C_i}p_j dx^j_p$. 
The transformation of $\Jring_i$ is evaluated as
\begin{align}
2\pi \hat\delta_\xi  \Jring_i &= \oint_{C_i}\left(\hat\delta_\xi p_j \frac{dx^j_p}{d\lambda} + p_j \frac{d(\hat\delta_\xi x^j_p)}{d\lambda}\right)d\lambda = 0,
\end{align}
where $\lambda$ is a parameter along $C_i$, and the integrand has immediately vanished upon substitution of Eqs.~\eqref{dxi x} and \eqref{dxi p}. One might be wary that this derivation has not allowed the contour $C_i$ to be altered by the gauge transformation; however, an infinitesimal change in a region of integration is proportional to an integral over the boundary of the region, which in this case trivially vanishes because $C_i$ has no boundary. Thus, we have verified that $\Jring_i$ is a gauge-invariant quantity.

We discuss gauge invariance more broadly, with dissipation, in Sec.~\ref{sec:dissipative sector}.

\subsection{Conserved quantities}
\label{sec:conserved quantities}

In the conservative sector, the canonical action variables $\Jring_i$ are all conserved. Here we identify two other important conserved quantities: the energy and azimuthal angular momentum (in the case of an axisymmetric background such as Kerr). We express all these quantities as functions of orbital frequencies, allowing us to identify them as quantities that were previously defined from the first law of black hole binary mechanics.

The energy $E$ is simply the on-shell value of the Hamiltonian ${}H_{\rm 6D}$: 
\begin{align}
  E(\Jring_i) = E_{(0)}(\Jring_i) + \frac{1}{2}\e\langle[\Hsix_{(1)}]\rangle. \label{E}
\end{align}
On the other hand, there are two distinct quantities we might define as the azimuthal angular momentum $L_z$. The first is the azimuthal component of the coordinate momentum: 
\begin{equation}
    \tilde p_\phi = \tilde g_{\phi\alpha}\frac{dx^\alpha_p}{d\tilde\tau} = z^{-1} \left(\tilde g_{t\phi} + \tilde g_{i\phi}\frac{dx^i_p}{dt} \right).
\end{equation}
This quantity is canonically conjugate to $\phi_p$ in the pseudo-Hamiltonian system: $d\phi_p/dt = [\partial\Hsix/\partial\tilde p_\phi]$ and $d\tilde p_\phi/dt = -[\partial\Hsix/\partial\phi_p]$. The fact that it is constant in the conservative sector was established in Eq.~\eqref{dp_phi/dt=0}. 
The alternative definition of azimuthal angular momentum is $\Jring_\phi$, which is canonically conjugate to $\phiring^\phi$ in the Hamiltonian system: $d\phiring^\phi/dt=\partial H_{\rm 6D}/\partial\Jring_\phi$ and $d\Jring_\phi/dt=-\partial H_{\rm 6D}/\partial\phiring^\phi$. To relate the two quantities, recall that $\Jstar_\phi = \tilde p_\phi$ by definition. Therefore $\Jstar_\phi$ is a constant, implying the oscillatory part of $\Jring_\phi^{(1)}$ vanishes.  
We therefore have 
\beq\label{Lz=pphi=Jstar}
\tilde p_\phi = \Jstar_\phi = \Jring_\phi -\e\left\langle \Jring^{(1)}_\phi\right\rangle + {\cal O}(\e^2).
\eeq
Note this is a function of the canonical actions rather than an independent constant of motion.

As we highlighted in the Introduction, the conserved energy $E$, equal to the on-shell Hamiltonian, is in perfect agreement with the binding energy obtained from the first law of binary mechanics~\cite{LeTiec:2011dp,Isoyama:2014mja}. To establish this, we first recall that the binding energy predicted by the first law has historically been written as an expansion in powers of $\e$ at fixed values of the orbital frequencies, which means that to make the comparison, we need to adjust our phase-space coordinates. Defining 
\beq\label{Jhat def}
\hat J_i = \Jring_i + \e \Delta \hat J_i(\Jring_j) 
\eeq
and re-expanding Eq.~\eqref{phidot conservative} at fixed $\hat J_i$, we obtain 
\beq
    \frac{d\phiring^i}{dt} = \Omega^i_{(0)}(\hat J_j) + \e\left(\frac{1}{2}\frac{\partial\left\langle\left[\Hsix_{(1)}\right]\right\rangle}{\partial\hat J_i}-\frac{\partial\Omega^i_{(0)}}{\partial\hat J_j}\Delta \hat J_j\right) + {\cal O}(\e^2),
\eeq
where functions on the right are evaluated at $\hat J_j$. We see that we can eliminate the frequency correction, making $\frac{d\phiring^i}{dt} = \Omega^i_{(0)}(\hat J_j)$, by setting
\begin{equation}\label{Delta Jhat}
\Delta \hat J_i = \frac{1}{2}\frac{\partial}{\partial\Omega^i_{(0)}}\left\langle\left[\Hsix_{(1)}\right]\right\rangle. 
\end{equation}
Since the physical frequencies are defined as $\frac{d\phiring^i}{dt}:=\Omega^i$,
the variables $\hat J_j$ are in one-to-one correspondence with the frequencies, as determined by the relation $\Omega^i_{(0)}(\hat J_j)=\Omega^i$. Hence, an expansion in powers of $\e$ at fixed $\hat J_i$ is also an expansion at fixed frequencies. Such a fixed-frequencies expansion is advocated for as part of the multiscale expansion of the field equations in Ref.~\cite{Pound:2021qin} and discussed further in Ref.~\cite{Mathews:2025nyb}; it can be extended to all orders.\footnote{This fails at critical surfaces where the Jacobian $\partial\Omega^i_{(0)}/\partial \hat J_j$ becomes singular~\cite{Warburton:2013yj}. The fixed-frequencies expansion is only defined away from these surfaces.} 

We can use the result~\eqref{Delta Jhat} to express the action variables as functions of the physical orbital frequencies: $\Jring_i=\hat J_i -\e\Delta \hat J_i$, which evaluates to
\begin{equation}\label{Jring(Omega)}
\Jring_i = J^{(0)}_i(\Omega^j) - \frac{\mu}{2}\frac{\partial\langle z_{(1)}\rangle}{\partial\Omega^i}.  
\end{equation}
Substituting this formula into Eq.~\eqref{E} and using Eq.~\eqref{H1=z1}, we similarly obtain the physical energy as a function of the physical frequencies:
\begin{align}\label{E first law}
  E(\Omega^i) = E_{(0)}(\Omega^i) + \frac{\mu}{2}\left(\langle z_{(1)}\rangle - \Omega^i\partial_{\Omega^i}\langle z_{(1)}\rangle\right). 
\end{align}
This is precisely the energy that was historically predicted by the first law, in the form given in Ref.~\cite{Isoyama:2014mja}. Similarly, if we take $L_z=\Jring_\phi$, we can write it as an expansion at fixed frequency using Eq.~\eqref{Jring(Omega)}:%
\begin{align}
    L_z(\Omega^i) = L_z^{(0)}(\Omega^i) - \frac{\mu}{2}\frac{\partial \langle z_{(1)}\rangle}{\partial\Omega^\phi}.
\end{align}
This formula also agrees with the angular momentum defined from the first law in the case of equatorial orbits in Ref.~\cite{LeTiec:2015kgg} and for circular orbits in Ref.~\cite{Isoyama:2014mja}. 

The fact that the first-law prediction~\eqref{E first law} is equal to the on-shell value of the Hamiltonian should also be derivable from the results of Blanco and Flanagan~\cite{Blanco:2022mgd}, though we have not verified this. Since our energy is equal to the first-law prediction, it is also equal to the mechanical energy defined by Fujita et al.~\cite{Fujita:2016igj} (see also Ref.~\cite{Isoyama:2014mja}), which was defined as the unique energy satisfying the first law in the conservative sector.

More broadly, our results have significant bearing on the formalism of Fujita et al.~\cite{Fujita:2016igj}. Our energy and angular momentum, and all of our canonical action variables in Eq.~\eqref{Jring(Omega)}, are in perfect agreement with the \emph{renormalized} variables (denoted $\tilde J_\mu$) in Ref.~\cite{Fujita:2016igj}; see Eq.~(8.11) therein, along with our Eq.~\eqref{<z>}. This clarifies that Fujita et al.'s renormalized variables, which were defined in order to satisfy the first law, are in fact the genuine canonical, invariant actions in the 6D Hamiltonian system. We explain this further in Appendix~\ref{sec:8D vs 6D AA variables}.

\section{Pseudo-Hamiltonian description of the dissipative sector}
\label{sec:dissipative sector}

We now consider the dissipative sector, which is governed by Eq.~\eqref{dJdt final}.

We first note that for any function ${\cal G}$ on the 6D phase space, we can write its time evolution as
\begin{equation}
    \frac{d{\cal G}}{dt} = \left[\{{\cal G},\Hsix\}_\star\right],
    \label{dGdt}
\end{equation}
where we have introduced the Poisson bracket
\begin{equation}
    \left\{f,g\right\}_\star:=\pdv{f}{\phistar^j}\pdv{g}{\Jstar_j}-\pdv{f}{\Jstar_j}\pdv{g}{\phistar^j}.
    \label{PBstar}
\end{equation}
Hence, 
\begin{equation}\label{dJdt star bracket}
    \frac{d\Jring_i}{dt} = \bigl[\bigl\{\Jring_i\,,\Hsix\bigr\}_\star\bigr].
\end{equation}
Since $(x^i_p,\tilde p_i)$ are related to $(\phistar^i,\Jstar_i)$ by a canonical transformation, we can also write the star bracket as
\begin{equation}
    \left\{f,g\right\}_\star:=\pdv{f}{x^j_p}\pdv{g}{\tilde p_j}-\pdv{f}{\tilde p_j}\pdv{g}{x^j_p}.
\end{equation}
In the remainder of this section, we describe several alternative forms of $d\Jring_i/dt$, focusing on how it can be formulated in terms of the Poisson bracket associated with the conservative (Hamiltonian) sector, 
\begin{equation}
    \left\{f,g\right\}:=\pdv{f}{\phiring^j}\pdv{g}{\Jring_j}-\pdv{f}{\Jring_j}\pdv{g}{\phiring^j}.
    \label{PBring}
\end{equation}

As a first step, we note the trivial identity $\{\Jring_i,g\}=-\partial g/\partial\phiring^i$ along with 
\begin{align}
\left\langle f \left\langle g\right\rangle\right\rangle &= \langle f \rangle \langle g \rangle,\\ 
f({\cal Q})\left[\,g({\cal Q},{\cal Q'})\,\right] &= \left[f g\right],\\
\partial_{\Jring_i}\langle [f]\rangle &= \langle\partial_{\Jring_i} [f]\rangle.
\end{align}
Using these, we can rewrite Eq.~\eqref{dJdt final} as
\begin{align}
    \frac{d\Jring_i}{dt}
    &= \bigl\langle\bigl[\bigl\{\Jring_i\,,\,\e\mathring\Hsix^{\rm rad}_{(1)}+\e^2\mathring\Hsix_{(2)}\bigr\}\bigr]\bigr\rangle + \e^2 K_i +{\cal O}(\e^3),\label{dJdt generic dissipative}
\end{align}
with 
\begin{equation}\label{eq:K}
    K_i  = \bigl\langle\bigl[\bigl\{\Jring_i^{(1)},\mathring\Hsix_{(1)}\bigr\}\bigr]\bigr\rangle
    +\left\langle\Jring_j^{(1)}\right\rangle\left\langle\pdv{}{\Jring_j}\!\left[\pdv{\mathring\Hsix^{\rm rad}_{(1)}}{\phiring^i}\right]\right\rangle.
\end{equation}
As in Eq.~\eqref{dJdt final}, all quantities on the right are functions of $(\phistar^i_{(0)},\Jring_i)$ (and functionals of $\gamma_{\rm 6D}$), and we have introduced square brackets as explicit reminders that derivatives are taken before evaluating at coincidence. 

Using Synge's rule once again, we can write the second term in $K_i$ as
\begin{equation} 
\left\langle\Jring_j^{(1)}\right\rangle\left\langle\pdv{}{\Jring_j}\!\left[\pdv{\mathring\Hsix^{\rm rad}_{(1)}}{\phiring^i}\right]\right\rangle = -\left\langle\left[\pdv{{\cal K}}{\phiring^i}\right]\right\rangle = \bigl\langle \bigl[\bigl\{\Jring_i\,,{\cal K}\bigr\}\bigr]\bigr\rangle,
\end{equation}
where
\begin{equation}\label{calK}
    {\cal K} := -\left\langle\Jring_j^{(1)}\right\rangle\left(\partial_{\Jring_j}\mathring\Hsix^{\rm rad}_{(1)}+\partial_{\Jring'_j}\mathring\Hsix^{\rm rad}_{(1)}\right).
\end{equation}
Given Eq.~\eqref{<J1> condition with T symmetry}, we can also put ${\cal K}$ in a more symmetrical form:
\begin{multline}
    {\cal K} = -\frac{1}{2}\left\langle\left[\partial_{\Jring_i}\mathring\Hsix^{\rm sym}_{(1)} - \partial_{\Jring'_i}\mathring\Hsix^{\rm sym}_{(1)}\right]\right\rangle \\
    \times\frac{\partial\Jring_i}{\partial\Omega^j}\!\left(\partial_{\Jring_j}\mathring\Hsix^{\rm rad}_{(1)}+\partial_{\Jring'_j}\mathring\Hsix^{\rm rad}_{(1)}\right).
\end{multline}

Combining $\cal K$ with the pseudo-Hamiltonian terms in Eq.~\eqref{dJdt generic dissipative}, we can write
\begin{multline}
    \frac{d\Jring_i}{dt}
    = \bigl\langle\bigl[\bigl\{\Jring_i\,,\, \mathring{\Hsix} \bigr\}\bigr]\bigr\rangle
    + \e\bigl\langle\bigl[\bigl\{\Jring_i^{(1)},\mathring{\Hsix}\bigr\}\bigr]\bigr\rangle +{\cal O}(\e^3),
\end{multline}
where
\begin{equation}
    \mathring{\Hsix} := \e\mathring{\Hsix}_{(1)} + \e^2\mathring{\Hsix}_{(2)} +\e^2{\cal K}.
\end{equation}
We can also express this result in terms of the complete pseudo-Hamiltonian in Eq.~\eqref{Hsix(phiring,Jring) v1}. Using the fact that $\langle \partial f({\cal Q})/\partial\phiring^i\rangle=0$ for any periodic $f$, we immediately find 
\begin{multline}
    \frac{d\Jring_i}{dt}
    = \bigl\langle\bigl[\bigl\{\Jring_i\,,\Hsix'+\e^2{\cal K}\bigr\}\bigr]\bigr\rangle \\
    + \e^2\Bigl\langle\Bigl[\Bigl\{\Jring_i, \Bigl(\phiring_{(1)}^i\partial_{\phiring^i} 
    +\Jring^{(1)}_i\partial_{\Jring_i}\Bigr) \mathring\Hsix_{(1)}\Bigr\}\Bigr]\Bigr\rangle \\
    + \e^2\bigl\langle\bigl[\bigl\{\Jring_i^{(1)},\mathring\Hsix_{(1)}\bigr\}\bigr]\bigr\rangle +{\cal O}(\e^3).
    \label{J dot as PB}
\end{multline}

We do not further belabor the dissipative evolution equations. In future work, starting from the equations presented here, we will aim to express $d\Jring_i/dt$ in terms of asymptotic fluxes, as was done at leading order in Ref.~\cite{Isoyama:2018sib}.

\section{Invariant 1PA waveform generation}\label{sec:gauge choices}

We now summarize the multiscale waveform generation framework in terms of our invariant action angles.

As reviewed in the Introduction, the multiscale waveform generation is described by Eq.~\eqref{multiscale waveform}. Reference~\cite{Mathews:2025nyb} established the invariance of this waveform under the phase-space transformation 
\begin{align}
\phiring^i&\to\phiring^i + \alpha^i_{(0)}(\piring_j) + \e \alpha^i_{(1)}(\piring_j) + \ldots,\\ \piring_i&\to \beta^{(0)}_i(\piring_j) + \e \beta^{(1)}_i(\piring_j) + \ldots
\end{align}
Here, we use that freedom to describe the waveform directly in terms of the canonical action variables. We then relate that formulation to the one in Ref.~\cite{Pound:2021qin}, in which the waveform is described in terms of the invariant frequencies $\Omega^i$.

Finally, in Sec.~\ref{sec:remarks} we remark on practical issues of implementation. To frame the discussion, we first write our evolution equations in their most compact form, bringing together Eqs.~\eqref{dphidt Hamilton} and~\eqref{dJdt star bracket}:
\begin{align}
    \frac{d\phiring^i}{dt} &= \frac{\partial H_{\rm 6D}}{\partial\Jring_i} , \label{dphiringdt compact}\\
    \frac{d\Jring_i}{dt} &= \bigl[\bigl\{\Jring_i\,,\Hsix\bigr\}_\star\bigr].\label{dJringdt compact}
\end{align}

\subsection{Waveforms in terms of canonical action-angle variables}

If we use the canonical action-angle variables $(\phiring^i,\Jring_i)$ as our coordinates on phase space, then the orbital evolution equations~\eqref{dphiringdt compact} and \eqref{dJringdt compact} can be written more explicitly as
\begin{align}
    \frac{d\phiring^i}{dt} &= \Omega^i_{(0)}(\Jring_k) + \frac{\mu}{2}\frac{\partial\langle z_{(1)}\rangle}{\partial \Jring_i} + \O(\e^2), \label{dphiringdt finalest}\\
    \frac{d\Jring_i}{dt} &= -\e\bigl\langle\bigl[\partial_{\phiring^i} \mathring\Hsix^{\rm rad}_{(1)}\bigr]\bigr\rangle \nonumber\\ 
    &\quad - \e^2\Bigl\langle\Bigl[\partial_{\phiring^i}\mathring\Hsix_{(2)} + \partial_{\phiring^i}{\cal K} - \bigl\{\Jring_i^{(1)},\mathring\Hsix_{(1)}\bigr\}\Bigr]\Bigr\rangle \nonumber\\
    &\quad + \O(\e^3),\label{dJringdt finalest}
\end{align}
with $\cal K$ given by Eq.~\eqref{calK} and $\Jring^{(1)}_i$ by Eqs.~\eqref{J1osc no modes} and~\eqref{<J1> solution}.
In this formulation, the Hamiltonian $H_{\rm 6D}$ determines the orbital frequencies $\Omega^i(\Jring_j,\e)$ on the tori of constant $\Jring_i$. Through Eq.~\eqref{dJringdt finalest}, the tori then slowly evolve. 

The waveform is still given by Eq.~\eqref{multiscale waveform}, where we define $\piring_i=\piring_i^{\rm FC}$ to be geodesically related to the action variables: $\Jring_i = J^{(0)}_i(\piring^{\rm FC}_j)$. Here we use ``FC'' to indicate the ``fixed constants of motion'' gauge on phase space, following Ref.~\cite{Piovano:2024yks}.

\subsection{Waveforms in terms of invariant frequencies}

In Refs.~\cite{Pound:2021qin,Mathews:2025nyb}, a prescription was given to obtain the evolution equations and waveform in terms of variables $\piring^{\rm FF}_i$ that are geodesically related to the physical orbital frequencies, meaning $\Omega^i=\Omega^i_{(0)}(\piring_j^{\rm FF})$. Here ``FF'' indicates ``fixed frequencies'', again following the notation of Ref.~\cite{Piovano:2024yks}.

This prescription is equivalent to using non-canonical coordinates $(\phiring^i,\hat J_i)$, where $\hat J_i = J^{(0)}_i(\Omega^j)$ are the ``action variables''~\eqref{Jhat def} that are geodesically related to the orbital frequencies. The orbital equations of motion then read 
\begin{align}
    \frac{d\phiring^i}{dt} &= \Omega^i_{(0)}(\hat J_k), \\
    \frac{d\hat J_i}{dt} &= -\e\bigl\langle\bigl[\partial_{\phiring^i} \mathring\Hsix^{\rm rad}_{(1)}\bigr]\bigr\rangle -\frac{\e^2}{2}\bigl\langle\bigl[\partial_{\phiring^j} \mathring\Hsix^{\rm rad}_{(1)}\bigr]\bigr\rangle\partial_{\hat J_j}\partial_{\Omega^i}\langle z_{(1)}\rangle \nonumber\\
    &\quad +\frac{\e^2}{2}\partial_{\hat J_j}\!\bigl\langle\bigl[\partial_{\phiring^i} \mathring\Hsix^{\rm rad}_{(1)}\bigr]\bigr\rangle\partial_{\Omega^i}\langle z_{(1)}\rangle  \nonumber\\
    &\quad -\e^2\Bigl\langle\Bigl[\partial_{\phiring^i}\mathring\Hsix_{(2)} + \partial_{\phiring^i}{\cal K} - \bigl\{\Jring_i^{(1)},\mathring\Hsix_{(1)}\bigr\}\Bigr]\Bigr\rangle \nonumber\\
    &\quad + \O(\e^3),\label{dJhatdt}
\end{align}
where all functions of action variables on the right are evaluated at $\hat J_i$. The second term in Eq.~\eqref{dJhatdt} arises from the time derivative of $\Delta\hat J$ in Eq.~\eqref{Jhat def}, and the third term arises from re-expanding the first term in Eq.~\eqref{dJringdt finalest}. 

In this formulation, the invariant tori are directly labeled with $\Omega^i$ rather than with $\Jring_i$. The waveform is still given by Eq.~\eqref{multiscale waveform}, where we now use $\piring_i=\piring_i^{\rm FF}$, which are geodesically related to $\hat J_i$ rather than to $\Jring_i$: $\hat J_i = J^{(0)}_i(\piring^{\rm FF}_j)$. Using the definition~\eqref{Jhat def}, we immediately see that the two sets of variables are related by
\begin{equation}
    \piring^{\rm FF}_i = \piring^{\rm FC}_i + \e\Delta\piring^{\rm FF}_i + \O(\e^2)
\end{equation}
with $\Delta\piring^{\rm FF}_j\partial_{\piring_j}J^{(0)}_i = \Delta\hat J_i$, or equivalently, $\Delta \piring^{\rm FF}_j = \partial_{\hat J_i}\piring^{(0)}_j\Delta\hat J_i$. The amplitudes in the waveform~\eqref{multiscale waveform} in the two gauges are therefore related by
\begin{equation}
    \mathring h^{(2)\rm {FF}}_{\ell m\bm{k}}(\mathring\pi^{\rm FF}_i) = \mathring h^{(2)\rm {FC}}_{\ell m\bm{k}}(\mathring\pi^{\rm FF}_i) -\Delta\mathring\pi^{\rm FF}_i\partial_{\mathring\pi^i}\mathring h^{(1)}_{\ell m\bm{k}}(\mathring\pi^{\rm FF}_i),
\end{equation}
where we have suppressed dependence on $\delta M_A$.

\subsection{Practical considerations}
\label{sec:remarks}

Building a 1PA waveform model requires calculation of several ingredients on a 4D grid of orbital parameters and black hole spin. As inputs to the evolution equations~\eqref{dphiringdt finalest} and \eqref{dJringdt finalest}, one requires the following:
\begin{enumerate}
    \item 0PA ``fluxes'' $     \left\langle\left[\partial\mathring \Hsix^{\rm rad}_{(1)}/\partial \phiring^i\right]\right\rangle$, which can be computed from solutions to the first-order Teukolsky equation. They are given explicitly by Eq.~(3) of Ref.~\cite{Isoyama:2018sib}.
    \item The averaged redshift $\langle z_{(1)}\rangle$. Methods of calculating this for equatorial orbits are standard~\cite{Barack:2011ed,Akcay:2015pza}, and results are reported for generic, inclined orbits in Ref.~\cite{Nasipak:redshift}.
    \item The linear transformations $\phiring^i_{(1)}$ and $\Jring^{(1)}_i$, given by Eqs.~\eqref{J1osc no modes}, \eqref{phi1osc no modes}, and \eqref{<J1> solution}. These explicitly enter the 1PA dissipative terms through the Poisson bracket $\{\Jring_i^{(1)},\mathring\Hsix_{(1)}\}$ and the quantity $\cal K$ defined in Eq.~\eqref{calK}. They also enter implicity through the field equations~\eqref{multiscale EFE 2}, since $\mathring{T}^{(2)}_{\mu\nu}$ involves terms of the form $\Jring_i^{(1)}\partial_{\Jring_i}\mathring{T}^{(1)}_{\mu\nu} + \phiring^i_{(1)}\partial_{\phiring^i}\mathring{T}^{(1)}_{\mu\nu}$. 
    
    If computed as written in these equations, these quantities require the mode expansion~\eqref{H1 modes} of $\mathring\Hsix_{(1)}$, along with integrals and derivatives thereof, evaluated at coincidence. Such coincidence limits can be numerically evaluated by (i) mode-expanding the leading-order stress-energy tensor as described in Secs.~7.1.1 and 7.1.2 of Ref.~\cite{Pound:2021qin}, and references therein, or in Refs.~\cite{Akcay:2013wfa,Osburn:2014hoa,Drasco:2005kz}; and (ii) constructing a Green's function as in Refs.~\cite{Galtsov:1982hwm,Sago:2005fn,Drasco:2005is} (for the radiative part of the Green's function) or \cite{Akcay:2013wfa,Osburn:2014hoa,Casals:2024ynr} (for the retarded Green's function) together with a subtraction scheme to remove the singular part (see, e.g., Ref.~\cite{Casals:2019heg} and references therein). 
    
    However, we note that the oscillatory parts of the transformation are necessarily equivalent to the transformation to $\piring_i$ outlined in Refs.~\cite{Pound:2021qin,Mathews:2025nyb}, since the oscillatory part of the transformation is fully determined by the requirement of eliminating oscillations from the equations of motion. References~\cite{VanDeMeent:2018cgn,Lynch:2021ogr,Lynch:2023gpu,Drummond:2023wqc,Lynch:2024ohd} demonstrate methods of calculating the order-$\e$ terms in such transformations. 
    
    In that sense, the only ``new'' ingredient is the averaged part of the transformation, $\langle \Jring^{(1)}_i\rangle$, which is equivalent to fixing the gauge freedom $\langle \pi^{(1)}_i\rangle$  (discussed in Sec.~\ref{subsubsec:gauge freedom}). This piece is given by Eq.~\eqref{<J1> solution}. 
    \item The 2SF dissipative term $\bigl\langle\bigl[\partial_{\phiring^i}\mathring\Hsix_{(2)}\bigr]\bigr\rangle$. In the 1PA evolution, this is the only term that requires solving the second-order field equation~\eqref{multiscale EFE 2}. We envision ultimately replacing it---or replacing the sum of \emph{all} terms in Eq.~\eqref{dJringdt finalest}---with asymptotic fluxes, as at 0PA. This replacement can be pursued by extending the methods from Refs.~\cite{Isoyama:2018sib,Grant:2024ivt}.
\end{enumerate}

\section{Conclusion}

In this paper, we have unified the pseudo-Hamiltonian formalism of Ref.~\cite{Fujita:2016igj} with the multiscale expansion of the Einstein equations in Refs.~\cite{Miller:2020bft,Pound:2021qin,Miller:2023ers,Mathews:2025nyb,Wei:2025lva}. There are several outcomes of this.

First, 1PA orbital evolution is formulated directly in terms of invariant action-angle variables. These variables are canonically conjugate in a 6D conservative Hamiltonian system that is embedded within the full, dissipative pseudo-Hamiltonian 1PA evolution. The actions can be written as integrals over the invariant tori in the 6D orbital phase space of the multiscale expansion. Solutions to the conservative, Hamiltonian equations remain on these invariant tori; dissipation then slowly, smoothly drives the system from one torus to the next. This formulation complements that of Ref.~\cite{Pound:2021qin}, where the waveform generation was instead formulated in terms of invariant frequencies. Both formulations are summarized in Sec.~\ref{sec:gauge choices}, and we have provided the transformation between the two.

In the multiscale expansion, the waveform itself is a function on the mechanical phase space, and each mode amplitude in the waveform (the amplitude of each frequency or ``voice''~\cite{Hughes:2021exa}) is itself an integral over an invariant torus. In Ref.~\cite{Mathews:2025nyb}, one of us characterized the residual phase-space gauge freedom within the multiscale expansion and showed that the multiscale waveform is invariant under this freedom. In this paper we highlighted the particularly simple, geometrical form that this invariant waveform generation takes when written in terms of canonical action-angle variables. 

As a key part of our formalism, we have shown that the pseudo-Hamiltonian and Hamiltonian 1PA orbital dynamics are manifestly \emph{local} on phase space. This is a nontrivial fact, given that the particle's orbit is coupled to the system's field degrees of freedom. Integrating out the field degrees of freedom leads to nonlocal-in-time equations of motion for the particle's orbit. However, the separation of time scales in an inspiral allows us to localize the dynamics. More precisely, we eliminate integrals over time and show that all time dependence reduces to dependence on the phase-space trajectory (and on the primary black hole's evolving mass and spin parameters).\footnote{Alternatively, we can say we have replaced integrals over time with integrals over invariant tori, given that our localization is a discrete Fourier representation that inherently involves such integrals.} This localization goes hand in hand with the multiscale expansion of the field equations. More broadly, we have shown how that multiscale expansion can be derived from the self-consistent formulation of self-force theory through an application of the stationary phase approximation, thereby reducing the metric perturbation to a local function on phase space. However, we note that at 2PA order, such localization-in-time breaks down as memory effects first enter the dynamics~\cite{Cunningham:2024dog}; we defer exploration of such memory effects to future work.

To facilitate our description of the pseudo-Hamiltonian dynamics, we have also provided a phase-space reformulation of the self-consistent expansion. Our reformulation clarifies precisely what is being held fixed in the expansion: the phase-space trajectory. It also more naturally links to the multiscale expansion and provides a more satisfactory treatment of the primary black hole's evolution.

Similarly, our treatment has clarified aspects of previous pseudo-Hamiltonian constructions in Refs.~\cite{Fujita:2016igj,Blanco:2022mgd}. Broadly speaking, we have shown how the formalism, when formulated in 6D, integrates into the multiscale expansion of the field equations and how to consistently account for dissipation in it through 1PA order. More specifically, our localization method has clarified the non-canonical nature of the transformations required in the multiscale expansion. It has also clarified the non-canonical transformations that are required to obtain canonical coordinates in the genuine (rather than pseudo-)Hamiltonian system governing the conservative dynamics. Our analysis highlights that the naive momenta are not canonically conjugate to the Boyer-Lindquist coordinates in the Hamiltonian system. On the other hand, we have shown that the ``renormalized'' action variables defined in Ref.~\cite{Fujita:2016igj} are, in fact, the geometrical action variables that are canonically conjugate to the angle variables in the 6D Hamiltonian system.

Importantly, unlike any previous accounts, we have shown how to construct the needed phase-space transformations in practical form, providing the expansions~\eqref{x(phi,J)} and~\eqref{p(phi,J)} of the Boyer-Lindquist trajectory in terms of action angles. This is necessary because those expansions enter explicitly into the source terms in the second-order field equation. 

Generically, the split of the 1PA dynamics into conservative and dissipative sectors is gauge dependent~\cite{Lynch:2021ogr}. However, we have established an invariant form of the division. Achieving this split required a nontrivial specification of the residual phase-space gauge freedom in the multiscale framework; this specification is made in Eq.~\eqref{<J1> solution}. In Secs.~\ref{sec:conservative sector} and \ref{sec:dissipative sector} we have elucidated the properties of the two sectors (conservative and dissipative).

Our work has also shed new light on aspects of current 1PA waveform models. These models utilize energy-balance arguments in which the gravitational binding energy is approximated by the value predicted by the first law of binary black hole mechanics~\cite{Wardell:2021fyy}. The accuracy of that approximation has not been precisely known, leading to substantial uncertainty in the model; see Ref.~\cite{Albertini:2022rfe} for a detailed discussion. We have clarified the interpretation of the first-law energy by showing it is equal to the on-shell value of the 6D Hamiltonian. Other very recent and forthcoming work~\cite{Trestini:2025nzr,Grant:FirstLaw} further clarifies that at 1PA order, this Hamiltonian mechanical energy differs from the particle's 1SF contribution to the Bondi mass: the two energies differ by a Schott term. This Schott term must be correctly accounted for in 1PA waveform models. Hence, our Hamiltonian description contributes to this emerging understanding of the first-law binding energy, its relevance in radiating binaries, and the correct form of 1PA energy-balance laws.

Other work, by Nasipak, will include the first calculation of the (6D) 1SF Hamiltonian for generic, inclined and eccentric orbits~\cite{Nasipak:redshift}. Our formulation shows how that result represents an invariant building block in the complete 1PA waveform.

In the future, building on Ref.~\cite{Isoyama:2018sib}, we will use our formulation of the 1PA dynamics to explore the possibility of deriving a complete set of flux-balance laws for the action variables. Such laws would further streamline the 1PA waveform generation by allowing the calculation of second-order dissipative effects directly from the solution to the second-order Teukolsky equation~\cite{Spiers:2023cip}.

\begin{acknowledgments}
JL and AP are indebted to Riccardo Gonzo for collaboration on a closely related project that proved invaluable in informing this paper. AP thanks Soichiro Isoyama for many helpful discussions about action-angle variables over the years, Alex Le Tiec for equally many helpful discussions of the first law of binary black hole mechanics, and Francisco Blanco, Eanna Flanagan, Tanja Hinderer, Jordan Moxon, and Paul Ramond for informative discussions of Hamiltonians, pseudo-Hamiltonians, and multiscale expansions. JL acknowledges the support of an STFC studentship. TK is supported by JST SPRING, Grant Number JPMJSP2110. AP acknowledges the support of a Royal Society University Research Fellowship and the ERC Consolidator/UKRI Frontier Research Grant GWModels (selected by the ERC and funded by UKRI [grant number EP/Y008251/1]). TT is supported by JSPS KAKENHI Grant Nos. JP23H00110  24H00963, 24H01809 and JP20K03928.
\end{acknowledgments}

\appendix

\section{Mass and spin evolution}
\label{app:BH perturbation}

In Eq.~\eqref{dg Kirchhoff} we present a formula for the metric perturbation that is self-consistently sourced by the evolving corrections to the central black hole's mass and spin. Here we discuss that perturbation and its derivation.

We first recall that in Ref.~\cite{Miller:2020bft}, one of us examined a problem that arose in earlier formulations of the self-consistent expansion used in self-force theory: they do not accurately track the long-term evolution of the primary black hole, and hence they do not admit a multiscale expansion.\footnote{This problem was pointed out to one of us (AP) by Eanna Flanagan and Jordan Moxon.} The black hole's mass and spin do evolve in these formulations, as they must by virtue of the Einstein equations, but they appear at second perturbative order and evolve exactly linearly in time, at a rate that is set by initial data on a Cauchy surface. This precisely linear behavior can be understood from the fact that the mass and spin corrections $\delta M_A$ evolve on the radiation-reaction time scale and therefore take the form $\delta M_A(\tilde v,\e)$, where $\tilde v = \e v$ is slow advanced time along the horizon. If the corrections $\delta M_A$ are not handled appropriately, then the perturbative Einstein equations force the expansion 
\begin{multline}\label{dMA Taylor}
\delta M_A(\e v,\e) = \delta M_A(0,0) + \e v\, \partial_{\tilde v}\delta M_A(0,0) \\
+ \e\partial_\e\delta M_A(0,0) + {\cal O}(\e^2v^2).    
\end{multline}
In a solution that accurately tracks the system's evolution, $\delta M_A(\e v,\e)$ must appear in the first-order metric perturbation, while in the expansion above, the time dependence first appears in the second-order perturbation---with spurious late-time growth. This restricts the method's accuracy to time scales much shorter than $1/\e$, on which Eq.~\eqref{dMA Taylor} is valid.

In Ref.~\cite{Miller:2020bft} we circumvented this problem by adding a field $\delta g_{\alpha\beta}=\frac{\partial g_{\alpha\beta}}{\partial M_A}\delta M_A(\e s,\e)$ to the first-order perturbation, where $s$ is a hyperboloidal time coordinate. This perturbation $\delta g_{\alpha\beta}$ does not satisfy the Lorenz-gauge field equation $E_{\alpha\beta}[\delta g]=0$, meaning we cannot freely add it to the point-particle solution $h^{\rm pp}_{\alpha\beta}$ and still satisfy the first-order field equation~\eqref{EFE1}. To counteract that failure, Ref.~\cite{Miller:2020bft} added another perturbation, call it $y_{\alpha\beta}$, defined as the retarded solution to $E_{\alpha\beta}[y]=-E_{\alpha\beta}[\delta g]$, such that the total linear perturbation due to the black hole's evolution became $h^{\rm BH}_{\alpha\beta}=\delta g_{\alpha\beta}+y_{\alpha\beta}$. 

That approach is unsatisfactory for two reasons. First, it is not entirely clear that $y_{\alpha\beta}$ contains no mass or spin content, meaning it could cancel the content in $\delta g_{\alpha\beta}$ that we wish to include in the first-order solution. Second, this solution involves instantaneous propagation: through $\delta g_{\alpha\beta}$, information about the state of the black hole is instantly transmitted across all of space on surfaces of constant $s$. 

Here we sketch a more satisfactory, manifestly causal alternative that mirrors the construction of $h^{\rm pp}_{\alpha\beta}$. This alternative is provided by the membrane paradigm~\cite{Price:1986yy}. We first consider how, even in principle, information about the black hole's state can propagate away from the black hole. Any information precisely on the event horizon cannot escape. However, if we place a stretched horizon $\Gamma$ slightly outside the event horizon, then information can propagate away from it out into the black hole's exterior. We next consider that as radiation falls into the black hole, the mass and angular momentum behind this stretched horizon grows. The resulting perturbation to the metric near the horizon acts as boundary data on $\Gamma$, or equivalently, as an effective source on $\Gamma$, which then re-radiates to the exterior.

More concretely, we imagine the growth of the black hole causes a perturbation $\delta g_{\alpha\beta}$ in a neighbourhood of the event horizon, and we place $\Gamma$ in that neighborhood. We then define $h^{\rm BH}_{\alpha\beta}$ as the causal solution to $E_{\alpha\beta}[\bar h^{\rm BH}]=0$ (i)~containing no incoming radiation from past null infinity, and (ii) consistent with the boundary data $\delta g_{\alpha\beta}$ on $\Gamma$. Straightforward manipulations following Ref.~\cite{Poisson:2011nh} show that the field satisfying these conditions can be written in the Kirchhoff form~\eqref{dg Kirchhoff}, which we rewrite here as
\begin{multline}\label{dg Kirchhoff 2}
    h^{\rm BH}_{\alpha\beta} = \frac{1}{16\pi}\int_{-\infty}^\infty dv'\int_{{\cal C}_{v'}} dS'\Bigl(G^{\rm ret}_{\alpha\beta}{}^{\,\alpha'\beta'}\nabla_{\gamma'}\delta g_{\alpha'\beta'} \\[-5pt]
    - \delta g_{\alpha'\beta'}\nabla_{\gamma'}G^{\rm ret}_{\alpha\beta}{}^{\,\alpha'\beta'}\Bigr) n^{\gamma'},
\end{multline}
where $v$ is advanced time, $n^\gamma$ is the spacelike unit normal to $\Gamma$, ${\cal C}_v$ is the  $v=\text{constant}$ cut of $\Gamma$, and $dS$ is the induced surface element on ${\cal C}_v$.  
This solution should survive in the limit where the stretched horizon asymptotes to the (perturbed~\cite{Bonetto:2021exn}) event horizon.

Within the integral, we can use $\delta g_{\alpha\beta}=\delta M_A \partial g_{\alpha\beta}/\partial M_A$, where $\delta M_A$ has an arbitrary time dependence. However, in addition to satisfying the wave equation $E_{\alpha\beta}[\bar h^{\rm BH}]=0$, $h^{\rm BH}_{\alpha\beta}$ must also satisfy the Lorenz gauge condition $\nabla^\beta \bar h^{\rm BH}_{\alpha\beta}={\cal O}(\e^2)$ once the true time dependence of $\delta M_A$ is specified through Eq.~\eqref{Mdot flux}. This can be achieved by enforcing that $\delta g_{\alpha\beta}$ satisfies $\nabla^\beta \bar {\delta g}_{\alpha\beta}\equiv0$ when $\delta M_A$ is independent of time. Since the wave equation preserves the gauge condition, this condition on the boundary data $\delta g_{\alpha\beta}$ suffices to enforce it on the perturbation everywhere in the black hole's exterior. One way to put $\delta g_{\alpha\beta}$ in this form is to solve the Lorenz-gauge field equation~\eqref{EFE1}, together with the gauge condition, analytically in a vicinity of the black hole horizon, effectively series-expanding for small distances from the horizon; such a near-horizon solution was constructed to derive boundary conditions for the calculations in Refs.~\cite{Pound:2019lzj,Warburton:2021kwk}, adapting methods from Refs.~\cite{Barack:1998bw,Barack:1999ma,Barack:1999st,Barack:1999ya}. In that solution, one can identify terms associated with perturbations of the mass and spin~\cite{Bonetto:2021exn}.  These terms provide a suitable local expression for $\delta g_{\alpha\beta}$. 

The resulting expression for $\delta g_{\alpha\beta}$ is linear in coefficients $\delta M_A$ that are arbitrary functions of advanced time along the horizon if the gauge condition is not enforced, constant in time if $\nabla^\beta \bar {\delta g}_{\alpha\beta}\equiv0$ is enforced, and satisfy Eq.~\eqref{Mdot flux} when the gauge condition~\eqref{Lorenz gauge} is enforced on the full perturbation. We refer back to Sec.~\ref{sec:summary} for discussion of this role of the gauge condition in the self-consistent expansion.

From Eq.~\eqref{dg Kirchhoff 2}, we can recover the multiscale form of $h^{\rm BH}_{\alpha\beta}$. Given Eq.~\eqref{Mdot flux}, we can consistently adopt the ansatz $\delta M_A=\delta M_A(\e v,\e)$, noting that here this is a function on the stretched horizon rather than across all of spacetime. We can also use the facts that (i) hyperboloidal time $s$ reduces to $v$ near the horizon, and (ii) the only time dependence in the integral, besides $\delta M_A$, enters through the Green's function, which is a function of $(s-s')$ by virtue of the background spacetime's stationarity. Employing these facts, we put the time integral in Eq.~\eqref{dg Kirchhoff 2} in the form $\int_{-\infty}^\infty ds' f(\e s')G(s-s')$, where $f$ represents the dependence on $\delta M_A$ and $G$ represents the dependence on the Green's function. A change of integration variable to $s''= s'-s$ reduces this to 
\beq
\int_{-\infty}^\infty ds'' f(\e s+\e s'')G(-s'')\approx f(\e s)\int_{-\infty}^\infty ds''G(-s'').
\eeq
The remaining integral over time picks out the zero-frequency mode of the Green's function, $\hat G^{\rm ret}_{\alpha\beta}{}^{\,\alpha'\beta'}(0,x^i,x'^i)= 2\pi\int ds' G(s',x^i,x'^i)$, where we adopt the notation of Eq.~\eqref{Gret Fourier s} for the Fourier transform of the Green's function. We then arrive at the slowly evolving metric perturbation
\begin{multline}
    h^{\rm BH}_{\alpha\beta} \approx \frac{1}{8}\delta M_A(\e s,\e)\int_{{\cal C}_v} dS'\Bigl\{\hat G^{\rm ret}_{\alpha\beta}{}^{\,\alpha'\beta'}(0,x^i,x'^i)\nabla_{\gamma'}k^A_{\alpha'\beta'} \\[-5pt]
    - k^A_{\alpha'\beta'}\nabla_{\gamma'}\hat G^{\rm ret}_{\alpha\beta}{}^{\,\alpha'\beta'}(0,x^i,x'^i)\Bigr\} n^{\gamma'},
\end{multline}
where we have written $\delta g_{\alpha\beta}=\delta M_A k^A_{\alpha\beta}$ for some $s$-independent tensors $k^A_{\alpha\beta}$.

In this solution, information about the state of the black hole is again instantaneously transmitted across hyperboloidal slices. However, that form is now a consistent approximation based on the long time scale $s\sim 1/\e$ over which the black hole evolves; the underlying solution~\eqref{dg Kirchhoff 2} is causal and not restricted to the particular separation of time scales in an inspiral.

A more thorough description of this construction of $h^{\rm BH}_{\alpha\beta}$ will be presented elsewhere. 

\section{Derivation of the multiscale expansion: stationary phase approximation}\label{sec:multiscale from self-consistent}

As alluded to throughout the body of the paper, the multiscale expansion of the metric perturbation can be derived from the self-consistent expansion by applying a stationary phase approximation. This effectively localizes the metric perturbation on phase space. Here we describe that localization.

The derivation mirrors the one for the pseudo-Hamiltonian in Sec.~\ref{sec:localization}. Like that derivation, we start by expressing the Green's function as an inverse Fourier transform,  
\begin{multline}\label{Gret Fourier s}
G^{\alpha\beta\alpha'\beta'}_{\rm ret}(s-s',x^i,x^{i'}) \\= \int_{-\infty}^\infty d\omega\, e^{-i\omega (s-s')}\hat G_{\rm ret}^{\alpha\beta\alpha'\beta'}(\omega,x^i,x'^{i}).
\end{multline}
This differs from Eq.~\eqref{GR Fourier} in that we use hyperboloidal time $s$ rather than Boyer-Lindquist time $t$, as appropriate for the multiscale expansion; see Sec.~\ref{sec:multiscale expansion}. In the derivation in Sec.~\ref{sec:localization}, all fields are evaluated on the particle's worldline, where $s$ reduces to $t$. In that case, we could more appropriately use $t$.

We next rewrite Eq.~\eqref{eq:hpp} as
\begin{equation}
   h^{\rm pp}_{\alpha\beta} = \frac{1}{\mu}\iint\! ds' d\omega\, e^{-i\omega(s-s')}\hat G^{\rm ret}_{\alpha\beta}{}^{\alpha'\beta'}(\omega,x^i,x_p'^i)\frac{\tilde p_{\alpha'}\tilde p_{\beta'}}{ds'/d\tilde\tau}.
\end{equation}
The factor $\hat G^{\rm ret}_{\alpha\beta}{}^{\alpha'\beta'}(\omega,x^i,x_p'^i)\frac{\tilde p_{\alpha'}\tilde p_{\beta'}}{ds'/d\tilde\tau}$ is triperiodic in the phases $\phiring^i(s',\e)=\xi^i(\e s')/\e$, where we again recall that $s=t$ on the particle's worldline. Hence, we can expand this factor in the discrete Fourier series 
\begin{multline}
\hat G^{\rm ret}_{\alpha\beta}{}^{\alpha'\beta'}(\omega,x^i,x_p'^i)\frac{\tilde p_{\alpha'}\tilde p_{\beta'}}{ds'/d\tilde\tau} \\
= \mu^2\sum_{\vec{k}}\hat G_{\alpha\beta}^{\vec{k}}[\omega,x^j,\Jring_j(\e s',\e),\e]e^{ik_i\xi^i(\e s')/\e}.
\end{multline}

Introducing $\tilde s:=\e s$ and $\tilde s':=\e s'$, we now write the integral as
\begin{multline}
   h^{\rm pp}_{\alpha\beta} = \mu\sum_{\vec{k}}\iint\! dt' d\omega\, e^{-i[\omega(\tilde s-\tilde s')-k_i\xi^i(\tilde s')]/\e}\\[-8pt]
   \times\hat G_{\alpha'\beta'}^{\vec{k}}[\omega,x^j,\Jring_j(\tilde s',\e),\e],
\end{multline}
which has a stationary point at $(\omega,\tilde s')=(\tilde\omega_{k}(\tilde s,\e),\tilde s)$, as in Sec.~\ref{sec:localization}. Applying the stationary phase approximation~\eqref{eq:2D stationary phase}, we obtain 
\begin{equation}
   h^{\rm pp}_{\alpha\beta} \approx 2\pi\mu\sum_{\vec{k}}\hat G_{\alpha'\beta'}^{\vec{k}}[\tilde\omega_k,x^j,\Jring_j(s),\e] e^{-ik_i\phiring^i(s)}
\end{equation}
plus higher-order terms with the same structure. 
If we now expand for small $\e$ at fixed $(\phiring^i,\Jring_i)$, we arrive at terms of the form~\eqref{multiscale h Fourier}. 

Similar analyses can be applied to higher-order fields and to $h^{\rm BH}_{\alpha\beta}$, thereby justifying the multiscale expansion of the metric and field equations. For the treatment of $h^{\rm BH}_{\alpha\beta}$, see Appendix~\ref{app:BH perturbation}. We then arrive at the identification 
\begin{equation}
\mathring h^{(1,\vec{k})}_{\alpha\beta}=2\pi\mu\,\hat G_{\alpha'\beta'}^{\vec{k}}(\Omega^i_{(0)}k_i,x^j,\Jring_j,0)+h^{\rm BH}_{\alpha\beta}.    
\end{equation}

\section{Stationary phase approximation of $\Hsix_{(1)}$}
\label{app:stationary phase H1}

In Sec.~\ref{sec:localized Hamiltonian for bound orbits} we sketch the derivation of a localized form of the pseudo-Hamiltonian. Here we detail a key step in the localization procedure: an application of the stationary-phase approximation.  

Our goal is to obtain an asymptotic approximation of the integrals (for each $k_i,k'_i$) appearing in the pseudo-Hamiltonian~\eqref{Ham modes}, which we rewrite here as 
\begin{multline}
    I_{\vec{k},\vec{k}'}=\int_{\mathbb{R}^2}d\tilde t'd\omega\, e^{-i[\xi_{k'}(\tilde t')+\omega (\tilde t-\tilde t')]/\e}\\[-8pt]
\times\hat G_{\vec{k}, \vec{k}'}[\omega,\Jring_i,\Jring'_i(\tilde t'),\e]. \label{stat phase integral}
\end{multline}
We specifically require an approximation to first subleading order in $\e$ as this will contribute to the pseudo-Hamiltonian at the same order as $\Hsix_2$. We closely follow the method of stationary phase for two-dimensional integrals presented in Ref.~\cite{Wong:doi:10.1137/1.9780898719260}. The phase function $f(\tilt',\omega)=\omega(\tilt'-\tilt)-\xi_{k'}(\tilt')$ has a single stationary point at $(\tilt',\omega)=(\tilt,\omega_{k'}(\tilt))$, which is a saddle point, giving this method particular utility.

The first step is to perform a change of variables such that the phase function is exactly quadratic in the integration variables. To achieve this, we expand $f$ in a Taylor series about the stationary point:
\begin{multline}
    f(\tilt',\omega)=-\xi_{k'}(\tilt)-\frac{1}{2}\xi''_{k'}(\tilt)(\tilt'-\tilt)^2+(\omega-\omega_{k'}(\tilt))(\tilt'-\tilt)\\-\sum_{n=3}^{\infty}\frac{\xi_{k'}^{(n)}(\tilt)}{n!}(\tilt'-\tilt)^n.
\end{multline}
Defining the variables
\begin{align}
    \zeta&:=(\tilt'-\tilt)-\frac{\omega-\omega_{k'}(\tilt)}{\xi''_{k'}(\tilt)},\\
    \eta&:=\omega-\omega_{k'}(\tilt)
\end{align}
brings our integral \eqref{stat phase integral} into the form
\begin{multline}
I_{\vec{k},\vec{k}'}=e^{-i\xi_{k'}(\tilt)/\e}\int_{\mathbb{R}^2}d\zeta\,d\eta\,\,g_{\vec{k},\vec{k}'}(\zeta,\eta,\e)\\\
    \times \exp\bigg\{\frac{i}{\e}\bigg(-\frac{1}{2}\xi_{k'}''(\tilt)\zeta^2+\frac{1}{2\xi_{k'}''(\tilt)}\eta^2 \\-\sum_{n=3}^{\infty}\frac{\xi_{k'}^{(n)}(\tilt)}{n!}\left(\zeta+\frac{\eta}{\xi_{k'}''(\tilt)}\right)^n\bigg)\bigg\},
\end{multline}
where $g_{\vec{k},\vec{k}'}(\zeta,\eta)=\hat{G}_{\vec{k},\vec{k}'}\bigl[\omega(\zeta,\eta),\Jring_i,\Jring'_i\left(\tilt'(\zeta,\eta)\right),\e\bigr]$. The dependence on $\Jring_i$, as well as on $\e$, is left implicit in the argument of $g_{\vec{k},\vec{k}'}$. Note that the new integration variables $\zeta$ and $\eta$ vanish at the stationary point.

We now make an additional change of variables such that the phase is an exact quadratic form. We define
\begin{align}
    u&:=\zeta\sqrt{1+P(\zeta,\eta)},\\
    v&:=\eta\sqrt{1+Q(\zeta,\eta)},
\end{align}
where $P(\zeta,\eta),\;Q(\zeta,\eta)$ are both regular double power series in $\zeta$ and $\eta$ satisfying $P(0,0)=Q(0,0)=0$. More explicitly, let $Z$ be the collection of terms in $f$ with a power of $\zeta$ greater than or equal to 2. Then 
\beq
P(\zeta,\eta)=-\frac{2Z(\zeta,\eta)}{\xi''_{k'}(\tilt)\zeta^2}-1, 
\eeq
and $Q(\zeta,\eta)$ is analogously defined in terms of the remainder of $f$, which is at least quadratic in $\eta$. After this change of variables our integral is given by
\begin{multline}
    I_{\vec{k},\vec{k}'}=e^{-i\xi_{k'}(\tilt)/\e}\int_{\mathbb{R}^2}du\,dv\;\bar{g}_{\vec{k},\vec{k}'}(u,v) e^{i F(u,v)/\e},
    \label{I of u,v}
\end{multline}
where 
\begin{equation}
  \bar{g}_{\vec{k},\vec{k}'}(u,v)=g_{\vec{k},\vec{k}'}\left(\zeta(u,v),\eta(u,v)\right)\left|\frac{\partial(\zeta,\eta)}{\partial(u,v)}\right|  
\end{equation}
and 
\beq
F(u,v)=-\frac{1}{2}\xi_{k'}''(\tilt)u^2 + \frac{v^2}{2\xi_{k'}''(\tilt)}. 
\eeq

In the above, we have allowed ourselves to perform Taylor series around the stationary point. The validity of this might be questioned because the integration range extends over all of $\mathbb{R}^2$. However, due to the oscillatory integrand, the integral \eqref{stat phase integral} receives a negligible contribution from the integrand outside a small neighborhood of the stationary point. To show this, we note that, by assumption (and assuming $\xi''_{k'}(\tilt)\neq 0$), both $\hat G_{\vec{k}, \vec{k}'}[\omega,\Jring_i,\Jring'_i(\tilde t'),\e]$ and $f(\tilt',\omega)$ are smooth with $\hat G_{\vec{k},\vec{k}'}$ decaying sufficiently rapidly to zero as $|\tilt'|,|\omega|\to\infty$. Given an open ball $B_{r'}(p^*)$ of radius $r'>0$ centered on the stationary point $p^*=(\tilt,\omega_{k'}(\tilt))$, we have that, for any $0<r<r'$, $p^*\in \overline{B_r(p^*)}\subset B_{r'}(p^*)$. We choose some such $r,r'$ and construct a smooth window function $W_r(\tilt',\omega)$ such that $W_r=1$ on $B_r(p^*)$ and $W_r=0$ on $\mathbb{R}^2\setminus B_{r'}(p^*)$. Then
\begin{align}
I_{\vec{k},\vec{k}'}&= \int_{\mathbb{R}^2}d\tilt'\,d\omega\;W_r(\tilt',\omega)\hat{G}_{\vec{k},\vec{k}'}(\tilt',\omega)e^{i f(\tilt',\omega)/\e}\nonumber\\
&+\underbrace{\int_{\mathbb{R}^2\setminus B_r}d\tilt'\,d\omega\;\tilde{G}_{\vec{k},\vec{k}'}(\tilt',\omega)e^{i f(\tilt',\omega)/\e}}_{R(\e)},
\end{align}
where $\tilde{G}=(1-W_r)\hat{G}$ is a smooth function that vanishes, along with its derivatives, on $\partial B_r(p^*)$, and as $|\tilt'|,|\omega|\to \infty$. We can approximate $R(\e)$ by integrating by parts. Since $\nabla f\neq0$ away from the stationary point, let $\vec{u}=\frac{\nabla f}{|\nabla f|^2}\tilde{G}$. Then 
\begin{equation}
    \nabla\cdot\left(\vec{u}\,e^{if/\e}\right)=\left(\nabla\cdot\vec{u}\right)e^{if/\e}+\frac{i}{\e}\tilde{G}e^{if/\e}.
\end{equation}
Hence using the divergence theorem
\begin{multline}
    R(\e)=-i\e\oint_{\partial B_r(p^*)}ds\;\vec{n}\cdot\vec{u}\,e^{if/\e}\\+i\e\int_{\mathbb{R}^2\setminus B_r(p^*)}\nabla\cdot\vec{u}\,e^{if/\e},
\end{multline}
where $\vec{n}$ is the \emph{inward}-pointing normal to $\partial B_r(p^*)$.
We now set $\tilde{G}_0=\tilde{G}$, $\vec{u}_0=\vec{u}$ and recursively define $\tilde{G}_n=\nabla\cdot \vec{u}_{n-1}$, $\vec{u}_{n}=\frac{\nabla f}{|\nabla f|^2}\tilde{G}_n$. Then we can repeatedly integrate $R(\e)$ by parts to obtain, for any $n\geq 1$,
\begin{equation}
    R(\e)=-\sum_{m=0}^{n-1}(i\e)^{m+1}\oint_{\partial B_r(p^*)}ds\;(\vec{n}\cdot \vec{u}_m)e^{if/\e}+R_n(\e),
\end{equation}
where
\begin{equation}
    R_n(\e)=(i\e)^n\int_{\mathbb{R}^2\setminus B_r(p^*)}\tilde{G}_n\,e^{if/\e}=\mathcal{O}(\e^n)\,\,\,\text{as}\;\e\to0.
\end{equation}
Moreover, $\tilde{G}$ and its derivatives all vanish on $\partial B_r(p^*)$, implying $R(\e)=R_n(\e)$. Since this is true  for all $n\geq1$, we conclude that $R(\e)$ is smaller than any power of $\e$ as $\e\to0$. From this argument we can conclude that, given any domain $D$ enclosing the stationary point at the origin (no matter how small: in fact we can give $\bar{g}$ as small a support as we require), we have
\begin{equation}
    e^{i\xi_{k'}(\tilt)}I_{\vec{k},\vec{k}'}=\int_{D}du\,dv\;\bar{g}_{\vec{k},\vec{k}'}(u,v)\,e^{i F(u,v)/\e}
\end{equation}
up to terms smaller than any power of $\e$, with the Taylor series about the stationary point now appropriately justified.

We can reduce this to an iterated integral via the co-area formula~\cite{EvansPDEs}: since $F(u,v)$ is smooth and for all $\tau\in \mathbb{R}\backslash\{0\}$ the set $\{(u,v)\in\mathbb{R}^2:F(u,v)=\tau\}$ is a smooth hyperbola in $\mathbb{R}^2$, assuming $\bar{g}_{\vec{k},\vec{k}'}$ is continuous and integrable  we have
\begin{equation}
    I_{\vec{k},\vec{k}'}= e^{-i\xi_{k'}(\tilt)}\int_{-\infty}^{\infty}h(\tau)e^{i\tau/\e}d\tau,
    \label{I_k}
\end{equation}
as $\e \to 0$, up to terms smaller than any power of $\e$, where
\begin{equation}
    h(\tau)=\int_{\{F(u,v)=\tau\}}\frac{\bar{g}_{\vec{k},\vec{k}'}(u,v)}{|\nabla F|}ds. \label{h of tau}
\end{equation}
Here, $ds$ is the element of arc length along the curve $\{F(u,v)=\tau\} $. In our justification via integration by parts, we noted that we can in general choose the support of $\bar{g}$ to be an arbitrarily small ball containing the origin. It thus suffices to obtain an asymptotic approximation for $h(\tau)$ as $\tau\to 0$, and use this to obtain an asymptotic expansion for \eqref{I_k}.

The curve of constant $\tau$ is a hyperbola in the plane, totally symmetric under reflection through the origin. We will perform the integral in the upper-right quadrant, noting the other portions of the hyperbola will contribute equally.

Let
\begin{align}
    u&=\left(-\frac{2\alpha}{\xi''_{k'}(\tilt)}\right)^{1/2}\cosh\beta, \\
    v&=\left(-2\xi_{k'}''(\tilt)\alpha\right)^{1/2}\sinh\beta,
\end{align}
such that $F(u,v)=\alpha=\tau$ along the curve (the sign of $\xi_{k'}''(\tilt)$ does not matter here, as we can simply change our parameterization above to account for this). We can then rewrite the integral \eqref{h of tau} as
\begin{equation}
    h(\tau)=\int_{\alpha=\tau}\bar{g}_{\vec{k},\vec{k}'}\left(u(\alpha,\beta),v(\alpha,\beta)\right)d\beta.
    \label{h alpha}
\end{equation}
Recall, by assumption we can restrict our domain of integration to a finite region enclosing the stationary point (which is now located at the origin), so we cut off this integral at $u=d$, which intersects the path of integration at  $\beta=\beta_{\text{max}}(\tau)=\cosh^{-1}\left[\left(-\xi''_{k'}(\tilt)/2\tau\right)^{1/2}d\right]$. We further expand $\bar{g}_{\vec{k},\vec{k}'}$ in a Taylor series in $(u,v)$, giving
\begin{multline}
    h(\tau)=4\sum_{m,n\geq 0} \frac{\bar{g}_{\vec{k},\vec{k}'}^{2m,2n}}{(\xi_{k'}''(\tilt)/2)^m(1/2\xi_{k'}''(\tilt))^n}\tau^{m+n} \\
    \times\int_0^{\beta_{\text{max}}(\tau)}d\beta\,
    \cosh^{2m}\beta\,\sinh^{2n}\beta,
    \label{h as a series}
\end{multline}
where $g_{i,j}$ is the coefficient of $u^iv^j$ in the Taylor expansion of $\bar{g}_{\vec{k},\vec{k}'}$. The factor of four on the right-hand side accounts for the fact that the integral in Eq.~\eqref{h of tau} includes the contribution from the pieces of the hyperbola lying in all four quadrants of the $(u,v)$-plane.
This also ensures we only receive contributions to $h(\tau)$ from the terms even in both $u$ and $v$ in the Taylor expansion of $\bar{g}$.

The integrals in Eq.~\eqref{h as a series} can be evaluated analytically~\cite{Gradshteyn:1943cpj}.  After some algebra and throwing away smooth terms that do not contribute to the asymptotic approximation, we have
\begin{equation}
 h(\tau)\sim 4\sum_{j=0}^{\infty}c_{j}\tau^j \log(\pm \tau)\;\;\;\; \text{as }\tau\to 0^{\pm}, 
 \label{h asym}
\end{equation}
where 
\begin{multline}
    c_{j}=\sum_{m+n=j}\frac{(-1)^{n+1}}{2^{2m+2n+1}}\frac{2m!2n!}{m!n!(m+n)!}\\
    \times\frac{\bar{g}_{\vec{k},\vec{k}'}^{2m,2n}}{(-\xi_{k'}''(\tilt)/2)^m(-1/2\xi_{k'}''(\tilt))^n}.
\end{multline}
Finally we consider the Fourier integral \eqref{I_k},
\begin{align}
   e^{i\xi_{k'}(\tilt)} I_{\vec{k},\vec{k}'}&= \int_{-\infty}^{\infty}h(\tau)e^{i\tau/\e}d\tau \nonumber\\
    &=\int_{0}^{\infty}h(\tau)e^{i\tau/\e}d\tau+\left(\int_0^{\infty}h(-\tau)e^{i\tau/\e}d\tau\right)^*\nonumber\\
    & \eqqcolon I_{+}(\e)+(I_{-}(\e))^*,
\end{align}
where 
\begin{multline}
    I_{\pm}(\e)\sim4\sum_{\nu=0}^{\infty}(\pm1)^\nu c_{\nu} \Gamma(\nu+1)e^{i(\nu+1)\pi/2} \\
    \times \left\{\frac{\Gamma'(\nu+1)}{\Gamma(\nu+1)}+i\frac{\pi}{2}+\log\e\right\}\e^{\nu+1},
\end{multline}
as $\e\to 0^+$, where $\Gamma'/\Gamma$ is the digamma function. Combining both terms we find that both the digamma and log terms cancel and we obtain the final asymptotic approximation for $I_{\vec{k},\vec{k}'}$:
\begin{equation}
    I_{\vec{k},\vec{k}'}\sim4\pi i e^{-i\xi_{k'}(\tilt)}\sum_{j=0}^{\infty}c_j e^{i(j+1)\pi/2}j!\e^{j+1}
    \label{final asymptotics}
\end{equation}
as $\e\to 0^+$, which is directly calculable in terms of derivatives of $\hat{G}_{\vec{k},\vec{k}'}$ and $f$.
\section{Averaged redshift}
\label{sec:averaged redshift}

Reference~\cite{Akcay:2015pza} contains a simple proof of Eq.~\eqref{<z>} for equatorial orbits. Here we show it remains true for generic orbits.

As noted below Eq.~\eqref{<z>}, it suffices to show that $\langle z_0\rangle=\langle z_{(0)}\rangle+{\cal O}(\e^2)$, where $z_0=d\tau/dt$ and $\tau$ is proper time as measured in the background metric. $z_{(0)}$ is also equal to $d\tau/dt$, but evaluated along the geodesic that has the same frequencies $\Omega^i$ as the accelerated orbit. The essential idea in our proof is that $z_0(x^i_p,\dot x^i_p)$, given in Eq.~\eqref{z0}, is the Lagrangian for a test mass in the background spacetime. It therefore satisfies the Euler-Lagrange equations when evaluated along a geodesic, a fact we can use in evaluating $\langle z_0\rangle$.

We expand $x^i_p$ and $\dot x^i_p$ around their geodesic values, 
\begin{align}
x^i_p(\mathring\varphi^j,\Omega^j,\epsilon) &= x^i_{(0)}(\mathring\varphi^j,\Omega^j) + \epsilon x^i_{(1)}(\mathring\varphi^j,\Omega^j)+{\cal O}(\e^2),\label{x expansion}\\
\dot x^i_p(\mathring\varphi^j,\Omega^j,\epsilon) &= \dot x^i_{(0)}(\mathring\varphi^j,\Omega^j) + \epsilon \dot x^i_{(1)}(\mathring\varphi^j,\Omega^j)+{\cal O}(\e^2).\label{xdot expansion}
\end{align}
Then 
\beq\label{<z0>=...}
\langle z_0\rangle = \langle z_{(0)}\rangle + \epsilon \left\langle x^i_{(1)}\frac{\partial z_{(0)}}{\partial x^i_{(0)}} + \dot x^i_{(1)}\frac{\partial z_{(0)}}{\partial\dot x^i_{(0)}}\right\rangle+{\cal O}(\e^2).
\eeq
Using 
\begin{equation}\label{integration by parts}
\left\langle \dot x^i_{(1)}\frac{\partial z_{(0)}}{\partial\dot x^i_{(0)}}\right\rangle = - \left\langle x^i_{(1)}\frac{d}{dt}\frac{\partial z_{(0)}}{\partial\dot x^i_{(0)}}\right\rangle,  
\end{equation}
we reduce the order-$\epsilon$ term in Eq.~\eqref{<z0>=...} to 
\beq
\left\langle x^i_{(1)}\left(\frac{\partial z_{(0)}}{\partial x^i_{(0)}} - \frac{d}{dt}\frac{\partial z_{(0)}}{\partial\dot x^i_{(0)}}\right)\right\rangle.
\eeq
This vanishes by virtue of the Euler-Lagrange equations, establishing the desired result, 
\beq
\langle z_0\rangle(\Omega^i,\e) = \langle z_{(0)}\rangle(\Omega^i) + {\cal O}(\e^2).
\eeq

One might initially presume that this proof is valid even if we do not expand at fixed values of the frequencies. However, the presumption is incorrect; the proof would \emph{not} yield $\langle z_0\rangle = \langle z_{(0)}\rangle + {\cal O}(\e^2)$ if we expanded at fixed $\Jring_i$ rather than fixed $\Omega^i$, for example. To see why the equality only holds for an expansion at fixed frequencies, note that Eq.~\eqref{integration by parts} relies on $\dot x^i_{(1)} = dx^i_{(1)}/dt$ (as well as on $d/dt = \Omega^j \partial/\partial \phiring^j$, which holds in the conservative sector). But in Eq.~\eqref{xdot expansion}, $\dot x^i_{(1)}$ is defined as the order-$\e$ term in $dx^i/dt$, not as $dx^i_{(1)}/dt$. Generically, differentiating Eq.~\eqref{x expansion} yields
\begin{equation}
\dot x^i_p = \Omega^j \frac{\partial x^i_p}{\partial\phiring^j} = \Omega^j \frac{\partial x^i_{(0)}}{\partial\phiring^j}+\e\Omega^j \frac{\partial x^i_{(1)}}{\partial\phiring^j}+{\cal O}(\e^2).
\end{equation}
If we expand at fixed $\Omega^i$, the first term here is $\dot x^i_{(0)}$ and the second is $\dot x^i_{(1)}$. But for any other expansion, $\Omega^i$ itself is expanded, as in  $\Omega^i_{(0)}(\Jring_j)+\e\Omega^i_{(1)}(\Jring_j)+\ldots$ (for example). Equation~\eqref{integration by parts} then picks up an additional term on the right-hand side, leading to an order-$\e$ term in $\langle z_0\rangle$.

\section{Action-angle variables in eight versus six phase-space dimensions}
\label{sec:8D vs 6D AA variables}

In this appendix we explain why the ``renormalized'' action variables in Ref.~\cite{Fujita:2016igj} correspond to our canonical action variables $\Jring_i$, as highlighted in Sec.~\ref{sec:conservative sector}. Our conclusion is that the renormalization arises from the fact that the 8D system is \emph{constrained}. Enforcing the constraint induces the renormalization.

Reference~\cite{Fujita:2016igj} formulated the Hamiltonian dynamics on the 8D phase space with coordinates $(x^\alpha,\tilde p_\alpha)$ and time variable $\tilde\tau$. On this phase space, they worked with action-angle variables $(w^\alpha,J_\alpha)$ satisfying
\begin{align}
    \frac{dw^\alpha}{d\tilde\tau} &= \frac{\partial H_{\rm 8D}}{\partial J_\alpha} = \omega^\alpha(J_\beta,\e),\\
    \frac{dJ_\alpha}{d\tilde\tau} &= -\frac{\partial H_{\rm 8D}}{\partial w^\alpha} = 0.    
\end{align}
with $H_{\rm 8D} = H_{(0)}(J_\alpha) + \frac{\mu}{2}\left\langle[H_{(1)}]\right\rangle_{\tau}$. 
However, this system also has the constraint that $H_{\rm 8D}=K(J_i,\e)$ on shell, where in analogy with Eq.~\eqref{Hamiltonian - final}, $K$ is given by~\cite{Fujita:2016igj}
\begin{equation}
    K = -\frac{\mu}{2}\left(1 +\e\left\langle[H_{(1)}]\right\rangle_{\tau} +{\cal O}(\e^2)\right)
\end{equation}
with $H_{(1)}$ given by Eq.~\eqref{H1(x,p)}. Here the $\tau$ average is related to our angle average by~\cite{Isoyama:2018sib}
\begin{equation}\label{tau average}
\langle\cdot\rangle_\tau = \frac{\langle z \cdot\rangle}{\langle z\rangle}.    
\end{equation}

The equations of motion above do not automatically enforce the on-shell condition $H_{\rm 8D}(J_t,J_i)=K(J_i)$. Instead, the condition specifies a submanifold on which $J_t=J_t(J_i,\e)$, and we must restrict the dynamics to that submanifold. As in Sec.~\ref{sec:6D pseudo-Hamiltonian}, we can obtain the equations of motion within the submanifold by differentiating the on-shell condition. Differentiating with respect to $J_i$, appealing to the chain rule, using the 8D Hamilton's equations, and multiplying by $z=d\tilde\tau/dt$, we obtain
\begin{equation}\label{dwdt}
    \frac{dw^i}{dt} = \beta^i z,
\end{equation}
with 
\begin{equation}
\beta^i:=\left(-\frac{\partial J_t}{\partial J_i}\omega^t + \frac{\partial K}{\partial J_i}\right).    
\end{equation}
We also trivially have $dJ_i/dt=0$.

If $K$ were independent of $J_i$, then we would be able to naturally identify $-J_t$ as the 6D Hamiltonian. This was the case in Sec.~\ref{sec:8D pseudo-Hamiltonian}, where the on-shell value of the 8D pseudo-Hamiltonian was a constant (equal to $-\mu/2$)  on the 8D phase space, and we were able to use $-\tilde p_t$ as a 6D pseudo-Hamiltonian. However, $K$ is \emph{not} independent of $J_i$. Comparing Eq.~\eqref{H1(x,p)} to Eq.~\eqref{H1=hRuu}, we can hence observe
\begin{equation}
    \left\langle[H_{(1)}]\right\rangle_{\tau} = \frac{1}{\langle z\rangle}\langle \Hsix_{(1)}\rangle.
\end{equation}
Noting that $\omega^t = \langle\frac{dt}{d\tilde\tau}\rangle_\tau$, we see that Eq.~\eqref{tau average} also implies 
\beq\label{omega^t}
\omega^t=1/\langle z\rangle.
\eeq

In addition to $-J_t$ not being the 6D Hamiltonian, the variables $w^i$ are also clearly not action angles in the 6D phase space. If they were, $dw^i/dt$ could not contain the oscillatory factor $z$ in Eq.~\eqref{dwdt}. To obtain action angles, we must perform an averaging transformation,
\begin{equation}\label{phi to w}
    \phiring^i = w^i - \Delta w^i(\phiring^j,\Jring_i,\e),
\end{equation}
where $\Delta w^i$ is $2\pi$-periodic in $\phiring^j$. We can find $\Delta w^i$ straightforwardly by differentiating Eq.~\eqref{phi to w} with respect to $t$, substituting Eq.~\eqref{dwdt}, decomposing $z$ into $z=\langle z\rangle +z_{\rm osc}$, and choosing $\Delta w^i$ to eliminate all oscillations from $d\phiring^i/dt$. We are then left with
\begin{equation}\label{dphidt from 8D}
    \frac{d\phiring^i}{dt} = \beta^i\langle z\rangle.
\end{equation}

It is now reasonably straightforward to show that Eq.~\eqref{dphidt from 8D} is equivalent to Hamilton's equation $d\phiring^i/dt = \partial H_{\rm 6D}/\partial\Jring_i$, where
\begin{align}
    H_{\rm 6D} &= -J_t\left(1 -\frac{\e}{2}\left\langle[H_{(1)}]\right\rangle_{\tau}\right),\label{H renorm}\\
    \Jring_i &= J_i\left(1 -\frac{\e}{2}\left\langle[H_{(1)}]\right\rangle_{\tau}\right).\label{J renorm}
\end{align}
These are the ``renormalized'' action variables from Ref.~\cite{Fujita:2016igj}, where they were denoted $-\tilde J_t$ and $\tilde J_i$. To confirm that $d\phiring^i/dt = \partial H_{\rm 6D}/\partial\Jring_i$, we note this is true if and only if $\beta^i\langle z\rangle = \partial H_{\rm 6D}/\partial\Jring_i$. Multiplying that equation by the Jacobian $\partial\Jring_i/\partial J_j$ and exploiting Eq.~\eqref{omega^t}, we establish that the equation is true if and only if
\begin{equation}
    \left(J_i\frac{\partial J^{(0)}_t}{\partial J_i}-\mu\langle z_0\rangle -J^{(0)}_t \right)\frac{\partial\left\langle[H_{(1)}]\right\rangle_{\tau}}{\partial J_i}=0.
\end{equation}
We can write this in terms of the geodesic energy $E_{(0)} =-J^{(0)}_t$ by appealing to the identity $\langle z_0\rangle = \partial E_{(0)}/\partial\mu$~\cite{Fujita:2016igj}. The above condition then becomes  
\begin{equation}\label{E condition}
    -J_i\frac{\partial E_{(0)}}{\partial J_i}-\mu\frac{\partial E_{(0)}}{\partial \mu} +E_{(0)} = 0.
\end{equation}
Finally, we note that $E_{(0)}(J_i,\mu) = \mu e_{(0)}(J_i/\mu)$, where $e_{(0)}$ is the specific geodesic energy; this is a trivial consequence of the fact that the specific geodesic energy can only depend on the specific actions. Hence, $E_{(0)}$ is a homogeneous function of order 1 in the variables $(J_i,\mu)$. Euler's homogeneous function theorem then implies that Eq.~\eqref{E condition} is satisfied, completing the proof that $d\phiring^i/dt = \partial H_{\rm 6D}/\partial\Jring_i$.

Here we have only shown that the renormalizations~\eqref{H renorm} and \eqref{J renorm} are a consequence of enforcing the constraint. We have not yet shown that Eqs.~\eqref{H renorm} and \eqref{J renorm} are equivalent to our expressions for $H_{\rm 6D}$ and $\Jring_i$ in the body of this paper. As noted in Sec.~\ref{sec:conserved quantities}, the equivalence between Eq.~\eqref{J renorm} and our Eq.~\eqref{Jring(Omega)} for $\Jring_i$ follows from Eq.~(8.11) of Ref.~\cite{Fujita:2016igj} and our Eq.~\eqref{<z>}. The equivalence between Eq.~\eqref{H renorm} and our Eq.~\eqref{E first law} for $H_{\rm 6D}$ follows from the same equations together with Eq.~(8.3) of Ref.~\cite{Fujita:2016igj}.

Although we have enforced the constraint by explicitly solving the constraint equation and eliminating redundant variables, we note that this is not the only way to deal with the constraint. One could instead consistently work in eight dimensions and formulate the 8D dynamics to properly account for the constraint. In that case we would find that, even in eight dimensions, the ringed variables $\Jring_\mu$ are the correct actions (with $\Jring_t = -E$). In this sense, the explicit reduction to 6D and the use of $t$ as the time parameter are not essential in reaching this conclusion. A future paper will analyze the 8D dynamics from a constrained-Hamiltonian perspective.


\bibliography{main.bib}

\end{document}